\documentstyle[prd,aps,epsf,floats,eqsecnum,amsfonts,amssymb,preprint,tighten]{revtex}
\newcommand{\beq}{\begin{equation}}\newcommand{\eeq}{\end{equation}}
\newcommand{\beqa}{\begin{eqnarray}}\newcommand{\eeqa}{\end{eqnarray}}
\newcommand{\bk}{{\bf k}}
\newcommand{\drm}{d }
\newcommand{\tx}{\text}
\draft
\begin{document}
\title{Calculation of the Casimir energy \\
at zero and finite temperature: some recent results}
\author{V.V.~Nesterenko$^a$\footnote{E-mail: nestr@thsun1.jinr.ru}}
\author{G.~Lambiase$^{b,c}$\thanks{E-mail: lambiase@sa.infn.it}}
\author{G.~Scarpetta$^{b,c,d}$\footnote{E-mail: scarpetta@sa.infn.it}}
\address{$^a$Bogoliubov Laboratory of Theoretical Physics \\
 Joint Institute for Nuclear Research - Dubna, Russia}
\address{$^b$Dipartimento di Fisica "E.R. Caianiello"
 Universit\'a di Salerno, 84081 Baronissi (Sa), Italy.}
\address{$^c$INFN - Gruppo Collegato di Salerno, Italy.}
 \address{$^d$International Institute for Advanced Scientific Studies, Vietri sul Mare (SA),
Italy}
\date{\today}
\maketitle

\begin{abstract}
This survey summarizes briefly results obtained recently in the
Casimir  energy studies devoted to the following subjects:
   i) account of the material characteristics of the media in
calculations of the vacuum energy (for example, Casimir energy of a
dilute dielectric ball);
   ii) application of the spectral geometry methods for investigating
the vacuum energy of quantized fields with the goal to gain some
insight, specifically, in the geometrical origin of the divergences
that enter the vacuum energy and to develop the relevant
renormalization procedure;
  iii) a universal method for calculating the high temperature
dependence of the Casimir energy  in terms of heat kernel
coefficients. A special attention is payed to the mathematical tools
applied in this field, namely, to the spectral zeta function method
and heat kernel technique.
\end{abstract}

\section{Introduction}
\setcounter{equation}{0}
In 1948 Casimir~\cite{Casimir} proceeding from the general
principles of quantum theory of electromagnetic field has shown
that two uncharged perfectly conducting plates, placed in vacuum,
should attract each other with the force per unit area
\begin{equation}
\label{1-1} F= -\frac{\pi^2c\hbar}{240 a^4}{,}
\end{equation}
where $a$ is the distance between the plates. Later this
prediction was verified experimentally~\cite{Sparnaay}. For
dielectric bodies the Casimir force was also measured with
reasonable agreement to the theory~\cite{dielectric}.

 The Casimir force is very weak, however it increases rapidly as the
separation $a$ decreases and it becomes measurable when $a\sim
1~\mu$m or less. For plates 1~{cm} in area with
$a=0.5\,\mu$m, the Casimir force  is about $0.2$~{dyn}.
Firstly  the Casimir  force has been measured, due to technical
reasons, between a conducting plane and a sphere  with progressively
higher precision  using torsion balances~\cite{Lamoreaux}, atomic
force microscopes~\cite{Mohideen}, and capacitance
bridges~\cite{chan1}. Only recently this force was measured between
parallel metallic surfaces in the range 0.5 -- 3.0~{$\mu$m} at the
15\% precision level using cantilevers~\cite{roberto}. The Casimir
force may be relevant in nanotechnology~\cite{chan1,Serry,chan2}. At
a separation of 10~{nm}, this force is  $\sim $ 1\,{atm}. Now the
temperature dependence of the Casimir forces becomes a feasible task
for experiment~\cite{thermal,confront}. The juxtaposition of the
experimental results and theoretical calculations with detailed
analysis of errors and precision achieved in this field can be found
in ref.~\cite{confront}. An interesting proposal was made recently to
measure {\it variations} of the Casimir energy by making use a
superconducting cavity~\cite{GEsp}.

 The Casimir effect is, in fact, the first {\it macroscopic} effect predicted
in the framework of the relativistic quantum field theory, namely,
in quantum electrodynamics. The macroscopic feature of the Casimir
effect implies that it is observed in the experiments with
macroscopic bodies, instead of the systems of atomic size. In this
respect, it is analogous to superconductivity and superfluidity.

This effect was always an interesting subject of both theoretical
and experimental studies. Now the Casimir effect is treated in a
more general way, namely, as the influence on the physical
characteristics of a quantum field system of the external
conditions (boundary conditions, background fields and so on) that
lead to the restriction of the configuration space of the system
under study.

Calculations of the Casimir forces for configurations more
complicated than two parallel plates proved to be difficult, and
the progress in this field is rather slow. Till now one has no
intuition even as to whether the force should be attractive or
repulsive for any given boundary geometry, not to mention the
estimation of its value.

In the case of perfectly conducting surfaces placed in vacuum the
Casimir energy is calculated exactly, safe for parallel
plates~\cite{Casimir}, for
spheres~\cite{Boyer,Davies,BD,MRS,piro,bowers} and for circular
infinite cylinders~\cite{MNN,deraad,romeo}.

Dielectric properties of the media separated by plane boundaries did
not add new mathematical difficulties~\cite{Lif,Kampen,SDeRM} as
compared with the Casimir pioneer paper~\cite{Casimir}. However the
first result on the calculation of the Casimir energy for the
non-flat boundaries was obtained only in 1968. By computer
calculations, lasted 3 years, Boyer found the Casimir energy of a
perfectly conducting spherical shell~\cite{Boyer}. Account of
dielectric and magnetic properties of the media in calculations of
the vacuum energy for nonflat interface leads to new
difficulties or, more precisely, to a new structure of divergencies.

     The calculation of the Casimir energy in a special case, when
both the material media have the same velocity of light, turns out
to be, from the mathematical standpoint, exactly the same as for
perfectly conducting shells placed in vacuum and having the shape
of the interface between these
media~\cite{MNN,BNP,NP-cyl,BC,Brevik+Nyland,BC1,BC2,BC3,BY,Klich,%
Klich-2,Klich-Romeo,Klich-FMR}.

    The first attempt to calculate  the Casimir energy of a
dielectric compact ball has been undertaken by Milton in
1980~\cite{Milton}. And only just recently the final result was
obtained for a {\it dilute} dielectric ball at
zero~\cite{MNg2,Barton,BMM,BM,LSN,Mar,HB-stat,HBA-stat-QFT} and
finite~\cite{NLS,Barton-T} temperature.

Situation with a dielectric compact cylinder proves to be much more
complicated. The main reason of this is that here there is no
decoupling between transverse-electric and transverse-magnetic
modes~\cite{MNN}. It was shown that the summation  of van der Waals
interactions inside a purely dielectric cylinder in the
dilute-dielectric approximation lead to a surprising null
result~\cite{MNN}. First it was derived independently  by Romeo and
Milonni (unpublished calculations) and by Barton in
ref.~\cite{Barton-cyl}. Bordag and Pirozhenko~\cite{BP} have found
the first few heat kernel coefficients for this configuration. Their
results imply, in particular, that zeta function technique should
supply a finite value for vacuum energy of a dielectric cylinder in
the {\it dilute} approximation. And only very recently Cavero-Pelaez
and Milton~\cite{C-P+Milton} have shown rigorously (in quantum field
theory approach) that the Casimir energy of an infinite circular
dielectric cylinder vanishes through the second order in the
deviation of the permittivity from its vacuum value. This result is
of a five-year-long effort! The same vanishing value for this energy
was derived anew in paper~\cite{RomeoMilton} by making use of the
mode summation method.

For the force between a plane and a sphere the exact result is not
known, and only an estimation valid when both are close enough is
available. This estimation is based on the so called proximity
theorem~\cite{derjaguin,Gies}. Similar estimations have been
recently obtained for the force between two
spheres~\cite{schaden1}. The problems of two concentric
spheres~\cite{HBA-stat-QFT,saha} and two concentric
cylinders~\cite{sahats} have also been considered recently. For
both cases expressions for the Casimir energy have been obtained
using the Abel-Plana formalism. However a detailed numerical
calculation and analysis of the results are still missing. The
authors of ref.~\cite{Ahmedov-close} studied the Casimir
interactions, due to the massless scalar field fluctuations, of
two surfaces which are close to each other. Two close co-axial
cylinders, co-centric spheres co-axial cones and two tori were
considered. The first correction to the parallel plates result was
calculated.

For dielectric-diamagnetic media with a continuous velocity of
light at the interface the exact results are known for a
sphere~\cite{BNP,BC,BC1,BC2,BC3,BY,Klich,Klich-2,Klich-Romeo,Klich-FMR}
and cylinder~\cite{MNN,NP-cyl,Klich-Romeo}.

For  arbitrary dielectric media the exact results are obtained for
parallel plates (Lifshitz formula~\cite{Lif,Kampen,SDeRM}), for a
dilute dielectric ball~\cite{MNg2,Barton,BMM,BM,LSN,Mar,NLS} and for
a dilute dielectric cylinder~\cite{C-P+Milton,RomeoMilton}. A single
surface immersed in an inhomogeneous medium also experiences the
Casimir force~\cite{Jaffe-1}.

Recently there appear  comprehensive
reviews~\cite{Milton-JPhys,Bordag} and excellent
book~\cite{Milton-book}  which are devoted to the Casimir effect.
Previous survey publications in this
field~\cite{MT,GMM,ElRomeo,Milonni-book,Kardar,PMG} remain valuable. Many
references concerning vacuum energy can be found
in~\cite{Roberts}. Of special interest are the Proceedings of
regularly organized, every 3 years since 1989, Workshops on
Quantum Field Theory Under the Influence of External
Conditions~\cite{Bordag-W3,Bordag-W4,Bordag-W5,Bordag-W6}.

In this situation a new survey in this field should be naturally
concerned with particular problems to investigation of which a
certain contribution has been done by present authors. First of
all, we are going to give a lucid and at the same time
comprehensive introduction into the mathematical methods that are
used in current Casimir calculations. We hope that the review will
provide a sufficient material in order to apply these methods to
coinciding problems in diverse branches of theoretical physics.

The layout of the review is as follows. In sect.\ \ref{ZPE} we
consider the basic  terms  used in the Casimir studies, specifically,
the notion of  zero point energy. To our mind employment of this term
is not only convenient but also physically justified in this field.
In sect.\ \ref{RVE} the mathematical methods applied in  Casimir
calculations are considered, namely, the spectral zeta functions and
heat kernel technique. In sect.\ \ref{SZF+VE} the spectral zeta
functions for simplest boundary configurations are calculated and the
relevant Casimir energies are found. Section~\ref{CEDB} is devoted to
calculation of the vacuum energy of electromagnetic field connected
with a dielectric ball. This problem proved to be nontrivial one. We
are trying to reveal the principal drawbacks encountered here and
summarize the experience obtained in these calculations. The physical
origin of the divergencies in vacuum energy of quantized field and
their  relation to the boundary properties in considered in sect.\
\ref{N-S}. In sect.\ \ref{HT} a universal method of constructing the
high temperature asymptotics in the Casimir calculations is
discussed. It is substantially based on the heat kernel technique,
more precisely, it uses the coefficients of the asymptotic expansion
of the heat kernel. In sect. \ref{Con} an attempt is undertaken to
summarize the lessons gained in the course of quiet long studies in
this field and to formulate the present status of our physical
understanding  of  the Casimir effect. In the Appendices \ref{appA},
\ref{appb} and \ref{appc} the mathematical details of the
calculations involved are placed.

\section{Zero point energy in quantum mechanics and in quantum field theory}

\setcounter{equation}{0}

\label{ZPE}

The quantum theory forbids the particle to posses
simultaneously a definite value of its coordinate and momentum.
This results, in particular, in the following. The particle placed
in a smooth potential with a local minimum cannot acquire the
value of energy which corresponds to the potential minimum. In
fact, smooth potential can be approximated around its minimum by
the potential of the harmonic oscillator with the known spectrum
\begin{equation}
\label{2-1} E_n=\hbar \omega\left ( n+\frac{1}{2}\right ){,} \quad
n=0, 1,2, \ldots {.}
\end{equation}
Thus the particle cannot posses the energy less than $E_0=\hbar
\omega /2$, which has obtained the name of {\it zero point energy}. It
is interesting that the notion of the zero point energy has first
appeared before the development of quantum mechanics, in Planck
paper of 1912, where he has suggested the ``second theory'' of the
black body spectrum~\cite{Planck}. By making use of zero point
energy, Einstein and Stern derived the Planck distribution
following practically the classical consideration~\cite{ES}.

  The prediction of zero point energy, obtained in the framework of
the quantum mechanics  dealing with the systems of a {\it finite} number of degrees
of freedom, is directly verified in experiment. For example, by
making use of the Roentgen rays or neutrons scattering by a
crystal lattice or by liquid helium one can convince that, when
temperature tends to zero, the atoms in these systems indeed
occupy the state with the energy $E_0$ and they cannot possess
lower energy. In actual fact these experiments provide the
Debye-Waller factor~\cite{DW}, calculation of which in the
framework of the quantum mechanics takes into account the zero
point oscillations of the atoms.

    A direct justification of the reality of zero point oscillations
(or zero point energy), in the case of a quantum mechanical
systems, is afforded by the observation of vibration spectra of
diatomic molecules. It is interesting that it has been also done
prior to the development of quantum mechanics~\cite{Mulliken}.

The zero point energy is also manifested in the temperature
dependence of the specific heat of rigid bodies. For example, in
the model suggested by Einstein~\cite{Einstein} the crystal
lattice is considered as the system of $3N$ independent
oscillators, $N$ being the total number of lattice sites. All the
oscillators have the same frequency (a single phenomenological
parameter of the model). In this model the area between the curve
$c(T)$ describing the temperature dependence of the specific heat
and the straight line $y=c(\infty)$ is proportional to the zero
point energy of all the oscillators. In a more realistic model of
specific heat of crystal bodies proposed by Debye~\cite{Debye} the
zero point energy is also taken into account without problem.

Drastically different situation concerning the zero point energy
arises when one considers the system of {\it infinite} number of
oscillators. It is this case that is encountered in  quantum field
theory. The Hamiltonian of the free (noninteracting) quantum field
is an infinite sum of the Hamiltonians of the harmonic
oscillators. When the field is considered in unbounded space this
sum is substituted by the integral
\begin{equation}
\label{2-2} H=\frac{1}{2}\int d^3{\bf k}\,
\omega(\bk)[a^\dagger (\bk)a(\bk)+ a(\bk) a^\dagger(\bf k)]\,{,}
\end{equation}
where the frequency $\omega(\bk)$ is equal to the energy of the
particle with momentum $\bk$. For simplicity we are considering
uncharged  spinless particles (bosons)
\begin{equation}
\label{2-2a} [a(\bk),a^\dagger(\bk')]= \delta^{(3)} (\bk -\bk'){.}
\end{equation}
 The equidistant energy spectrum of the harmonic oscillator
(\ref{2-1}) enables one to interpret the quantum field dynamics in
terms of particles, namely, the transition of one of the
oscillators in the Hamiltonian (\ref{2-2}) to the next higher
(lower) level is treated as creation (annihilation) of the
particle with the energy $\omega(\bk)$. In the framework of this
interpretation, the zero point energy of quantum field $E_0$ is
treated as the {\it vacuum energy}, because in this case there are no
excited oscillators (all them occupy  the energy level $E_0$).
Hence there are no real particles. For any physically acceptable
dispersion law~$\omega(\bk)$ this energy is, obviously, unbounded
\begin{equation}
\label{2-3} E_0=\frac{1}{2}\int d^3\bk \, \omega (\bk) = \infty{.}
\end{equation}
Naturally,  question arises here how to treat this energy.

Divergences are typical for any realistic quantum field theory
because the Hamiltonian of  quantum field (see, for example, the
free field Hamiltonian (\ref{2-2})) involves the oscillators with
arbitrary large frequencies  $\omega (\bk )$ when $k\to \infty $.
Physically it is clear that such quanta cannot be relevant in any
concrete calculation. However, quantum field theory  does not
provide us with the mechanism that could suppress the contribution
of such quanta, and the problem of divergencies in
relativistic quantum field theory requires suitable renormalization procedures.

If the energy in the system under consideration is conserved (for
example, the scattering of elementary particles possessing the
same total energy in the initial and final states is considered)
then the infinite vacuum energy $E_0$ can be ignored by shifting
the origin of the energy scale by $E_0$. Formally this subtraction
is accomplished by transition to the {\it normal ordering} of the
operators in the field Hamiltonian~(\ref{2-2})
\begin{equation}
\label{2-4} H= \frac{1}{2}\int d^3\bk  \,
\omega(\bk)[:a^\dagger(\bk) a(\bk):+:a(\bk)a^\dagger(\bk):]=\int d
^3\bk\;\omega (\bk)\,a^\dagger(\bk) a(\bk){.}
\end{equation}

However when the  energy of the quantum field is not conserved
then such a procedure of removing $E_0$ cannot be justified. This
issue is of a special importance  in the gravitation theory
because the energy of matter fields is the source of the
gravitation field.

In quantum field theory there is only one receipt of a consistent,
from the mathematical standpoint, treatment of the divergencies.
It is the renormalization procedure, i.e., transition from the
initial parameters of the theory, which  turn out to be, as a
rule, infinite and therefore unobservable, to finite  physical
parameters~\cite{MatthewsSalam,Gunn}.  On principle, the
renormalization procedure in  quantum field theory should be done
always independently of the presence of divergencies. If the
theory does have the infinities then they are removed in the
course of renormalization ``by the way''. Certainly, it is true only
for renormalizable field theories.

A consistent renormalization procedure is developed only for the
scattering processes in unbounded Minkowski space (for example,
the Bogoliubov $R$ operation). If the field occupies a bounded
region in space, the relevant renormalization scheme is absent.
For example, the renormalizability of quantum electrodynamics
considered in a compact space region, by imposing the
corresponding boundary conditions, is not proved explicitly.
However there is no reason to suspect this renormalizability.
Indeed, the boundary conditions, confining the fields to a compact
region, arise due to transition from a microscopic description of
the problem at hand (in terms of charges and electromagnetic field
by making use of standard renormalizable quantum electrodynamics
in unbounded space) to a macroscopic description by imposing
boundary conditions on electromagnetic field and introducing
phenomenological characteristics (permittivity and permeability)
specifying the materials the boundary made up.

With  lacking  general renormalization algorithm for nontrivial
boundary conditions, one has to develop this procedure for each
concrete configuration a new. In fact, it is the main drawback in
the Casimir calculations for boundaries of diverse geometries.
Undoubtedly, comparison of the theoretical predictions with the
experimental observations should play here the crucial part. By
now only the simplest configuration,   plane plates, is feasible
for experimental study (see Introduction).  In the next sect.\ we
consider different mathematical methods for removing the
divergencies in the Casimir calculations.

It is interesting to note that in the case of fermions the vacuum
energy has opposite sign in comparison with  bosons. Really, the
Hamiltonian function for noninteracting fermionic oscillators is
given by~\cite{Dirac}
\begin{equation}
\label{2-5} H_F=\frac{1}{2}\int d^3\bk  \,
\omega(\bk)[b^\dagger(\bk) b(\bk)-b(\bk)b^\dagger(\bk)]{.}
\end{equation}
In quantum theory the amplitudes $b(\bk)$ and $b^\dagger(\bk)$
obey the anticommutation relation
\begin{equation}
\label{2-6} [b(\bk),b^\dagger(\bk)]_+=\delta^{(3)}(\bk -\bk'){,}
\end{equation}
which enables one to cast eq. (\ref{2-5}) to the form
\begin{equation}
\label{2-7} H_F=\int d^3\bk  \, \omega(\bk)\,b^\dagger(\bk)
b(\bk)+ E_{0F}{,}
\end{equation}
where $E_{0F}$ is the fermionic zero point energy
\begin{equation}
\label{2-8} E_{0F}= -\frac{1}{2}\int d^3\bk  \, \omega(\bk){.}
\end{equation}

As known, the opposite signs in eqs. (\ref{2-3}) and (\ref{2-8})
result in substantial conciliation of divergencies in field
theories symmetric with respect to permutation of bosons and
fermions, i.e., in supersymmetric theories.

 The Casimir effect can be considered
without introducing the notion of the vacuum fields and its
zero-point energy, for example, in the framework of the
Schwinger's source
theory~\cite{Schwinger-LMP,Milonni-PRA,Milonni-PS}. In this
connection one can meet the opinion~\cite{Milonni-book}, according
to which there is no sense to insist that ``... observable
phenomena like the Casimir effect strongly suggest that the vacuum
electromagnetic field and its zero-point energy are real physical
entities and not mere artifices of the quantum formalism''. And it
is largely ``matter of taste'' to use for the interpretation of
this effect vacuum or source fields.

A similar assertion has been done in a recent paper~\cite{Jaffe} by
R.~L.~Jaffe: ``... no known phenomenon, including the Casimir effect,
demonstrates that zero point energies are ``real''. The author
stressed once more the well known fact that the Casimir force
originates in the forces between charged particles in the metal
plates or, more precisely, between fluctuating dipoles inside the
different plates~\cite{Lif,Milton-book,Barash}. In view of this, it
is obvious that in the microscopic theory, describing the atomic
structure of the plates, the Casimir force vanishes as $\alpha$, the
fine structure constant, goes to zero. The independent of $\alpha$
result (\ref{1-1}) obtained in macroscopic approach corresponds to
the limit $\alpha \to \infty $ in the microscopic consideration. It
is clear that in microscopic description  the notion of zero point
energy does not appear.

  However we think that denying the reality of the zero point energy
of {\it physical fields}  mentioned above is not sufficiently
justified. Here it is worthy to recall that the usual classical
mechanics may be reformulated without using the notion of the force,
as has been proposed by Hertz~\cite{Hertz} (see the evaluation of
this approach due to Sommerfeld~\cite{Sommerfeld}). However, now
nobody suspects the reality of forces, as the physical reasons that
cause the alteration of the velocity of a test particle. Therefore we
believe that it is well-grounded to treat the experimentally observed
Casimir forces as the direct manifestation of the zero point energy
of the relevant vacuum fields~\cite{Milton-book} because the use of
this term enables one to accomplish the theoretical analysis of the
Casimir effect in the most simple and clear way.

Closing this section it is interesting to note that the notion of
the zero point energy was  also considered and used in the
framework of the classical theory of electromagnetic field (see
the papers~\cite{Marshall,Boyer-1,MB-exp} and references therein).

\section{Renormalization of the vacuum energy}

\setcounter{equation}{0} \label{RVE}

\subsection{Physical background for removing the divergencies}

In the general case one cannot assign the physical meaning  to the
vacuum energy of quantum field (\ref{2-3}) unlike to the zero point
energy of quantum mechanical systems with finite degrees of freedom.
However if the configuration space ${\cal M}$ is bounded (the Casimir
effect), then one can get a finite value for the vacuum energy
subtracting appropriate counter terms, i.e., \ by making the
renormalization.

Certainly, in the experiment one can observe only the Casimir
forces that are derived from the  vacuum energy by differentiation
with respect to relevant distance. We shall always bare in mind
this when considering the vacuum energy problem.

When the field occupies a bounded region ${\cal M}$, its
eigenfrequencies are discrete and their values are determined by
the geometry of the boundary $\partial {\cal M}$. As a result, the
vacuum energy also depends on the form of the boundary
\begin{equation}
\label{3-1} E_0(\partial {\cal M})= \frac{1}{2}\sum_{n}\omega_n{.}
\end{equation}
Following the basic line of the quantum field theory formalism,
one can expect that an observable value of the vacuum energy is
obtained  by subtracting from eq. (\ref{3-1}) the same expression
calculated for the boundary of a special form. In the simplest
case, when electromagnetic field is confined between two parallel
perfectly conducting plates, it is sufficient to subtract from eq.
(\ref{3-1}) the contribution of unbounded Minkowski space~$E_0(0)$
\begin{equation}
\label{3-2} E_0= E_0(\partial {\cal M}) - E_0(0){.}
\end{equation}
Preliminary both  expressions for $E_0$ should be regularized,
for example, by introducing a smooth function, that suppresses the
contributions due to large momenta. In this way the Casimir energy
for parallel plates is calculated, for example, in the text book
on quantum field theory~\cite{IZuber}.

 The prescription (\ref{3-2}) for obtaining the physical value of the
vacuum energy may be interpreted in the following way. In the case of
infinite number of degrees of freedom (quantum field theory) the
observable quantity is not the zero point energy itself, but only
its excess,  caused by boundaries or inhomogeneities of the space,
compared with  the zero point energy of quantum field occupying
unbounded homogeneous space. It is the main point that differs
treating the zero point energy in quantum mechanics and in quantum
field theory.

However, the subtraction of the Minkowski space contribution turns
out to be insufficient in order to obtain   finite vacuum energy
in the case of more complicated boundaries, for example, for
perfectly conducting sphere. Unfortunately, there are no general
rules that could enable one to  construct  the subtraction
procedure for the boundary of arbitrary form, i.e., , to find the
corresponding counter terms. In view of this, in calculation of
vacuum energy the analytical regularization, or zeta
regularization, proved to be very useful because it simultaneously
accomplishes the renormalization without introducing the relevant
counter terms. Here we are dealing with renormalization of the
field energy. This problem is central one in quantum field theory
on nonflat background~\cite{Birrell-Davies}.

  There is a few surveys and books concerned with the spectral zeta
function technique and its application to the Casimir
calculations~\cite{Bordag,Kirsten,Od,Ten,Vassilevich,Santangelo,BVW,%
Bytsenko,Bytsenko-1,Od-3}. Therefore we present here only the basic
formulae and stress the points that are not sufficiently elucidated
in the mathematical literature.

\subsection{Spectral zeta functions}

Let the field $\varphi (t,x)$ is governed by the following
equation
\begin{equation}
\label{3-3} \left ( L +\frac{\partial ^2}{c^2 \partial t^2} \right
)\varphi (t,x)=0{,}
\end{equation}
where $L$ is an elliptic differential operator of the second order
acting only on the spatial variables $x$. For example, in the case
of a free scalar massless field $L=- \Delta$, where $\Delta$ is the
Laplace operator.

Upon separating the time dependence $\varphi (t,x)=e^{\pm i \omega
t}\varphi_n(x)$ the field equation  (\ref{3-3}) and the relevant
boundary conditions with respect the variable $x$ generate the
spectral problem
\begin{equation}
\label{3-4} L\varphi_n (x)=\lambda_n \varphi_n (x) \quad \text{or}
\quad L |n > =\lambda _n |n>, \quad \lambda _n=\omega _n^2/c^2 {.}
\end{equation}
It is convenient to use here the Dirac bracket notation. We assume
that the spectral problem is well posed, i.e., , all the equations
written below have sense.

The completeness relation of the  vector set $\{ |n>\}$ can be
represented (formally) in the following way
\begin{equation}
\label{3-5} I=\sum_n |n><n| {,}
\end{equation}
where $I$ is a unity operator acting in the linear space of the
vectors $|n>$. In view of this we have for the inverse operator
$L^{-1}$
\begin{equation}
\label{3-6} L^{-1} =\sum_n \frac{|n><n|}{\lambda _n} {.}
\end{equation}
It can be easily checked with allowance for eq. (\ref{3-4}). In
the same way we have for the  $s$th power of the inverse
operator~$L^{-1}$
\begin{equation}
\label{3-7} L^{-s} =\sum_n \frac{|n><n|}{\lambda _n^s} {.}
\end{equation}
The {\it local} spectral zeta function $\zeta_L (s;x)$ of the operator
$L$ is a diagonal element of the operator $L^{-s}$
\begin{equation}
\label{3-8} \zeta_L (s;x) =\sum_n \frac{<x|n><n|x>}{\lambda
_n^s}=\sum_n\lambda_n ^{-s}\varphi ^*_n(x)\, \varphi _n(x){.}
\end{equation}
The {\it global} spectral zeta function~\cite{DC,Hawking} is defined by
\begin{equation}
\label{3-9} \zeta_L(s)= \text{Tr}\, L^{-s} =\sum_n\lambda_n^{-s}{.}
\end{equation}
It is obtained by integration of $\zeta_L(s;x)$ over the whole
space
\begin{equation}
\label{3-10} \zeta_L(s)= \int\zeta_L (s;x)\, d x{.}
\end{equation}

The definition of the spectral zeta function (\ref{3-9}) is a direct
extension of the Riemann  zeta function~\cite{WW}
\begin{equation}
\label{3-10a} \zeta_{\text{R}}(s)= \sum_{n=1}^\infty\frac{1}{n^s},
\quad \text{Re}\;s >1
\end{equation}
to the spectrum of the operator $L$.

In the case of the $d$-dimensional space the index $n$ in eq.
(\ref{3-9}) stands for the set of $d$ indices $n=\{ n_1,n_2,
\ldots, n_d\}$ and summation should be carried out with respect to
each of them in respective ranges. In the majority of the problems,
having  physical application,
\begin{equation}
\label{3-10b} \lambda_n\sim n_1^2+n_2^2+\ldots +n_d^2, \quad
n_i\to \infty, \quad 1\leq i\leq d{.}
\end{equation}
In this case the series (\ref{3-9}), defining the global zeta
function, converges in the domain $\text{Re}\;s>d/2$, where it
represents an analytical function $\zeta_L(s)$ of a complex
variable $s$. This function can be analytically continued into the
left half-plane $\text{Re}\;s<d/2$ except  for simple separate
poles, for example, at the point $s=d/2$. This fact is a
mathematical basis for employment of the spectral zeta functions
with the aim of obtaining finite expressions when treating the
divergencies.

In the framework of the zeta function technique the vacuum energy
$E_0$ of the quantum field $\varphi (t,x)$ is defined
by~\cite{BVW}
\begin{equation}
\label{3-11} E_0=\frac{1}{2}\zeta_L\left( s=-\frac{1}{2}\right
){.}
\end{equation}
In view of eqs. (\ref{3-10}) and (\ref{3-11}), the quantity
$(1/2)\zeta_L(-1/2;x)$ can be interpreted as the vacuum energy
density. Besides this, it is easy to show that the vacuum
expectation value of the canonical energy-momentum tensor
$T_{00}(x)$ for the field $\varphi(t,x)$ can be represented in the
form
\begin{equation}
\label{3-12}
<0|T_{00}(x)|0>=\frac{1}{2}\sum_n\lambda^{1/2}_n\varphi^*_n(x)
\varphi_n(x)=\frac{1}{2}\zeta_L\left
(-\frac{1}{2};x
 \right) {.}
\end{equation}

In order to construct the global zeta function, the spectrum of
the problem under consideration should be known. Certainly, it is
feasible only for boundaries with high symmetry, for example, with
spherical or cylindrical symmetry.  In the Casimir calculations
with $d=3$ and $n=\{n_1,n_2,n_3\}$, the configurations are usually
considered when, for fixed values of $n_1$ and $n_2$, the
eigenfrequencies $\omega_{n_3}$ are the roots of the equation
\begin{equation}
\label{3-13} f_{n_1n_2}(\omega)=0
\end{equation}
with a known function $f_{n_1n_2}(z)$. The sum of these roots $
\omega _{n_3}$ can be found by making use of a contour integration
in complex frequency plane~\cite{WW}. It leads to the following
compact representation for the spectral zeta function
\begin{equation}
\label{3-14} \zeta (s)=\frac{1}{2\pi i}\sum_{n_1,n_2}\oint_C
z^{-s}\frac{d }{d z} f_{n_1n_2}(z)\,d z{,}
\end{equation}
where the contour $C$ encloses, counterclockwise,  all the positive
roots of eq.~(\ref{3-13}). This representation proves to be
convenient for analytical continuation of the right hand side of eq.
(\ref{3-14}) to the left half plane, specifically, at the point
$s=-1/2$. However, it may happens that the zeta function $\zeta (s)$
has a pole at this point. It will imply that the zeta regularization
does not  afford a finite value for the vacuum in the problem at
hand. There is a rigorous mathematical criterion that enables one to
predict when this happens. In order to formulate it we need another
spectral function, {\it heat kernel}.

Before to go over to consideration of this function it is worthy
to make a general comment concerning the spectral functions. In
the theory of differential operators these functions play in fact
the same part as characteristic polynomials do for matrix
operators of a finite dimension, namely, they store  all the
information about the spectrum of a given operator in an
``encoded'' form. When the spectrum consists of a finite number of
eigenvalues $\lambda _n,\quad n=1,2,\ldots, N$ , it is natural  to
construct on this basis the polynomial of $N$th degree the roots
of which give this spectrum: $P_N(\lambda)=\prod
_{n=1}^{N}(\lambda -\lambda_n), \quad \lambda \in \Bbb{C}$. The
matrix equation $P_N(L_N)=0$ is used, as known, for constructing
the functions of the matrix $L_N$. In the case of an infinite
spectrum a concrete form of the relevant spectral function is not
so definite. Ultimately, all is determined by the problem under
study. For example, when dealing with the spectral sums, it is
convenient to use the zeta function~(\ref{3-9}).

The procedure of constructing the local spectral zeta function
(\ref{3-8}) proves to be much more complicated as compared with that
for the global zeta function. Besides the spectrum of the operator
$L$ we have to use the respective natural modes. As far as we know
the local zeta function has been derived only for the scalar Laplace
operator defined on a cone or in wedge. This function was used in
studies of black hole physics~\cite{blackholes,Od-1}, cosmic
strings~\cite{cosmicstr} and in calculation of the Casimir effect for
wedge~\cite{wedge}.

The relation of the zeta function technique to more transparent
regularizations is discussed, for example, in refs.~\cite{CVZ,BS}.

\subsection{Heat kernel technique}

\label{RVE-hkt} In Casimir calculations also useful is another
spectral function of the operator $L$ introduced in eq.
(\ref{3-3}), namely, the heat kernel of this operator
\begin{equation}
\label{3-15} K(\tau)=\text{Tr}\,(e^{-\tau L})=\sum_n e^{-\lambda _n
\tau }{,}
\end{equation}
where $\tau$ is an auxiliary variable ranging from 0 to $+\infty$.
Such a name of this function is due to the following. By making
use of the unity operator (\ref{3-5}) one can write
\begin{equation}
\label{3-16} e^{-\tau L}=\sum_n e^{-\tau \lambda_n}|n><n|{.}
\end{equation}
The matrix element of this operator
\begin{eqnarray}
\label{3-17} K(x,y;\tau)&\equiv&<x|e^{-\tau L}|y>\nonumber
\\&=&\sum_n e^{-\tau \lambda_n}<x|n><n|y>=\sum_n e^{-\tau
\lambda_n}\varphi^*_n(x)\varphi_n(y)
\end{eqnarray}
is the Green function of the heat conduction equation with the
operator~$L$
\begin{equation}
\left ( L_x+\frac{\partial}{\partial \tau}\right )K(x,y;\tau)=0{,}
\label{3-18}
\end{equation}
 \begin{equation} K(x,y;\tau)= \delta
(x,y), \quad \tau \to +0{.}\label{3-19}
\end{equation}
For the functions $K(\tau)$ and $K(x,y;\tau)$ the relation
analogous to~(\ref{3-10}) holds
\begin{equation}
\label{3-20} K(\tau) =\int d x\,K(x,x;\tau){.}
\end{equation}

The integrated heat kernel (\ref{3-15}) and the spectral zeta
function (\ref{3-9}) are connected by the Mellin transform.
Indeed, from the definition of the gamma function~\cite{GR} it
follows that
\begin{equation}
\label{3-21}
\frac{1}{\lambda^s_n}=\frac{1}{\Gamma(s)}\int_{0}^{\infty}d
\tau \,\tau^{s-1}e^{-\lambda_n \tau}, \quad \text{Re}\, s>0{.}
\end{equation}
Upon summation in both sides of this equation we obtain
\begin{equation}
\label{3-22} \zeta(s)=\frac{1}{\Gamma(s)}\int_0^\infty\drm\tau\,
\tau^{s-1}K(\tau){,}\quad \text{Re}\,s >\frac{d}{2}{.}
\end{equation}
To the left half-plane $\text{Re}\, s<d/2$ eq. (\ref{3-22}) should
be continued analytically.

  Thus, from the mathematical  point  of view,  both the spectral functions $\zeta (s)$
and $K(\tau)$ are equivalent. However in practical use it is
difficult to succeed in finding  the spectral zeta function or the
heat kernel in an exact form, and one has to deal with the
approximate expressions. Therefore  different initial formulae
either (\ref{3-9}) or (\ref{3-15}) and (\ref{3-18}) -- (\ref{3-20})
prove to be useful.

In physical applications are important the coefficients in the
asymptotic expansion of the heat kernel, when $\tau \to +0$
\begin{equation}
\label{3-23} K(\tau)= \sum_ne^{-\lambda _n \tau}= (4 \pi
\tau)^{-d/2}\sum_{n=0,1,2,\ldots}^\infty\tau ^{n/2}B_{n/2}+
\text{ES}{.}
\end{equation}
In this expansion $d$ is the dimension  of the configuration space
manifold ${\cal M}$ in the problem at hand,  ES stands for the
exponentially small corrections as $\tau\to +0$.  The first two
coefficients in expansion (\ref{3-23}) are determined by the
volume $V$ of the manifold ${\cal M}$ and by the area of the
boundary $\partial {\cal M}$. For example, if $L=-\Delta$, where
$\Delta $ is the Laplacian, acting on a massless scalar field,
then
\begin{equation}
\label{3-23a} B_0=V, \quad B_{1/2}=\mp\frac{\sqrt {\pi}}{2} S{.}
\end{equation}
The upper sign in this equation is for the Dirichlet boundary
conditions and the  lower sign is for the Neumann conditions.

For flat manifolds all the coefficients $B_{n/2}, \;n\geq 1$ are
due to the boundary contributions, namely, they are defined by the
integrals over the boundary with the integrands expressed in terms
of geometrical invariants of the
boundary~\cite{Bordag,Kirsten,Od,Vassilevich}. The boundary is
expected to be sufficiently smooth.

 We give here the properties of the heat kernel coefficients which
are especially important when calculating the vacuum energy. The
proof of these statements can be found in the literature (see, for
example~\cite{Bordag,Kirsten,Od}). The first few coefficients
$B_{n/2}$ yield the ultraviolet divergencies of the vacuum energy
(\ref{3-11}). For $d=3$ it is the coefficients up to $B_{2}$. If the
coefficients $B_{2}$ is equal to zero, then the zeta regularization
gives a finite value for the vacuum energy according to the formula
(\ref{3-11}). For $d=2$ the role of such an indicator plays the
coefficient $B_{3/2}$ (see below). The heat kernel coefficients also
specify conformal anomalies taking place in a concrete field theory
model, the high temperature behavior of the thermodynamic functions
(see sect.~\ref{HT}) and so on~\cite{Kirsten}.

  The heat kernel coefficients are determined by the residua of
the product $\Gamma(s)\zeta (s)$ at corresponding points. Indeed,
substituting the expansion (\ref{3-23}) into eq. (\ref{3-22}) and
integrating over $\tau$ in the vicinity of the origin we obtain
\begin{equation}
\label{3-24} \frac{B_{n/2}}{(4 \pi)^{d/2}}=
\lim_{s\to\frac{d-n}{2}} \left ( s+\frac{n-d}{2} \right )\Gamma
(s) \zeta(s),\quad n=0,1,2, \ldots \,{.}
\end{equation}
In order to use this formula the spectral zeta function $\zeta
(s)$ should be known in the vicinity of the following points
\begin{equation}
\label{3-25} s=\frac{d}{2}, \frac{d-1}{2}, \ldots \,{.}
\end{equation}
For this one has to do analytical continuation of the initial formula
(\ref{3-9}) to   the left half-plane Re~$s<d/2$.

  If $d=3$ and $B_2\neq 0$ or for a given $d$ the coefficient
$B_{(d+1)/2}$  does not vanish  then according to eq.\ (\ref{3-24})
the zeta function has a pole at the point $s=-1/2$. As a result in
these cases the renormalization  by zeta function does not provide a
finite value for the vacuum energy $E_0$ (see the definition
(\ref{3-11})).

There is another technique for calculating the heat kernel
coefficients that enables one to avoid the procedure of analytical
continuation~\cite{Moss,Esposito,GEKK,NPD}. It is close to the method
presented above but sometimes calculations become simpler.

  Let us consider the spectral zeta function depending on a parameter $x^2$
\begin{equation}
\label{eq3-47} \zeta(s,x^2)=\sum_{n}(\lambda_n+x^2)^{-s}{.}
\end{equation}
It may be regarded as an extension
of the Epstein-Hurwitz zeta function
\[
\zeta_{\text{EH}}(s,a^2)=\sum_{n=1}^{\infty}(n^2+a^2)^{-s}
\]
to the general spectral problem (\ref{3-4}).

It turns out that the heat kernel coefficients $B_{n/2}$ can be
found from the expansion of the  function $\zeta(s,x^2)$ in terms
of inverse powers of $x$ developed for a certain value of $s$. It
is convenient to chose this value to be equal to $1+d/2$. In fact,
from the definition of the  gamma function (\ref{3-21}) it follows
that
\begin{equation}
\label{eq3-48} \Gamma
(s)\,(\lambda+x^2)^{-s}=\int_{0}^{\infty}\!\!d  t\,t^{s-1}
e^{-(\lambda+x^2)t}, \quad \text{Re}\, s>0{.}
\end{equation}
For $s=1+d/2$ eq.\  (\ref{eq3-48}) gives
\begin{eqnarray}
\label{eq3-49}  \Gamma\left( 1+\frac{d}{2} \right )
\sum_{n}(\lambda_n+x^2)^{-1-d/2}&=&\int_{0}^{\infty}d
t\,t^{d/2}e^{-x^2t} \sum_{n}e^{-\lambda_n t}\nonumber \\
&=&\int_{0}^{\infty}d  t\,t^{d/2}e^{-x^2t}K(t){.}
\end{eqnarray}
On substituting  the asymptotic expansion (\ref{3-23}) in eq.\
(\ref{eq3-49}) we obtain
\begin{eqnarray}
\Gamma\left ( 1+\frac{d}{2}\right  )\zeta \left (
1+\frac{d}{2},x^2 \right )
&\simeq&\sum_{n=0}^{\infty}\frac{B_{n/2}}{(4\pi)^{d/2}} \Gamma
\left ( 1+\frac{n}{2}
\right)      x^{-n-2} \nonumber\\
 &=&\frac{1}{(4\pi)^{d/2}}\left [
\frac{B_0}{x^2}+\frac{B_{1/2}\Gamma(3/2)}{x^3}+\frac{B_1\Gamma(2)}{x^4}
+\frac{B_{3/2}\Gamma(5/2)}{x^5}\nonumber  \right .\\
&& \left .
+\frac{B_2\Gamma(3)}{x^6}+\frac{B_{5/2}\Gamma(7/2)}{x^7}+\frac{B_3\Gamma(4)}{x^8}
+{\cal O} (x^{-9}) \right ]{.} \label{eq3-50}
\end{eqnarray}

Let $  F(z)=0$ be the frequency equation which determines the
spectrum $\lambda_n$ in the problem under consideration. We also
suppose that the function $F(z)$ allows one to rewrite this
equation in the form
\begin{equation}
\label{eq3-52} \prod_{n}(\lambda_n-z^2)=0\,{.}
\end{equation}
Taking into account the equality
\begin{equation}
\label{eq3-53}
\frac{1}{(\lambda_n+x^2)^m}=-\frac{(-1)^m}{\Gamma(m)}\left(
\frac{d }{2x\, d  x} \right )^m    \ln
(\lambda_n+x^2)\,{,}\quad z=ix\,{,}
\end{equation}
we recast the left-hand side of eq.\ (\ref{eq3-50})  to the form
\begin{equation}
\label{eq3-54} \Gamma \left (1+\frac{d}{2} \right )\zeta \left (
1+\frac{d}{2}, x^2 \right )=-\left (-\frac{1}{2x}\frac{d }{d
x} \right )^{1+d/2}\ln F(ix)\,{.}
\end{equation}
Obviously formula (\ref{eq3-54}) is applicable only to the manifolds
of even dimension. Now we have to substitute the left-hand side of
eq.\ (\ref{eq3-50}) by the expansion of eq.\ (\ref{eq3-54}) for large
$x^2$. After that the heat kernel coefficients $B_{n/2}$ are obtained
by comparing the coefficients of $x^2$ in both sides of eq.\
(\ref{eq3-50}).

This technique can be extended to the odd values of $d$ by
calculating the zeta function $\zeta(s,x^2)$ at the point
$s=(1+d)/2$. Instead of eqs.\ (\ref{eq3-50}) and (\ref{eq3-54}) we
have now
\begin{eqnarray}
\label{eq3-55}
 \Gamma\left ( \frac{d+1}{2} \right ) \zeta \left (
\frac{d+1}{2},x^2 \right )&=&-\left (-\frac{1}{2x}\frac{d }{d  x}
\right )^{(d+1)/2}\ln F(ix) \nonumber \\
&\simeq&\sum_{n=0}^{\infty}\frac{B_{n/2}}{(4\pi)^{d/2}} \Gamma \left
( \frac{n+1}{2} \right)      x^{-n-1} {.}
\end{eqnarray}
Again comparing the coefficients of $x$ in both sides of eq.\
(\ref{eq3-55}) we obtain the heat kernel coefficients for odd~$d$.

\section{Spectral zeta functions and vacuum energy for simplest boundaries}

\setcounter{equation}{0} \label{SZF+VE} Practically every problem on calculation of the
Casimir energy (or force) has been considered often with
employment of more and more effective and elaborated mathematical
methods. For example, the first calculation of the Casimir energy
of a perfectly conducting spherical shell carried out by
T.H.~Boyer in 1968~\cite{Boyer} has required computer calculations
during 3 years~\cite{Milton-3}. Later this problem was considered
in many papers~\cite{Davies,BD,MRS}. By making use of the modern
methods~\cite{NP} it can be solved almost without numerical
calculations (with a precision of a few percent). It requires only
the application of the uniform asymptotic expansion for the Bessel
functions. The zeta function regularization enables one to
calculate the vacuum energy in a consistent way without dealing
with  the manifestly infinite expressions.

\subsection{Parallel perfectly conducting plates}

\label{SZF+VE-plates}

  As known, for example, from the theory of waveguides and
resonators
 \cite{HdP1} the
vectors of electric and magnetic fields in the problem at hand are
expressed in terms of  the electric ($\bf{\Pi}'$) and magnetic
($\bf{\Pi}''$) Hertz vectors, each having only one nonzero
component $\Pi^{'}_z$ and $\Pi^{''}_z$ satisfying, respectively,
Dirichlet and Neumann conditions on the internal surface of the
plates. The functions $\Pi^{'}_z$ and $\Pi^{''}_z$ obey  the
equations
\begin{equation}
\left(\frac{\partial^2}{\partial
z^2}+\bbox{\nabla}^2\right)\,\Pi^{'}_z=
\frac{\omega^2}{c^2}\Pi^{'}_z,\quad
\left(\frac{\partial^2}{\partial
z^2}+\bbox{\nabla}^2\right)\,\Pi^{''}_z=
\frac{\omega^2}{c^2}\Pi^{''}_z, \label{eq3_1}
\end{equation}
where $\omega$ is the frequency of electromagnetic oscillations,
$\bbox{\nabla}^2$ stands for the two-dimensional Laplace
operator for the variables $(x,y)={\bf x}$. The separation of
variables results in the following solution
\begin{eqnarray}
\Pi^{'}_{z}({\bf x},z)&= &\exp(i{\bf k x} )\,\sin\left(\frac{n\pi
z}{a}\right), \quad
n=1,\,2,\dots {,}\nonumber\\
\Pi^{''}_{z}({\bf x},z)&= &\exp(i{\bf k x} )\,\cos\left(\frac{n\pi
z}{a}\right), \quad
n=0,\,1,\,2,\dots {,}\nonumber\\
\omega^2_n({\bf k})&=& c^2\left[{\bf k}^2+
\left(\frac{n\pi}{a}\right)^2\right], \label{eq3_2}
\end{eqnarray}
where $a$ is the distance between the plates. Hence, the states of
electromagnetic field with the energy $\hbar\omega_n,\;\;n\geq 1,$
are doubly degenerate, while the state with the energy
$\hbar\omega_0=\hbar c k$ is nondegenerate.

With allowance for this the zeta function in the problem under
consideration is given by
\begin{equation}
\zeta(s)=\frac{L_x\,L_y}{c^{2s}}\int\frac{d ^2 {\bf
k}}{(2\pi)^2}\left\{2\sum_{n=1}^{\infty} \left[{\bf
k}^2+\left(\frac{n\pi}{a}\right)^2 \right]^{-s}+\left( {\bf
k}^2+\mu^2\right)^{-s} \right\}, \label{eq3_3}
\end{equation}
where $L_x$ and $L_y$ are the dimensions of the plates.

For a correct definition of the integral in this formula in the
small ${\bf k}$ region the photon mass $\mu$ is introduced
(infrared regularization). At the final step of calculations one
should put $\mu=0$.  On integrating in eq.\ (\ref{eq3_3}) and
substituting the sum over $n$ by the Riemann zeta function one
arrives at the result
\begin{equation}
\zeta(s)=\frac{L_x\,L_y}{2\pi\,c^{2s}} \left [
\left(\frac{\pi}{a}\right)^{2-2s}\frac{\zeta_{\text
{R}}(2s-2)}{s-1}+ \frac{1}{2}\frac{\mu^{2-2s}}{s-1}\right ].
\label{eq3_4}
\end{equation}

The zeta function~(\ref{eq3_4}) leads to the well-known value for
the Casimir energy
\begin{equation}
 \label{eq3_6}
E_{0}=\frac{\hbar}{2}\,\zeta\left(-\frac{1}{2}\right)=-c\,\hbar\,
\frac{\pi^2}{720}\,\frac{L_x L_y}{a^3}
\end{equation}
or for its density
\begin{equation}
\frac{E_{0}}{V}=-\frac{c \hbar\pi^2}{720 a^4},\quad \mbox{where}
\quad V=a\,L_x \,L_y. \label{eq3_7}
\end{equation}
Differentiation of the vacuum energy (\ref{eq3_6}) with respect to
the distance $a$ gives the Casimir force~(\ref{1-1}).

\subsection{Sphere}

\label{SZF+VE-sph}

 Let us consider a solid ball of radius $a$,
consisting of a material which is characterized by permittivity
$\varepsilon_1$ and permeability $\mu_1$. The ball is assumed to
be placed in an infinite medium with permittivity $\varepsilon_2$
and permeability $\mu_2$. In the case of spherical symmetry the
solutions to Maxwell equations are expressed in terms of two
scalar Debye potentials $\psi$ (see, for example,~\cite{Jackson})
 \begin{eqnarray}
 {\bf E}^{\text{TM}}_{lm}&= &\bbox{\nabla} \times
 \bbox{\nabla} \times(\bbox{r}\psi^{\text{TM}}_{lm})\,,
 \quad \bbox{ H}^{\text{TM}}_{lm}=-i\omega \bbox{\nabla}\times
 (\bbox{r}\psi^{\text{TM}}_{lm}) \qquad
 (E\text{-modes}){,} \nonumber \\
 {\bf E}^{\text{TE}}_{lm}& =& i\omega \bbox{\nabla} \times( {\bf r}\psi_{lm})\,,
 \qquad {\bf H}^{\text{TE}}_{lm}=\bbox{\nabla}\times \bbox{\nabla}\times
(\bbox{r}\psi^{\text{TE}}_{lm}) \qquad
 (H\tx{-modes}){.}
 \label{4-7}
\end{eqnarray}
These potentials obey the Helmholtz equation, have the indicated
angular dependence, and are regular at the origin \beq \label{4-8}
(\nabla^2+k_i^2)\psi_{lm}=0, \quad k_i^2=\varepsilon
_i\mu_i\frac{\omega ^2 }{c^2},\quad i=1,2,\quad (r\neq a);\quad
\psi_{lm}(\bbox{r})=\phi (r) Y_{lm}(\Omega){.} \eeq

Here we have to make a note concerning the formulation of the
spectral problem for differential elliptic operators defined on
unbounded manifolds. First of all, the question arises what
conditions should be imposed on the eigenfunctions at the spatial
infinity. In order for the uniqueness theorem to hold for the
solutions looked for, the {\it radiation conditions} at infinity should
be introduced. It is a well known fact in classical mathematical
physics (see, for example,~\cite{RR}). However, in this case there
are no solutions with {\it real frequencies} and one has to consider the
{\it complex frequencies} and respective natural modes, i.e., \
{\it quasi-normal modes}. Certainly, it is absolutely unclear  how
to use these modes when quantizing the fields. Here one can
proceed in two different ways. The quantum field under
consideration can be placed inside a sphere of a large radius $R$
 imposing on field functions reasonable conditions on the
internal surface of the sphere~\cite{bowers}. All outside the
sphere is ignored. The spectrum of oscillations in this case is
discrete, infinite and real because we are dealing  simply with
standing waves. On this basis the spectral zeta function can be
constructed in a standard way and ultimately the radius of an
auxiliary sphere $R$ should be allowed to go to infinity.

More appealing, from the physical point of view, is an approach
that uses at infinity the conditions on field functions taken from
the {\it scattering problem} instead of the radiation conditions. The
frequency spectrum in this case is real, positive but continuous.
The drawbacks due to unnormalizable  natural modes can be overcome
by introducing into consideration the relevant spectral density.
This density is expressed in terms of the phase shifts, or more
precisely, through the Jost function of the corresponding
scattering problem.

Both  these approaches lead to the same final results.
Furthermore, it turns out, that one can also deal with the
radiation conditions at infinity and consequently  with complex
frequencies, ignoring the problem of treating the corresponding
quasi-normal modes. Point is that the  equation for real
frequencies in the case with additional large sphere reduces, when
$R\to \infty$, to the frequency equation in the approach which
uses the radiation condition at infinity. And further, the Jost
function mentioned above is determined by the left-hand side of
the equation for complex frequencies.

In view of all this, we shall impose  at infinite the radiation
conditions on the function $\phi (r)$ in eq. (\ref{4-8}) and deal
with the frequency equation with complex roots keeping in mind
that it is in fact the relevant Jost function.

At the ball surface the tangential components of electric and
magnetic fields are continuous. As a result, the eigenfrequencies
of electromagnetic field for this configuration are determined by
the frequency equation for the TE--modes \beq\label{TE}
\Delta_l^{\text{TE}}(a\omega)\equiv \sqrt{\varepsilon_1\mu_2}
\,\tilde{s}^{\prime}_l (k_1a)\,\tilde{e}_l(k_2a)-
\sqrt{\varepsilon_2\mu_1}\,\tilde{s}_l(k_1a)\,\tilde{e}^{\prime}_l(k_2a)=0\,{,}
\eeq and the analogous equation for the TM--modes \beq\label{TM}
\Delta_l^{\text{TM}}(a\omega)\equiv \sqrt{\varepsilon_2\mu_1}
\,\tilde{s}^{\prime}_l (k_1a)\,\tilde{e}_l(k_2a)-
\sqrt{\varepsilon_1\mu_2}\,\tilde{s}_l(k_1a)\,\tilde{e}^{\prime}_l(k_2a)=0\,{,}
\eeq where $k_i=\sqrt{\varepsilon_i\mu_i}\,\omega,\; i=1,2$ are
the wave numbers inside and outside the ball,
respectively~\cite{Stratton}. Here $\tilde{s}_l(x)$ and
$\tilde{e}_l(x)$ are the Riccati--Bessel functions \beq\label{set}
\tilde{s}_l(x)=\sqrt{\frac{\pi}{2x}}\, J_{l+1/2}(x)\,{,} \qquad
\tilde{e}_l(x)=\sqrt{\frac{\pi}{2x}}\,H^{(1)}_{l+1/2}(x)\,{,} \eeq
and prime stands for the differentiation with respect to their
arguments, $k_1a$ or $k_2a$. The orbital momentum $l$ in eqs.
(\ref{TE}) and (\ref{TM}) assumes the values $1,2, \ldots$. The
roots of the frequency equations (\ref{TE}) and (\ref{TM}),
including the complex ones, have been investigated by Debye in his
PhD thesis concerned with the light pressure on a material
ball~\cite{Debye-ball}.

The material presented below is based in main part on the
paper~\cite{LNB}. When considering the boundaries with spherical
and cylindrical symmetries, it is convenient to subtract in the
definition of the spectral zeta function (\ref{3-9}) the
contribution of the same configuration under the condition $a\to
\infty $ with frequencies $\bar \omega _p$
\begin{equation}
\label{zeta} \zeta
(s)=\sum_{p}(\omega_p^{-s}-\bar{\omega}_p^{-s})\,{.}
\end{equation}

As usual when one is dealing with an analytic continuation, it is
useful to represent the sum (\ref{zeta}) in terms of the contour
integral by making use of eq.~(\ref{3-14})
\begin{equation}
\label{zeta_C}
\zeta_{\text{ball}}(s)=\sum_{l=1}^\infty\frac{2l+1}{2\pi
i}\lim_{\mu\to 0}\oint_C \frac{d  z}{(z^2+\mu^2)^{s}} \frac{d
}{d  z}\ln \frac{\Delta_l^{\text{TE}}(az)\Delta_l^{\text{TM}}
(az)}{\Delta_l^{\text{TE}}(\infty) \Delta_l^{\text{TM}} (\infty)}\,{,}
\end{equation}
where the contour $C$ surrounds, counterclockwise, the roots of
the frequency equations (\ref{TE}) and (\ref{TM}) in the right
half-plane. For brevity we write in (\ref{zeta_C}) simply
$\Delta_l(\infty)$ instead of $\lim_{a\to\infty}\Delta_l(az)$.
Transition to the complex frequencies $z$ in eq. (\ref{zeta_C}) is
accomplished by introducing the unphysical photon mass~$\mu$
\begin{equation}\label{mu} \omega \rightarrow
(z^2+\mu^2)^{1/2}\vert_{\mu\to 0}\,{.} \end{equation}
 Extension to
the complex $z$-plane of the frequency equations
$\Delta_l^{\text{TE}}(az)$ and $\Delta_l^{\text{TM}}(az)$ should be
done in the following way. In the upper (lower) half-plane the
Hankel functions of the first (second) kind $H_{\nu}^{(2)}(az)$
($H_{\nu}^{(1)}(az)$) must be used.
 If we are considering only outgoing
            waves it is natural to use the solutions of the Maxwell equations which
           are finite
          (or at least do not grow infinitely) in the future. Such solutions should
          be proportional to $\exp{(-i\omega t)} H_{\nu}^{(1)}(\omega r)$
          when $\omega$ lies
          in the lower half-plane. For the upper half-plane $\omega$, the solutions
          describing outgoing waves and finite in the future should contain the factor
          $\exp{(i\omega t)} H_{\nu}^{(2)}(\omega r)$.

Location of the roots of eqs. (\ref{TE}) and (\ref{TM}) enables
one to deform the contour $C$ into a segment of the imaginary axis
$(-i\Lambda, i\Lambda)$ and a semicircle of radius $\Lambda$ in
the right half-plane. When $\Lambda$ tends to infinity the
contribution along the semicircle into $\zeta_{\text{ball}} (s)$
vanishes because the argument of the logarithmic function in the
integrand tends in this case to 1. As a result we obtain
\begin{equation}\label{zetaim}
\zeta_{\text{ball}}(s)=-a^{2s}\sum_{l=1}^\infty\frac{(2l+1)}{2\pi
i}\lim_{\mu\to 0} \int_{-i\infty}^{+i\infty}\frac{d  z}{
(z^2+\mu^2)^{s}} \frac{d }{d
z}\ln\frac{\Delta_l^{\text{TE}}(z)\Delta_l^{\text{TM}}(z)}
{\Delta_l^{\text{TE}}(i\infty) \Delta_l^{\text{TM}}(\infty)}\,{.}
\end{equation}
 Now we impose the condition that the velocity of light
inside and outside the ball is the same \beq\label{vel}
\varepsilon_1\mu_1=\varepsilon_2\mu_2=c^{-2}\,{.} \eeq

The physical motivation for this is the following. The constancy
condition for the velocity of gluonic field when crossing the
interface between two media is used, for example, in a dielectric
vacuum model (DVM) of quark confinement~\cite{PMG,Adler,FGK}. This
model has many elements in common with the bag models~\cite{Bag},
but among the other differences, in DVM  there is no explicit
condition of the field vanishing outside the bag. It proves to be
important for calculation of the Casimir energy contribution to
the hadronic mass in DVM. The point is that in the case of
boundaries with nonvanishing curvature there happens a
considerable (not full, however)  mutual cancellation of the
divergences from the contributions of internal  and external (with
respect to the boundary) regions. If only  the field confined
inside the cavity is considered, as in the bag
models~\cite{Milton-bag,Bender,BEKL,EBK}, then there is  no such a
cancellation, and one has to remove some divergences by means of
renormalization of the phenomenological parameter in the model
defining the QCD vacuum energy density.

>From a physical point of view the vanishing of the field or its
normal derivative precisely on the boundary is an unsatisfactory
condition, because, due to quantum fluctuations, it is impossible
to measure the field as accurately as desired at a certain point
of the space~\cite{LP}.

Under condition (\ref{vel}) the argument of the logarithm in
(\ref{zetaim}) can be simplified considerably \cite{BNP} with the
result \beq\label{zetareal}
\zeta_{\text{ball}}(s)=\left(\frac{a}{c}\right)^{2s}\frac{\sin(\pi
s)}{\pi}\sum_{l=1}^{\infty}(2l+1)\, \int_0^{\infty}\frac{d  y}{
y^{2s}}\frac{d }{d  y} \ln[1-\xi^2\sigma_l^2(y)]\,{,} \eeq
where \beq\label{xisigma} \xi^2=\left (
\frac{\varepsilon_1-\varepsilon_2}{\varepsilon_1+\varepsilon_2}\right
)^2=\left ( \frac{\mu_1-\mu_2}{\mu_1+\mu_2}\right )^2{,}
 \quad \sigma_l (y)=\frac{d }{d
y}[s_l(y)\,e_l(y)]\,{.} \eeq Here $s_l(y)$ and $e_l(y)$ are the
modified Riccati--Bessel functions \beq\label{se}
s_l(x)=\sqrt{\frac{\pi x}{2}}I_{\nu}(x),\quad e_l(x)=\sqrt{\frac{2
x}{\pi}}K_{\nu}(x), \quad \nu=l+1/2\,{.} \eeq More details
concerning the contour integral representation of the spectral
$\zeta$ functions can be found in \cite{BEKL,Bordag-95,ELR,LR}.

Further the analytic continuation of eq.\ (\ref{zetareal}) is
accomplished by expressing the sum over $l$ in terms of the
Riemann zeta function. This cannot be done in a closed form.
Making use of the uniform asymptotic expansion (UAE)  for the
Bessel functions\footnote{In physical literature this expansion is
usually referred to as the Debye expansion. However we did not
find these formulae in Debye  papers. Furthermore in mathematical
handbook~\cite{Magnus} the Debye type expansions for the Bessel
functions are defined as those for large argument $x$ and large
order $\nu$ both real and positive such that $\nu/x$ is fixed and
simultaneously $|x-\nu|={\cal O}(x^{1/3})$. The uniform asymptotic
expansions were derived in ref.~\cite{Olver}} in increasing powers
of $1/\nu$ enables one to construct the analytic continuation
looked for in the form of the series, the terms of which are
expressed through the Riemann zeta function. The problem of the
convergence of this series does not arise because its terms go
down very fast, at least when calculating the vacuum energy.

We demonstrate this keeping only two terms in UAE for the product
of the modified Bessel functions $I_{\nu}(\nu z)K_{\nu}(\nu z)$~\cite{AS}
\beq\label{UAE} I_{\nu}(\nu z)K_{\nu}(\nu z)\simeq \frac{t}{2\nu}
 \left[1+\frac{t^2(1-6t^2+t^4)}{8\nu^2}+\ldots
\right],\quad t=\frac{1}{\sqrt{1+z^2}}\,{.} \eeq After changing
the integration variable $y=\nu z$ in eq. (\ref{zetareal}) we
substitute (\ref{UAE}) into this formula and expand the logarithm
function up to the order $\nu^{-3}$ keeping at the same time only
the terms linear in $\xi^2$.
 The last assumption is not crucial.
It is introduced for simplicity and in order to have possibility
of a direct comparison with the results of ref.~\cite{BNP}. Thus
we have \beq\label{exp} \frac{d }{d
z}\ln\left\{1-\xi^2\left[\frac{d }{d  z}(zI_{\nu}(\nu
z)K_{\nu} (\nu z))\right]^2\right\} = \eeq
$$
=\frac{3}{2}\frac{\xi^2}{\nu^2}\,zt^8+\frac{\xi^2}{16\nu^4}\,zt^8(-12+216t^2
-600t^4+420t^6) +{\cal O}(\nu^{-6})\,{.}
$$
Integration over $z$ can be done by making use of the formula~\cite{GR}
\begin{equation}
\label{beta} \int_0^{\infty}z^{-\alpha-1}t^{\beta}d
z=\frac{1}{2} \frac{\Gamma \left(\displaystyle
\frac{\alpha+\beta}{2}\right) \Gamma \left(-\displaystyle
\frac{\alpha}{2}\right)}{\Gamma \left(\displaystyle
\frac{\beta}{2}\right)}{,}\quad \text{Re}\,\alpha <0,\quad
\text{Re}\,(\alpha+\beta) >0{.}
\end{equation}
 Also the properties of the $\Gamma$ function
\begin{equation}
\label{gamma} \Gamma (z)\Gamma (1-z)=\frac{\pi}{\sin\pi z}, \qquad
\Gamma (1+z)=z\Gamma (z)
\end{equation}
prove to be useful. After
simple calculations we arrive at the result
\begin{equation}\label{in} \zeta_{\text{ball}}(s)\simeq
\frac{\xi^2}{4}\left(\frac{a}{c}\right)^{2s}s(1+s)(2+s)\left(
\sum_{l=1}^{\infty}\nu^{-1-2s}+p(s)\sum_{l=1}^{\infty}\nu^{-3-2s}+\ldots
\right){,}
\end{equation}
where \begin{equation} \label{pol}
p(s)=-\frac{1}{2}\left[1-\frac{9}{2}(3+s)
+\frac{5}{2}(3+s)(4+s)-\frac{7}{24}(3+s)(4+s)(5 +s) \right]\,{.}
\end{equation}

The first term in eq. (\ref{in}) is defined in the region
$0<\text{Re}\, s<1/2$ and the second one exists, when $-1<\text{Re}\,
s<1/2$, the left boundaries being  given by the sums over the
values of the angular momentum $l$ and the right limits are due to
the integral formula (\ref{beta}). Thus, this expression can be
used for its analytical continuation outside these regions.
Obviously, it can be done by accepting the integration formula
(\ref{beta}) outside the domain indicated there and by expressing
the sums over angular momentum $l$ trough the Riemann zeta
function according to the formula~\cite{GR}
\begin{equation}\label{hur}
\sum_{l=1}^{\infty}\nu^{-s}=(2^s-1)\zeta_{\text{R}}
(s)-2^s\,{,}\qquad \nu=l+1/2\,{.}
\end{equation}
As a result one gets
\begin{eqnarray}
 \zeta_{\text{ball}}(s)&\simeq &
\frac{\xi^2}{4}\left(\frac{a}{c}
\right)^{2s}s(1+s)(2+s)\{(2^{1+2s}-1)
\zeta_{\text{R}}(1+2s)-2^{1+2s}\nonumber \\
&& +p(s)[(2^{3+2s}-1)\zeta_{\text{R}}(1+2s)-2^{3+2s}]+\ldots \}\,{.}
\label{con}
\end{eqnarray}
The singularities in eq. (\ref{in}) are transformed in (\ref{con})
into the poles of the Riemann zeta functions at the points $s=-k,
\quad k=0,1,2,\ldots$
\begin{eqnarray} \zeta_{\text{R}}(1+2s) & \simeq
& \displaystyle \frac{1}{2s}+\gamma+\ldots, \quad s\to 0
\,{,} \nonumber \\
\zeta_{\text{R}}(3+2s) & \simeq &
\displaystyle\frac{1}{2s+2}+\gamma+\ldots,\quad s\to -2
\,{,}\label{poles} \\
\zeta_{\text{R}}(5+2s) & \simeq &
\displaystyle\frac{1}{2s+4}+\gamma+\ldots, \quad s\to -4
\,{,} \nonumber \\
.........  & ...... & ...............................
{\,},\nonumber
\end{eqnarray}
where $\gamma$ is the Euler constant. The first three poles are
annihilated by the multipliers in front of the curly brackets in
eq.\ (\ref{con}). The nearest surviving singularity (simple pole)
appears only at the point $s=-3$. Thus  formula (\ref{con})
affords the required analytic continuation of the function
$\zeta_{\text{ball}}(s)$ into the region Re~$s< 0$. In view of eq.\
(\ref{3-11}) we are interested in the point $s=-1/2$ where
$\zeta_{\text{ball}}(s)$ given by (\ref{con}) is regular
\beq\label{final-ball} E_{\text{ball}}=\frac{1}{2}\zeta_{\text{ball}}
\left (-\frac{1}{2}\right
)=\frac{3\xi^2c}{64a}\left[1+\frac{9}{128}\left(\frac{\pi^2}{2}
-4\right)+ \ldots \right]\,{.} \eeq It is exactly the first two
terms in eq. (3.10) of ref.~\cite{BNP}. The procedure of analytic
continuation presented above can be extended in a straightforward
way to the arbitrary order of the uniform asymptotic expansion
(\ref{UAE}). Certainly in this case analytical calculations should
be done by making use of {\it Mathematica} or {\it Maple}.

In sect.\ \ref{CEDB} the {\it exact} in the $\xi^2$-approximation
expression for the Casimir energy (\ref{final-ball}) will be
obtained by making use of another approach (see eq.\
(\ref{exact})).

The problem under consideration with $\xi=1$ is of a special
interest because in this case it gives the Casimir energy of a
perfectly conducting spherical shell. As it was noted above, this
configuration has been considered by many authors. We present here
the basic formulae which afford the analytical continuation of the
corresponding spectral zeta function. We again content ourselves
with two terms in the UAE (\ref{UAE}).  It is impossible to put
simply $\xi=1$ in the next formula (\ref{exp}). One has to do the
expansion here anew keeping all the terms $\sim 1/\nu^4$. This
gives
\[ \frac{d }{d
z}\ln\left\{1-\left[\frac{d }{d  z}(zI_{\nu}(\nu z)K_{\nu}(\nu
z))\right]^2\right\} =
\]
\begin{equation}
=\left[\frac{3}{2\nu^2}zt^8+\frac{3}{4\nu^4}zt^8
\left(-1+18t^2-50t^4+35t^6\right)+ {\cal O}(\nu^{-6})\right]\,{.}
\label{shell1}
\end{equation}
After integration and elementary simplifications we arrive at the
following result for the spectral function in hand
\begin{equation}\label{shell2} \zeta_{\text{shell}}(s)\simeq \frac{a^{2s}}{4}
s(1+s)(2+s) \left[\sum_{l=1}^{\infty}\nu^{-1-2s}+
q(s)\sum_{l=1}^{\infty}\nu^{-3-2s}+\ldots \right]\,{,}
\end{equation}
where \beq\label{qpol} q(s)=
\frac{1}{480}\,(60+217s+252s^2+71s^3)\,{.} \eeq The analytic
continuation of eq. (\ref{shell2}) is accomplished in the same way
as it has been done with eq. (\ref{in}), i.e., , by making use of
eqs.~(\ref{beta}) and (\ref{hur})
 \begin{eqnarray}
\zeta_{\text{shell}}(s)&\simeq&\frac{a^{2s}}{4}\,
s(1+s)(2+s)\{(2^{1+2s}- 1)\zeta_{\text{R}} (1+2s)-2^{1+2s}+
\nonumber\\
&&+q(s)[(2^{3+2s}-1)\zeta_{\text{R}} (3+2s)-2^{3+2s}]+\ldots\}\,{.}
\label{shell3}
\end{eqnarray}
The nearest singularity in this formula is simple pole at $s=-3$.
As above it is originated in the term $\sim 1/\nu^7$ in the UAE
(\ref{UAE}). At the point $s=-1/2$ the spectral zeta function
$\zeta_{\text{shell}}(s)$ is regular and gives the following value
for the Casimir energy of a perfectly conducting spherical shell
\beq\label{shell4}
E_{\text{shell}}=\frac{1}{2}\,\zeta_{\text{shell}}\left (
-\frac{1}{2}\right )=\frac{3}{64a}\left[1-\frac{3}{256}
\left(\frac{\pi^2}{2}-4\right)+\ldots \right]=\frac{1}{a}\,
0.046361\ldots \,{.} \eeq Without considering the analytic
continuation and do not carrying out the analysis of the
singularities in the complex $s$ plane this result has been
obtained in~\cite{NP}. For nanometer size, that is for
$a=10^{-7}$cm, the energy (\ref{shell4}) is (in $\hbar=c=1$
unit, 1{eV} $\simeq 0.5 \cdot 10^{5}\rm{cm}^{-1}$)
$E_{\text{shell}}\simeq 10\rm{eV}$ which is of a considerable
magnitude.

Unlike the Casimir energy for parallel conducting plates
(\ref{eq3_6}) both  the Casimir energies for
dielectric-diamagnetic ball (\ref{final-ball}) and for perfectly
conducting spherical shell (\ref{shell4}) prove to be positive. It
implies that the Casimir forces in these cases are {\it
repulsive}. In view of this  the Casimir idea~\cite{Cas-electron} to construct an
extended model for electron failed because the vacuum forces cannot stabilize it.

In principle, it is possible to get a negative Casimir energy for
a sphere, however for this purpose  we have to consider the scalar
massless field obeying on inner and outer surfaces of the sphere
the Neumann boundary conditions~\cite{NP}
\begin{equation}
\label{N-sphere} E^{\text{N}}=-\frac{1}{a}\,0.223777\ldots {.}
\end{equation}
The same field obeying the Dirichlet conditions on a sphere has
positive vacuum energy~\cite{NP}
\begin{equation}
\label{D-sphere} E^{\text{D}}=\frac{1}{a}\,0.002819\ldots {.}
\end{equation}
When obtaining the Casimir energies (\ref{N-sphere}) and
(\ref{D-sphere}) the summation over the values of the angular
momentum $l$ starts from $l=0$, unlike the electromagnetic field,
where $l$ assumes the values $1,2,\ldots$.

Undoubtedly, the calculation of the Casimir energy for a material
ball with $\varepsilon_1\mu_1\neq \varepsilon_2\mu_2$ by a
rigorous zeta function method is also of a special interest.
However, in this case the very definition of the spectral zeta
function should probably be changed in order to incorporate  the
contact terms which seem to be essential in this
problem~\cite{MNg2,BMM,BM,MNg1} (see sect.~\ref{CEDB}).

\subsection{Cylinder}

\label{SZF+VE-cyl}

Calculation of the Casimir energy of an circular infinite
cylinder~\cite{deraad,MNN} proves to be a more involved problem in
comparison with that for sphere. Here the spectral zeta function
$\zeta_{\text{cyl}}(s)$ for this configuration will be constructed,
its analytical continuation into the left half-plane of the
complex variable $s$ will be carried out, and relevant
singularities will be analyzed.

Thus we are considering an infinite cylinder of radius $a$ which
is placed in an uniform unbounded medium. The permittivity and the
permeability of the material making up the cylinder are
$\varepsilon_1$ and $\mu_1$, respectively, and those for
surrounding medium are $\varepsilon_2$ and $\mu_2$. We assume
again that the  condition (\ref{vel}) is fulfilled. In this case
the electromagnetic oscillations can  be divided into the
transverse-electric (TE) modes and transverse-magnetic (TM) modes.
In terms of the cylindrical coordinates $(r, \theta, z)$ the
eigenfunctions of the given boundary value problem contain the
multiplier \beq\label{mult} \exp{(\pm i\omega t+ik_zz+in\theta)}
\eeq and their dependence on $r$ is described by the Bessel
functions $J_n$ for $r< a$ and by the Hankel functions $H_n^{(1)}$
or $H_n^{(2)}$ for $r> a$. The frequencies of TE- and TM-modes are
determined, respectively, by the equations~\cite{Stratton}
\begin{eqnarray} \Delta_n^{\text{TE}}(\lambda , a)\equiv & \lambda
a[\mu_1J_n^{\prime}(\lambda a)H_n(\lambda a)-
\mu_1J_n(\lambda a)H_n^{\prime}(\lambda a)]=0 \label{fr1}\,{,} \\
\Delta_n^{\text{TM}}(\lambda , a)\equiv & \lambda a[\varepsilon_1
J_n^{\prime}(\lambda a)H_n(\lambda a)- \varepsilon_1 J_n(\lambda
a)H_n^{\prime}(\lambda a)]=0 \label{fr2}\,{,}
\end{eqnarray}
where $n=0,\pm 1,\pm 2, \ldots $. Here $\lambda^2$ is the eigenvalue
of the corresponding transverse (membrane-like) boundary value
problem \beq\label{lambda} \lambda^2=\frac{\omega^2}{c^2}-k_z^2\,{.}
\eeq Decoupling of TE- and TM-modes in this problem is due to the
condition (\ref{vel}).

 In a complete
analogy with the ball the Casimir energy per unit length of the
cylinder is defined by the relevant spectral zeta function
according to  eq. (\ref{3-11}). Let $\lambda_{nr}$ be the roots of
the frequency equations (\ref{fr1}) and (\ref{fr2}), then the
function $\zeta_{\text{cyl}} (s)$ is introduced in the following way
\begin{equation}\label{zcyl1}
\zeta_{\text{cyl}}(s)=c^{-2s}\int_{-\infty}^{+\infty}\frac{d
k_z}{2\pi}\sum_{n,r}
[(\lambda_{nr}^2(a)+k_z^2)^{-s}-(\lambda_{nr}^2(\infty)+k_z^2)^{-s}]\,{.}
\end{equation} In terms of the contour integral (see eq. (\ref{3-14}))
it can be represented in the form
\begin{equation}\label{zcyl2}
\zeta_{\text{cyl}}(s)=c^{-2s}\int_{-\infty}^{+\infty}\frac{d
k_z}{2\pi} \sum_{n=-\infty}^{+\infty} \oint_C
(\lambda^2+k_z^2)^{-s}d _{\lambda}\ln\frac{\Delta_n^{\text{TE}}
(\lambda a)\Delta_n^{\text{TM}} (\lambda
a)}{\Delta_n^{\text{TE}}(\infty)\Delta_n^{\text{TM}}(\infty)}\,{.}
\end{equation}
 Again we can take the contour $C$ to consist of the
imaginary axis $(+i\infty , -i\infty)$ closed by a semicircle of
an infinitely large radius in the right half--plane. Continuation
of the expressions $\Delta_n^{\text{TE}}(\lambda a)$ and
$\Delta_n^{\text{TM}}(\lambda a)$ into the complex plane $\lambda$
should be done in the same way as in the preceding section, i.e., \
by using $H_n^{(1)}(\lambda)$ for Im~$\lambda < 0$ and
$H_n^{(2)}(\lambda)$ for Im~$\lambda > 0$. On the semicircle the
argument of the logarithm in eq. (\ref{zcyl2}) tends to 1. As a
result this part of the contour $C$ does not give any contribution
to the zeta function $\zeta_{\text{cyl}} (s)$. When integrating
along the imaginary axis we choose the branch line of the function
$f(\lambda )=(\lambda^2+k_z^2)^{-s}$ to run between $-ik_z$ and
$+ik_z$, where $k_z=+\sqrt{k_z^2}>0$ and use further that branch
of this function which assumes real values when $\vert y\vert <
k_z$, where $y=$ Im~$\lambda$. More precisely we have \beq f
(iy)=\left\{\begin{array}{ll}
              e^{-i\pi s}(y^2-k_z^2)^{-s},  & y>k_z, \nonumber \\
  (k_z^2-y^2)^{-s},               & \vert y\vert <k_z,\label{phi} \\
              e^{i\pi s}(y^2-k_z^2)^{-s},   & y<-k_z.\nonumber
              \end{array}  \right.
\eeq Employment of the Hankel functions $H_n^{(1)}(\lambda)$ and
$H_n^{(2)}(\lambda)$ by extending the expressions
$\Delta_n^{\text{TE}}(\lambda )$ and $\Delta_n^{\text{TM}}(\lambda )$
into the complex plane $\lambda$, as it was noted above, gives
rise to the argument of the logarithm function depending only on
$y^2$ on the imaginary axis. It means that the derivative of the
logarithm is odd function of $y$. As a result the segment of the
imaginary axis $(-ik_z, +ik_z)$ gives zero, and eq.\ (\ref{zcyl2})
acquires the form
\begin{equation}\label{3.9}
\zeta_{\text{cyl}}(s)=\frac{\sin{\pi
s}}{c^{2s}\pi^2}\sum_{n=-\infty}^{+\infty}\int_0^{\infty} d
k_z\int_{k_z}^{\infty}(y^2-k_z^2)^{-s}d _y
\ln\frac{\Delta^{\text{TE}}(ay)_n\Delta^{\text{TM}}_n(ay)}
{\Delta^{\text{TE}}_n(i\infty)\Delta^{\text{TM}}_n(i\infty)}\,{.}
\end{equation}
 Changing the order of integration of $k_z$ and $y$ and
taking into account the value of the integral~\cite{GR} \beq\label{3.10}
\int_0^{y}d  k_z(y^2-k_z^2)^{-s}=\frac{\sqrt{\pi}}{2}\,
y^{1-2s}\, \frac{\Gamma\displaystyle
\left(1-{s}\right)}{\Gamma\left(\displaystyle\frac{3}{2}-s
\right)}\,{,}\qquad \text{Re}~s<1\,{,} \eeq we obtain after the
substitution $ay\to y$
\begin{equation}\label{3.11}
\zeta_{\text{cyl}}(s)=\frac{1}{2\sqrt{\pi}\, a\Gamma(s)\Gamma
\left(\displaystyle\frac{3}{2}-s\right)}
\left(\frac{a}{c}\right)^{2s}\sum_{n=-\infty}^{+\infty}\int_0^{\infty}
d  y\, y^{1-2s}\frac{d }{d  y}\ln[1-\xi^2\mu_n^2(y)]\,{,}
\end{equation}
where \beq\label{3.12} \mu_n(y)=y[I_n(y)K_n(y)]^{'}, \eeq and the
parameter $\xi^2$ was defined in eq. (\ref{xisigma}).
 We shall again content ourselves with the first
two terms in the uniform asymptotic expansion (\ref{UAE}) and take
into account only the terms linear in $\xi^2$. In this
approximation, upon changing the integration variable $y=nz,
n=\pm1, \pm2, \ldots$, we have \beq\label{3.13}
\ln\left\{1-\xi^2\left[z\frac{d }{d
z}(I_n(nz)K_n(nz))\right]^2\right\}= \eeq
$$
=
-\xi^2\,\frac{z^4t^6}{4n^2}\left[1+\frac{t^2}{4n^2}(3-30t^2+35t^4)
+{\cal O}(n^{-4})\right]\,{.}
$$
Now we substitute (\ref{3.13}) into all the terms in (\ref{3.11})
with $n\neq 0$. The term with $n=0$ in this sum will be treated by
subtracting and adding to the logarithmic function the quantity
\beq\label{3.14} -\frac{\xi^2}{4}\frac{y^4}{(1+y^2)^3}\,{.} \eeq
As a result the zeta function $\zeta_{\text{cyl}}(s)$ can be
presented now as the sum of three terms \beq\label{3.15}
\zeta_{\text{cyl}}(s)=Z_1(s)+Z_2(s)+Z_3(s)\,{,} \eeq where
\begin{eqnarray} Z_1(s)&=& \displaystyle \frac{\displaystyle
\left(\frac{a}{c}\right)^{2s}}{2\sqrt{\pi}a \Gamma(s)
\Gamma\left(\displaystyle\frac{3}{2}-s\right)}
\displaystyle\int_0^{\infty}d  y\,y^{1-s} \frac{d }{d
y}\left\{\ln [1-\xi^2\mu_0^2(y)]+
\frac{\xi^2}{4}\,y^4t^6\right\}, \label{3.16} \\
Z_2(s)&=& -\xi^2\displaystyle\left(\frac{a}{c}
\right)^{2s}\frac{\displaystyle 2\sum_{n=1}^{+\infty}n^{-2s-1} +1}
{\displaystyle 8\sqrt{\pi}\,a\,\Gamma
(s)\Gamma\left(\displaystyle\frac{3}{2}-s \right)}
\displaystyle\int_0^{\infty}d  z\,z^{1-s}\frac{d  }{d  z}
 (z^4t^6) \,{,} \label{3.17} \\
Z_3(s)&=& -\xi^2\frac{\displaystyle
\left(\frac{a}{c}\right)^{2s}\displaystyle
\sum_{n=1}^{+\infty}n^{-3-2s}} {\displaystyle 32\sqrt{\pi}a\,
\Gamma(s) \Gamma\left(\displaystyle\frac{3}{2}-s
\right)}\displaystyle\int_0^{\infty}d  z\,z^{1-2s}
\frac{d }{d  z}[z^4t^8(3-30t^2+35t^4)]\,{.} \label{3.18}
\end{eqnarray}
 In these equations $Z_1(s)$ has accumulated the term
with $n=0$ from eq.\ (\ref{3.11}) subtracted by (\ref{3.14});
$Z_2(s)$ involves the contribution of the term of order $1/n^2$ in
expansion (\ref{3.13}) and the added expression (\ref{3.14});
$Z_3(s)$ is generated by the terms of order $1/n^4$ in the
expansion (\ref{3.13}).

Taking into account that \beq\label{3.19} \mu_0^2(y)\vert_{y\to
0}\to 1\quad \mbox{and}\quad \mu_0^2(y)\vert_{y\to \infty}\to
\frac{1}{4y^2}+\frac{3}{16y^4}\,{,} \eeq the integration by parts
in eq.\ (\ref{3.16}) can be done  for $-3/2<\text{Re}~s<1/2$ with
the result \begin{equation} \label{3.20}
Z_1(s)=\frac{(2s-1)a^{2s-1}}{2\sqrt{\pi}\,c^{2s} \Gamma(s)
\Gamma\left(\displaystyle\frac{3}{2}-s\right)}
\int_0^{\infty}\frac{d  y}{y^{2s}} \left\{\ln
[1-\xi^2\mu_0^2(y)]+ \frac{\xi^2}{4}\,y^4t^6(y)\right\}\,{.}
\end{equation}
 With allowance for (\ref{3.19}) one infers easily that the
function $Z_1(s)$ is an analytic function of the complex variable
$s$ in the region $-3/2<\mbox{Re}~s<1/2$. In the linear order of
$\xi^2$ it reduces to \beq\label{Z1lin}
Z_1^{\text{lin}}(s)=\xi^2\displaystyle\frac{2s-1}{2\sqrt{\pi}a
\Gamma(s)\Gamma
\left(\displaystyle\frac{3}{2}-s\right)}\left(\frac{a}{c}\right)^{2s}
\int_0^{\infty}\frac{d
y}{y^{2s}}\left[\frac{y^2}{4(1+y^2)^3}-\mu_0^2(y)\right]\,{.} \eeq
This function is also analytic in the region
$-3/2<\mbox{Re}~s<1/2$. Integration in eq.\ (\ref{3.17}) can be
accomplished exactly by making use of the formula~\cite{GR}
\begin{equation}
\label{3.21} \int_0^{\infty}d  z\,z^{1-2s}\frac{d }{d
z}(z^4t^6)=\frac{2s-1}{4}
\Gamma\left(\displaystyle\frac{1}{2}+s\right)
\Gamma\left(\displaystyle\frac{5}{2}-s\right)\,{,}
-\frac{1}{2}<\mbox{Re}~s<\frac{5}{2}\,{.}
\end{equation}
This gives for $Z_2(s)$ in (\ref{3.17}) \beq\label{Z2}
Z_2(s)=\xi^2\left(\frac{a}{c}\right)^{2s}
\frac{(1-2s)(3-2s)}{64\sqrt{\pi}\,a}\left(
2\sum_{n=1}^{\infty}n^{-2s-1}+1\right)
\frac{\Gamma\left(\displaystyle\frac{1}{2}+s\right)}{
\Gamma(s)}\,{.} \eeq In view of the sum over $n$ in (\ref{Z2}) the
function $Z_2(s)$ is defined only for $s>0$.

For simplicity we apply in eq.\ (\ref{3.18}) the integration by
parts which is correct for $-3/2<\mbox{Re}~s<1$ and leads to the
result \beq\label{Z3}
Z_3(s)=\xi^2\left(\frac{a}{c}\right)^{2s}\frac{(1-2s)(3-
2s)(28s^2-8s-27)}{6144\sqrt{\pi}\,a}
\frac{\Gamma\left(\displaystyle\frac{3}{2}+s\right)}{
\Gamma(s)}\,\sum_{n=1}^{\infty}n^{-2s-3}\,{.} \eeq Again the sum
over $n$ in (\ref{Z3}) gives the restriction Re~$s>-1$ for the
definition of the function $Z_3(s)$.

Thus the spectral zeta function $\zeta_{\text{cyl}}(s)$ in the
linear approximation with respect to $\xi^2$ and with allowance
for the first two terms in the UAE (\ref{3.13}) is given by
\beq\label{3.25}
\zeta_{\text{cyl}}(s)=Z_1^{\text{lin}}(s)+Z_2(s)+Z_3(s)\,{,} \eeq
where the $Z$'s are presented in eqs. (\ref{Z1lin}), (\ref{Z2})
and (\ref{Z3}), respectively. Summing up all the restrictions on
the complex variable $s$ which have been imposed when deriving
these equations we infer that
$\zeta_{\text{cyl}}(s)$ is defined in the strip $0<\mbox{Re}~s<1/2$.
In order to continue this function into the surroundings of the
point $s=-1/2$, it is sufficient to express the sum in eq.\
(\ref{Z3}) in terms of the Riemann zeta function and treat the
right-hand side of eq.\ (\ref{3.21}) as an analytic continuation
of its left-hand side over all the complex plane $s$. This gives
\begin{equation}\label{Z3con}
Z_2(s)
=\xi^2\left(\frac{a}{c}\right)^{2s}\frac{(1-2s)(3-2s)}{64\sqrt{\pi}\,
a}\,[2\zeta_{\text{R}} (2s+1)+1]
\displaystyle\frac{\Gamma\left(\displaystyle\frac{1}{2}+s\right)}
{\Gamma(s)}\,{.}
\end{equation}
 It is left now to take the limit
$s\to -1/2$ in eqs.\ (\ref{Z1lin}), (\ref{Z2}) and (\ref{Z3con}).
A special care should be paid when calculating this limit in eq.\
(\ref{Z3con}) in view of the poles of the function $\Gamma(1/2+s)$
at this point. Using the formulae \beq\label{3.27} \zeta_{\text{R}}
(0)=-\frac{1}{2}, \quad \zeta_{\text{R}}^{\prime}(0)=-\frac{1}{2}\ln
2\pi\,{,} \quad \Gamma(x)=\frac{1}{x}-\gamma +{\cal O}(x)\quad
x\to 0 \,{,} \eeq one derives \beq\label{3.28} \lim_{s\to
-1/2}[2\zeta_{\text{R}} (1+2s)+1]\Gamma\left(\frac{1+2s}{2}\right)=
\eeq
$$
=\lim_{s\to -1/2}[2\zeta_{\text{R}}
(0)+2\zeta_{\text{R}}^{\prime}(0)(1+2s)+ {\cal O}
((1+2s)^2)+1]\left[\frac{2}{1+2s}+ \gamma+{\cal O}
(1+2s)\right]=-2\ln (2\pi )\,{.}
$$
Making allowance for this, we obtain from (\ref{Z3con})
\beq\label{Z2fin} Z_2(-1)=\frac{\ln (2\pi)}{8\pi}\frac{c\xi^2}{
a^2}\,{.} \eeq The appearance of the finite term proportional to
$\ln (2\pi )$ is remarkable for the problem under consideration.
It is derived here in a consistent way by making use of an
analytic continuation of the relevant spectral zeta function. In
ref.~\cite{MNN} it was calculated  in a more transparent though not
rigorous way.

Gathering together eqs. (\ref{Z1lin}), (\ref{Z3}) with $s=-1/2$
and eq. (\ref{Z2fin}) we obtain \beqa \zeta_{\text{cyl}}(-1/2) & = &
\frac{c\xi^2}{2\pi a^2}\left\{\int_0^{\infty} y\,d
y\left[\frac{y^4}{4(1+y^2)^3}-\mu_0^2(y)\right]+\frac{1}{48}
\sum_{n=1}^{+\infty}\frac{1}{n^2}+\frac{1}{4}\ln (2\pi )\right\}
\nonumber \\
                & = & (-0.490878+0.034269+0.459469)\frac{c\xi^2}{2\pi
                a^2} =  0.002860 \frac{c\xi^2}{2\pi a^2}\,{.}
                \label{cyl3}
\eeqa This result is not the final answer in the problem in hand.
Point is that in view of severe cancellations in eq.\ (\ref{cyl3})
the contribution of the next term in the UAE (\ref{3.13}) proves
to be essential. Its account gives~\cite{MNN}
 \beq\label{3.31}
\zeta_{\text{cyl}}(-1/2)=0\,{.} \eeq
 Thus the Casimir energy of a
compact cylinder possessing the same speed of light inside and
outside proves to be zero in the $\xi^2$-approximation. The
consideration presented in this section can be extended to the
next term of order $\sim 1/n^6$ in the UAE (\ref{3.13}) in a
straightforward way. Therefore we shall not present here these
rather cumbersome expressions.  More precisely, in ref.\
                 \cite{MNN} not only the next term in UAE (\ref{3.13}) has been
                 taken into account but also the first 5 terms in the sum in
                 (\ref{3.18}) were taken exactly instead of using their asymptotics
                 $n^{-3-2s}$. This turns out to be essential for reaching the
                 required precision. However this point is not critical for our
                consideration because $Z_3(s)$ does not need analytic
                continuation. The exact zero result for this
                energy was obtained in paper~\cite{Klich-Romeo} by
                making use of the addition formulae for the Bessel
                functions instead of the UAE.

   However the Casimir energy of a compact cylinder
possessing the same speed of light inside and outside does not
vanish in higher approximations with respect to the parameter
$\xi^2$~\cite{NP-cyl,Klich-Romeo}. For example, up to the $\xi^4$
term it is defined by the formula
\begin{equation}
E(\xi^2)=-0.0955275\frac{c\xi^4}{4\pi a^2}\,
=-0.007602\,\frac{c\xi^4}{a^2}. \label{e-2-26}
\end{equation}
In contrast to the Casimir energy of a compact ball with the same
properties (see eq.(\ref{final-ball}))
\begin{equation}
E_{\text{ball}}\simeq \frac{3}{64 a}c \xi^2 =0.046875 \,\frac{c
\xi^2}{a} \label{e_2_27}
\end{equation}
the Casimir energy of a cylinder under consideration turned out to
be  negative. Consequentially, the Casimir forces strive to
contract the cylinder. The numerical coefficient in
eq.~(\ref{e-2-26}) proved really to be small, for example, in
comparison with the analogous coefficient in eq.~(\ref{e_2_27}).
Probably it is a manifestation of the vanishing of the Casimir
energy of a pure dielectric  cylinder noted in the Introduction.
Numerically the Casimir energy $E(\xi^2)$ was calculated for
arbitrary values of the parameter $\xi^2$ in paper~\cite{NP-cyl}.

Now we address to the consideration of a special case when $\xi
=1$. It corresponds to a perfectly conducting cylindrical shell
\cite{MNN}. Instead of the expansion (\ref{3.13}) we have
\beq\label{3.32} \ln\left\{1-\left[z\frac{d }{d
z}(I_n(nz)K_n(nz))\right]^2\right\}= \eeq
$$
= -\frac{z^4t^6}{4n^2}\left[1+\frac{t^2}{4n^2}\left(3-
30t^2+35t^4+\frac{1}{2}\,z^4t^4 \right)+{\cal
O}(n^{-4})\right]\,{.}
$$
Proceeding in the same way as above we obtain for the spectral
zeta function concerned \beq\label{zetasc}
\zeta_{\text{cyl}}^{\text{shell}}(s)=Z_1(s)+Z_2(s)+Z_3(s)\,{,} \eeq
where $Z_1(s)$ is given by eq. (\ref{3.20}) with $\xi =1$,
$Z_2(s)$ is the same as in eq. (\ref{Z3con}), and $Z_3(s)$ now is
\beq\label{Z3N}
Z_3(s)=\frac{(1-2s)(3-2s)(284s^2-104s-235)}{61440\sqrt{\pi}a^{1-2s}}\,\frac{
\displaystyle \Gamma\left(\frac{3}{2}+s\right)}{\displaystyle
\Gamma(s)} \sum_{n=1}^{+\infty}n^{-3-2s}\,{.} \eeq At the point
$s=-1/2$ it acquires the value \beqa \zeta_{\text{cyl}}^{\text{shell}}\left
(-\frac{1}{2}\right )&=&-0.6517\frac{1}{2\pi a^2}+ \frac{1}{2\pi
a^2}\frac{7}{480} \sum_{n=1}^{+\infty}\frac{1}{n^2}+\frac{1}{8\pi
a^2}\,\ln (2\pi )
\nonumber \\
                       &=&(-0.6517+0.0240+0.4595)\frac{1}{2\pi
                       a^2} = - 0.0268\frac{1}{a^2}{.} \label{zetasc1}
\eeqa This exactly reproduce the contribution of the first two
terms in calculations of the Casimir energy for perfectly
conducting circular cylindrical shell in ref.~\cite{MNN}. With
higher accuracy this energy is given by~\cite{MNN,deraad}
\beq\label{deraad} E_{\text{cyl}}^{\text{shell}}=-
0.01356\frac{1}{a^2}{.} \eeq

The Casimir energy of a massless field subjected to Dirichlet and
Neumann conditions on internal and external surfaces of an
infinite circular cylindrical shell are also finite but of
opposite signs~\cite{cylinder+circle}
\begin{equation}
\label{4-72} E^{\text{D}}_{\text{cyl}}=0.000606\frac{1}{a^2}{,}
\end{equation}
\begin{equation}
\label{4-73} E^{\text{N}}_{\text{cyl}}=-0.01417\frac{1}{a^2}{.}
\end{equation}
The sum of these two energies gives the vacuum energy of
electromagnetic field for this configuration~(\ref{deraad}).

However the zeta regularization does not give a finite value for
the Casimir energy in the case of the  two-dimensional version of
the configuration under consideration, i.e., \ for a circle of
radius $a$ placed on a plane. In paper~\cite{cylinder+circle}
these energies were calculated by making use of the relation
between the zeta functions for a cylinder and a circle of the same
radius
\begin{equation}
\label{relation} \zeta_{\text{cir}}(s)=2\sqrt
\pi\frac{\Gamma(s+1/2)}{\Gamma(s)}\zeta_{\text{cyl}}(s+1/2){.}
\end{equation}
Having calculated the zeta function for an infinite cylinder in
the region $-1/2 \leq \tx{Re}\,
 s\leq 0$ one obtaines immedeatly
the Casimir energy for a cylinder via $\zeta_{\text{cyl}}(-1/2)$ and
the Casimir energy for a circle in terms of $\zeta_{\text{cyl}}(0)$
\begin{equation}
\label{4-74} E^{\text{D}}_{\text{cir}}=\frac{1}{a}\left( 0.0023595
-\left .\frac{1}{256} \frac{1}{s}\right |_{s\to 0} \right ){,}
\end{equation}
\begin{equation}
\label{4-75} E^{\text{N}}_{\text{cir}}= -\frac{1}{a}\left( 0.3131
+\left .\frac{5}{256} \frac{1}{s}\right |_{s\to 0} \right ) {.}
\end{equation}
The Casimir energy for the sum of these two fields is also
infinite
\begin{equation}
\label{4-76} E^{\text{D+N}}_{\text{cir}}=- \frac{1}{a}\left( 0.2895
+\left .\frac{3}{128} \frac{1}{s}\right |_{s\to 0} \right ) {.}
\end{equation}
On a plane electromagnetic field is reduced to a scalar massless
field obeying the Neumann boundary conditions~\cite{LesRom}. It was
shown, that the zeta function regularization  does not lead to a
finite answer for the vacuum energy for spheres  in spaces of
arbitrary even dimensions~\cite{LesRom,Sen,BenderMilton,CEK}. As
usual the coefficients in front of the pole-like contributions
calculated by different methods coincide, but the finite parts of
the answers differ~\cite{LesRom,Sen}. Thus for obtaining here
physical results additional renormalization is needed.

In ref.~\cite{romeo} the vacuum energy of a perfectly conducting
cylindrical surface has been calculated to much higher accuracy by
making use of another version of the zeta function technique. By
integrating over $d  k_z$ directly in eq.\ (\ref{zcyl1}) the
authors reduced this problem to investigation of the zeta function
for circle, which has been considered earlier by introducing the
partial wave zeta functions for interior and exterior region
separately. In this respect our approach dealing only with one
spectral zeta function for given boundary conditions proves to be
more simple and straightforward.

Many results on calculation of the Casimir energies for
spherically symmetric cavities by making use of the zeta function
technique can be found in ref.~\cite{CEK}.

\section{Casimir energy of a dielectric ball}

\setcounter{equation}{0} \label{CEDB} Calculation of the vacuum energy of electromagnetic
field connected with a dielectric ball proves to be a quite
involved problem. The first attempt to calculate  this
energy  has been undertaken by K.~A.\
Milton in 1980~\cite{Milton}.  And only just recently the vacuum
electromagnetic energy of a dilute dielectric ball was
found at zero~\cite{MNg2,Barton,BMM,BM,LSN,Mar,HB-stat,HBA-stat-QFT} and
finite~\cite{NLS,Barton-T} temperature. It is worth noting that
the first rough and not rigorous estimation of the
Casimir energy of a dilute dielectric ball has been done in
ref.~\cite{BNP}. The difficulties encountered here are due to the
following mathematical peculiarities of the problem in question.

As it was noted in subsect.~\ref{SZF+VE-sph} the solutions to the
Maxwell equations with spherical symmetry are expressed  in terms
of two scalar Debye potentials obeying Helmholtz equations
(\ref{4-8}). In the Casimir calculations these equations should be
treated as one eigenvalue problem, hence we have to rewrite them
in the form
\begin{equation}
\label{5-1} -c^2_i \Delta \psi_{lm}(r)= \omega ^2\psi_{lm}(r),
\quad c_i^2=\frac{c^2}{\varepsilon_i \mu_i},\quad i=1 \tx{ for }
r<a,\quad i=2 \tx{ for }r>a{.}
\end{equation}
Here $c_1$ and $c_2$ are the velocities of light respectively
inside and outside the ball. Thus the coefficient in front of the
differential operator $-\Delta$ is {\it discontinuous}, it has a jump at
the surface of the ball. It turns out that the heat kernel
coefficient $B_2$ for the spectral problem (\ref{5-1}) does not
vanish but is a function of $c_1-c_2$~\cite{Bordag-db}. If this
difference is small (a dilute dielectric ball) it was shown that
\begin{equation}
\label{5-2} B_2\sim(c_1-c_2)^3{.}
\end{equation}
Thus the coefficient $B_2$ vanishes up to terms $(c_1-c_2)^2$. It
is this approximation that have been used in all the calculations
of the Casimir energy of a dielectric ball.

In sect.~\ref{RVE} we have tried to show that the zeta function
technique is the most consistent, from the mathematical standpoint,
method for calculation of vacuum energy. However one did
not manage to use this technique in a direct way for finding the
Casimir energy of a dielectric ball. Here we describe briefly the
methods practically  used in these calculations.

In ref.~\cite{MNg2} the Casimir energy of a dilute dielectric ball
was derived  by summing up the van den Wails forces between the
individual molecules inside the compact ball. Barton~\cite{Barton}
found this energy in the framework of a special perturbation
theory, in which the dielectric ball was treated as a {\it
perturbation} when treating the electromagnetic field in unbounded
empty space. Milton, Brevik and Marachevsky~\cite{BMM,BM,Mar}
started from the Green's function of the quantized Maxwell field
with an explicit account of the so called {\it contact terms}, on
the stage of the numerical calculations the uniform asymptotic
expansion for the Bessel functions and the zeta regularization
technique being applied.  We have considered  this
problem~\cite{LSN} by direct mode summation method with use of
the addition theorem for the Bessel functions. In calculations
without employment of  the uniform asymptotic expansions for the
Bessel functions~\cite{MNg2,Barton,LSN} the exact (in the $\Delta
n^2$ approximation) result for the Casimir energy at hand was
obtained.  Brevik {\it et al}~\cite{HB-stat,HBA-stat-QFT} considered
this problem in a statistical approach.

First we present the calculation of this Casimir energy at  zero
temperature and further we  extended this approach to a finite
temperature.

\subsection{Zero temperature}

\label{ddb-0} As in subsect.~\ref{SZF+VE-sph},  a solid ball of a
radius $a$ placed in an unbounded uniform medium is considered but
the condition (\ref{vel}) is not  imposed now. Unfortunately, the zeta
regularization cannot be used here in a direct way. The reason for
this will be seen below.

     We shall proceed from the standard definition of the vacuum
energy as the sum over the eigenfrequencies of electromagnetic
oscillations
\begin{equation}\label{def}
  E=\frac{1}{2}\sum_s(\omega_s-\overline{\omega}_s)\,.
\end{equation}
Here $\omega_s$ are the  frequencies defined by eqs.~(\ref{TE})
and (\ref{TM}), and the frequencies $\bar{\omega}_s$ correspond to
a certain limiting boundary conditions that will be specified
below.

     In the case of the plane geometry of boundaries or when
considering the Casimir effect for  distinct bodies it is
sufficient to subtract in eq.\ (\ref{def}) the contribution of the
Minkowski space. In the problem at hand it implies  taking the
limit $a\to \infty$, i.e., \ that the medium 1 tends to fill the
entire space. But it turns out that this subtraction is not
sufficient because the linear in $\varepsilon_1-\varepsilon_2$
contribution into the vacuum energy retains.  Further, we assume
that the difference $\varepsilon_1-\varepsilon_2$ is small and
content ourselves only with the
$(\varepsilon_1-\varepsilon_2)^2$-terms.

     The necessity to subtract the contributions to the vacuum
energy linear in  $\varepsilon_1-\varepsilon_2$ is justified by
the following consideration.  The Casimir energy of a dilute
dielectric ball can be thought of as the net result of the van der
Waals interactions between the molecules making up the ball
\cite{MNg2}. These interactions are proportional to the dipole
momenta of the molecules, i.e., to the quantity
$(\varepsilon_1-1)^2$. Thus, when a dilute dielectric ball is
placed in the vacuum, then its Casimir energy should be
proportional to $(\varepsilon_1-1)^2$. It is natural to assume
that when such a dielectric ball is surrounded by an infinite
dielectric medium with permittivity $\varepsilon_2$, then its
Casimir energy should be proportional to
$(\varepsilon_1-\varepsilon_2)^2$. The physical content of the
contribution into the vacuum energy linear in
$\varepsilon_1-\varepsilon_2$ has been investigated in the
framework of the microscopic model of the dielectric media (see
paper~\cite{MSS} and references therein). It has been shown that
this term represents the self-energy of the electromagnetic field
attached to the polarizable particles or, in more detail, it is
just the sum of the individual atomic Lamb shifts. Certainly this
term in the vacuum energy should be disregarded when calculating
the Casimir energy which is originated in the electromagnetic
interaction between {\it different} polarizable particles or atoms
\cite{Barton,BMM,HB-stat,HBA-stat-QFT,Barton-dis}.

     Further, we put for sake of symmetry
\begin{equation}\label{dn}
  \sqrt{\varepsilon_1}=n_1=1+\frac{\Delta n}{2}\,,\quad
 \sqrt{\varepsilon_2}=n_2=1-\frac{\Delta n}{2}\,.
\end{equation}
Here $n_1$ and $n_2$ are the refractive indices of the ball and of
its surroundings, respectively, and it is assumed that $\Delta
n<<1$. From here it follows, in particular, that
\begin{equation}\label{epn}
  \varepsilon_1-\varepsilon_2=(n_1+n_2)(n_1-n_2)=2\Delta n\,.
\end{equation}
Thus, using the definition (\ref{def}) we shall keep in mind that
really two subtractions should be done: first the contribution,
obtained in the limit $a\to \infty$, has to be subtracted and then
all the terms linear in $\Delta n$ should also be removed.

We present the vacuum energy defined by eq.\ (\ref{def}) in terms
of the contour integral in the complex frequency plane. Upon the
contour deformation one gets
\begin{equation}\label{integral}
  E=-\frac{1}{2\pi}\sum_{l=1}^{\infty}(2l+1)\int_0^{K}d  y\,y\,\frac{d }{d  y}
  \ln\frac{\Delta_l^{\text{TE}}(iay)\Delta_l^{\text{TM}}(iay)}
  {\Delta_l^{\text{TE}}(i\infty)\Delta_l^{\text{TM}}(i\infty)}\,,
\end{equation}
where $\Delta_l^{\text{TE}}(iay)$ and $\Delta_l^{\text{TM}}(iay)$ are
defined in eqs. (\ref{TE}) and (\ref{TM}).

In eq.\ (\ref{integral}) we have introduced cutoff $K$  in
integration over the frequencies. This regularization is natural
in the Casimir problem because physically it is clear that the
photons  of a very short wave length do not contribute to the
vacuum energy since they do not ``feel'' the boundary between the
media with different permittivities $\varepsilon _1$ and
$\varepsilon _2$. In the final expression the regularization
parameter $K$ should be put to tend to infinity, the divergencies,
that may appear here, being  cancelled by appropriate counter
terms.

 Taking into account the
asymptotics of the Riccati-Bessel functions we obtain
\begin{equation}\label{asymp2}
 \Delta_l^{\text{TE}}(i\infty)\Delta_l^{\text{TM}}(i\infty)=
 -\frac{(n_1+n_2)^2}{4}\,e^{2(n_1-n_2)y}\,.
\end{equation}
Upon substituting eqs.\ (\ref{TE}),  (\ref{TM}) and (\ref{asymp2})
into eq.\ (\ref{integral}) and changing the integration variable
$ay\to y$, we cast eq.\  (\ref{integral}) into the form (see ref.
\cite{BNP})
\begin{eqnarray}\label{final}
 E&=&-\frac{1}{2\pi a}\sum_{l=1}^{y_0}(2l+1)\int_0^{y_0}d  y
 \,y\,\frac{d }{d  y}
  \ln\left\{ \frac{4e^{-2(n_1-n_2)y}}{(n_1+n_2)^2} \right. \\
  & & \left. [n_1n_2((s_l^{\prime}e_l)^2+
  (s_le_l^{\prime})^2)-(n_1^2+n_2^2)s_ls_l^{\prime}e_le_l^{\prime}]\right\}\,,
  \nonumber
\end{eqnarray}
where $s_l\equiv s_l(n_1y)$, $e_l\equiv e_l(n_2y)$, $y_0=a K$.

It should be noted here that in eq.\ (\ref{final}) only the first
subtraction is accomplished, which removes the contribution to
the vacuum energy obtained when $a\to\infty$. As noted above, for
obtaining the final result all the terms linear in $\Delta n$
should be discarded also.

Further it is convenient to rewrite eq.\  (\ref{final}) in the
form
\begin{equation}\label{E1E2}
 E=E_1+E_2
\end{equation}
with
\begin{eqnarray}\label{E1}
 E_1&=&\frac{\Delta n}{2\pi
 a}\sum_{l=1}^{\infty}(2l+1)\int_0^{y_0}y\,d  y\,, \\
 E_2&=&-\frac{1}{2\pi a}\sum_{l=1}^{\infty}(2l+1)\int_0^{y_0}\!\!d  y\,
 y\,\frac{d  }{d  y}\ln\left[W_l^2(n_1y, n_2y)-\frac{\Delta n^2}{4}\,
 P_l^2(n_1y, n_2y)\right],
 \label{E2}
\end{eqnarray}
where
\begin{eqnarray}
  W_l(n_1y, n_2y)&=&s_l(n_1y)e_l^{\prime}(n_2y)-s_l^{\prime}(n_1y)e_l(n_2y)\,,
  \label{W} \\
  P_l(n_1y, n_2y)&=&s_l(n_1y)e_l^{\prime}(n_2y)+s_l^{\prime}(n_1y)e_l(n_2y)\,.
  \label{P}
\end{eqnarray}
The term $E_1$ accounts for only the expression $\exp (-2\Delta n
\,\,y)$ in the argument of the logarithm function in eq.\
(\ref{final}) and it appears as a result of subtracting the
Minkowski space contribution to the Casimir energy (the sum with
$\bar \omega_s$ in eq.\ (\ref{def}) and the denominator in eq.\
(\ref{integral})).

It is worth noting that the term $E_1$ is exactly  the Casimir
energy considered by Schwinger in his attempt to explain the
sonoluminescence~\cite{sono}. Really, introducing the cutoff $K$
for frequency integration and the cutoff $y=\omega/a$ for the
angular momentum summation we arrive at the result
\begin{equation}\label{Schwinger}
  E_1=\frac{\Delta n}{\pi a}\int_0^{aK}y\,d  y\, \sum_{l=1}^{\infty}
  \left(l+\frac{1}{2}\right)\sim \frac{\Delta n}{2\pi a}\int_0^{aK}y^3\,d  y=
  \Delta n\,\frac{K^4a^3}{8\pi}\,.
\end{equation}
We have substituted here the summation over $l$ by integration. Up
to the multiplier $(-2/3)$ it is exactly the Schwinger value for
the Casimir energy of a ball ($\varepsilon_1=1$) in water
($\sqrt{\varepsilon_2}\simeq 4/3$)~\cite{MNg1}.  The term linear
in $\Delta n$ and of the same structure was also derived in papers
 \cite{Barton,HB-stat,Barton-dis}. As it was explained
above the energy $E_1$  should be discarded.

In our calculation, we content ourselves with the $\Delta
n^2$-approximation. Hence, in eq.\  (\ref{E2}) one can put
$P^2_l(n_1y, n_2y)\simeq P^2_l(y,y)$ and keep in expansion of the
logarithm function only the terms proportional to $\Delta n^2$. In
this approximation, the contributions of $W_l^2$ and $P_l^2$ into
the vacuum energy are additive
\begin{equation}\label{EWP}
  E^{\text{ren}}=E_W^{\text{ren}}+E_P^{\text{ren}}\,.
\end{equation}
In the Appendix \ref{appA} it is shown that for obtaining the
$\Delta n^2$-contribution into the Casimir energy of the function
$W_l^2$ in the argument of the logarithm in eq.\  (\ref{E2}), it
is sufficient to calculate the $\Delta n^2$-contribution of the
function $W_l^2$ alone but changing the sign of this contribution
to the opposite one (see eq.\ (\ref{AW20})). Hence,
\begin{equation}\label{EW} E_W=\frac{1}{2\pi
  a}\sum_{l=1}^{\infty}(2l+1)\int_0^{y_0}d  y \,y\,\frac{d }{d  y}
 W_l^2(n_1y, n_2y)\,,
\end{equation}
and only the $\Delta n^2$-term being preserved in this expression.

For $E_P$ we have
\begin{equation}\label{EP}
  E_P=\frac{\Delta n^2}{8\pi
  a}\sum_{l=1}^{\infty}(2l+1)\int_0^{y_0}\,d  y\,
 y\frac{d  }{d  y} P_l^2(n_1y, n_2y)\,.
\end{equation}
Usually, when calculating the vacuum energy in the problem with
spherical symmetry, the uniform asymptotic expansion of the Bessel
functions is used (see sect.\ \ref{SZF+VE}). As a result, an
approximate value of the Casimir energy can be derived, the
accuracy of which depends on the number of terms preserved in the
asymptotic expansion.

We shall persist in another way employing the technique of the
paper~\cite{Klich}. By making use of the addition theorem for the
Bessel functions~\cite{AS}, we first do the summation over the
angular momentum $l$ in eq.\  (\ref{E2}) and only after that we
will integrate over the imaginary frequency~$y$. As a result, we
obtain an exact (in the $\Delta n^2$-approximation) value of the
Casimir energy in the problem involved.

The addition theorem for the Bessel functions is given by the formula~\cite{AS}
\begin{equation}\label{addition}
  \sum_{l=0}^{\infty}(2l+1)s_l(\lambda
  r)e_l(\lambda\rho)P_l(\cos\theta)=\frac{\lambda r\rho}{R}\,
e^{-\lambda R}
  \equiv {\cal D}\,,
\end{equation}
where
\begin{equation}\label{R}
  R=\sqrt{r^2+\rho^2-2r\rho\cos\theta}\,.
\end{equation}
Differentiating the both sides of eq.\  (\ref{addition}) with
respect to $\lambda r$ and squaring the result we deduce
\begin{equation}\label{Dsqr}
  \sum_{l=0}^{\infty}(2l+1)[s_l^{\prime}(\lambda
  r)e_l(\lambda\rho)]^2=\frac{1}{2r \rho}\int_{r-\rho}^{r+\rho}
  \left(\frac{1}{\lambda}\,\frac{\partial{\cal D}}{\partial r}
 \right)^2R\,d  R\,.
\end{equation}
Here the orthogonality relation for the Legendre polynomials
\[
  \int_{-1}^{+1}P_l(x)P_m(x)d  x=\frac{2\delta_{lm}}{2l+1}
\]
has been taken into account. Now we put
\begin{equation}\label{ndel}
  \lambda=y\,,\quad r=n_1=1+\frac{\Delta n}{2}\,,\quad
  \rho=n_2=1-\frac{\Delta n}{2}\,.
\end{equation}
Applying eq.\ (\ref{Dsqr}) and analogous ones, we derive
\begin{eqnarray}\nonumber
  \sum_{l=1}^{\infty}(2l+1)W_l^2(n_1y, n_2y)&=&
  \frac{1}{2r\rho\lambda^2}\int_{r-\rho}^{r+\rho}R\,d  R\left({\cal D}_r-
  {\cal D}_{\rho}\right)^2-e^{2\Delta n y} \\
     &=& \frac{\Delta n^2}{8}\int_{\Delta n}^{2}\frac{e^{-2yR}}{R^5}
     \left(4+R^2+4yR-yR^3\right)^2d  R-e^{2\Delta n y}\,, \label{WD} \\
  \sum_{l=1}^{\infty}(2l+1)P_l^2(y,
  y)&=&\frac{1}{2}\int_0^2\left[\frac{\partial}{\partial y}
  \left(\frac{y}{R}\,e^{-yR}\right)\right]^2R\,d  R-e^{-4y}\,. \label{PD}
\end{eqnarray}
Here ${\cal D}_r$ and ${\cal D}_{\rho}$ stand for the results of
the partial differentiation of the function ${\cal D}$ in eq.\
(\ref{addition}) with respect to the corresponding variables and
with the subsequent substitution of (\ref{ndel}). The last terms
in eqs.\ (\ref{WD}) and (\ref{PD}) are $W_0^2(n_1y, n_2y)$ and
$P_0^2(y, y)$, respectively. As it was stipulated before, in eq.\
(\ref{WD}) we have to keep only the terms proportional to $\Delta
n^2$ and in eq.\  (\ref{PD}) we have put $\Delta n=0$.

The calculation of the contribution $E_P$ to the Casimir energy is
straightforward.  Upon differentiation of the right-hand side of
eq.\ (\ref{PD}) with respect to $y$, the integration over $d  R$ can
be done here. Substitution of this result into eq.\ (\ref{EP})
gives
\begin{equation}\label{EP1}
  E_P=-\frac{\Delta n^2}{2\pi
  a}\left(-\frac{1}{4}\right)\int_0^{y_0}d  y\,\left[e^{-4y}\left(2y^2+2y+
  \frac{1}{2}\right)-\frac{1}{2}\right]\,.
\end{equation}
The term $(-1/2)$ in the square brackets in eq.\  (\ref{EP1})
gives rise to the divergence\footnote{This divergence has the same
origin as those arising in summation over $l$ when the uniform
asymptotic expansions of the Bessel functions are used
\cite{BMM,BM}. The technique employed here is close to the
multiple scattering expansion \cite{BD}, where these divergencies
are also subtracted.} when  $y_0 \to \infty$
\begin{equation}
\label{ct1} E_P^{\text{div}} = - \frac{\Delta n^2}{16\pi a}y_0\,{.}
\end{equation}
Therefore we have to subtract it with the result
\begin{equation}\label{EPF}
  E_P^{\text{ren}}=E_P-E_P^{\text{div}}
=\frac{5}{128}\,\frac{\Delta n^2}{\pi a}\,.
\end{equation}

It is worth noting here that eq. (\ref{EPF}) after substitution
$\Delta n^2/4=\xi^2$ gives the {\it exact}, in the
$\xi^2$-approximation, value for the Casimir energy of a compact
ball with the same velocity of light inside and outside the
ball~\cite{Klich,LSN,NLS,BD}
\begin{equation}
\label{exact} E_{\text{ball}}=\frac{5}{32}\frac{\xi^2c}{\pi a}
\end{equation}
(compare this formula with eq. (\ref{final-ball})).  In subsect.\ \ref{SZF+VE-sph} this
energy has been calculated by making use of the zeta function
technique (see eq. (\ref{final-ball}) ). As this zeta function is
not known exactly, we have derived there only an {\it
approximation} for the exact result (\ref{T0}).

As far as the expression (\ref{WD}), it is convenient to
substitute it into eq.\  (\ref{EW}), to do the integration over
$y$ and only after that to address the integration over $d  R$
\begin{eqnarray}
 \lefteqn{\frac{\Delta n^2}{8}\int_{\Delta n}^2 dR \int_0^{\infty}
  d  y\,y\,\frac{d }{d  y}\left[\frac{e^{-2yR}}{R^5}\left(4+R^2+4yR-yR^3
  \right)^2\right]=}
  \nonumber \\
  &=&-\frac{\Delta n^2}{4}\int_{\Delta
  n}^2\left(\frac{10}{R^6}+\frac{1}{R^4}+\frac{1}{8R^2}\right)\,d  R
  \nonumber \\
  &=&\frac{1}{8}\left(\frac{\Delta n^2}{3}-\frac{4}{\Delta n^3}-
  \frac{2}{3\Delta n}
  -\frac{\Delta n}{4}\right)\,.\label{long}
\end{eqnarray}
We have put here $y_0=\infty$ without getting the divergencies. As
it is explained in the Appendix \ref{appA}, in eq.\ (\ref{long})
we have to pick up only the term proportional to $\Delta n^2$.
Remarkably  this term is finite. It is an essential advantage
of our approach. The rest of the terms in this equation are
irrelevant to our consideration. Thus the counter term for $E_W$
vanishes due to the regularizations employed (see the Appendix
\ref{appA}). In view of this we have
\begin{equation}
\label{EWF} E_W^{\text{ren}}= E_W=\frac{1}{2\pi
a}\,\frac{1}{8}\,\frac{\Delta
  n^2}{3}=\frac{1}{48}\,\frac{\Delta n^2}{\pi a}\,.
\end{equation}
Finally we arrive at the following result for the Casimir energy
of a dilute dielectric ball
\begin{equation}\label{EFN}
  E^{\text{ren}}=E_W^{\text{ren}} +E_P^{\text{ren}}
=\frac{\Delta n^2}{\pi a}\left(\frac{1}{48}+
  \frac{5}{128}\right)=\frac{23}{384}\,\frac{\Delta n^2}{\pi a}\,.
\end{equation}
Taking into account the relation (\ref{epn}) between
$\varepsilon_i$ and $n_i$, $i=1,2$, we can write
\begin{equation}\label{F}
  E^{\text{ren}}=\frac{23}{1536}\,\frac{(\varepsilon_1-
 \varepsilon_2)^2}{\pi a}\,.
\end{equation}

     At the first time, this value for the Casimir energy of a dilute
dielectric ball has been derived in ref. \cite{MNg2} by summing up
the van der Waals interactions between individual molecules making
up the ball ($\varepsilon_2 =1$). The result (\ref{F}) was
obtained also by treating a dilute dielectric ball as a
perturbation in the complete Hamiltonian of the electromagnetic
field for relevant configuration~\cite{Barton}. In papers
\cite{BMM,BM}, the value close to the exact one, eq.\  (\ref{F}), has been
obtained by employing the uniform asymptotic expansion of the
Bessel functions.

     In ref.~\cite{BNP} the estimation of the Casimir energy of a
dilute dielectric ball has been done taking into account, as it is
clear now, only the second term in eq.\ (\ref{EFN}). And
nevertheless it was not so bad having the accuracy about $35\%$.

\subsection{Finite temperature}
\label{ddb-T} An essential advantage of the calculation of the
Casimir energy of a dilute dielectric ball, carried out above by
the mode summation method, is  the possibility for its
straightforward generalization to the finite temperature.   The
employment of the addition theorem for the Bessel functions  again
enables one to carry out the summation over the angular momentum
in a closed form. As a result, the exact (in the $\Delta n^2$
approximation) value for  internal and free energies, as well as
for entropy  of a dilute dielectric ball are derived for finite
temperature also. The divergencies, inevitable in such studies,
are removed by making use of the renormalization procedure
developed for calculation of the relevant Casimir energy at zero
temperature (see subsect.~\ref{ddb-0}). The thermodynamic
characteristics are presented as the sum of the respective
quantity for a compact ball with uniform velocity of light and an
additional term  which is specific only for a pure dielectric
ball. The behavior of the thermodynamic characteristics in the low
and high temperature limits is investigated.

Practically the extension to finite temperature $T$ is
accomplished by substituting the $y$-integration in
eqs.~(\ref{E2}) by summation over the Matsubara frequencies $
\omega_n=2\pi nT$. Doing in this way we obtain the {\it internal}
energy of a dielectric ball
\begin{eqnarray}
\label{U} U(T)&=& -T \sum_{l=1}^\infty (2l+1)  \sum_{n=0}^\infty
{}^{'} w_n \frac{d }{d  w_n}\ln  \bigg[ W^2_l(n_1w_n,n_2w_n)
\nonumber \\
&&\left .- \frac{\Delta n^2}{4} P_l^2(n_1w_n, n_2 w_n) \right ],
\end{eqnarray}
where
\begin{eqnarray}
  W_l(n_1w_n, n_2w_n)&=&s_l(n_1w_n)e_l^{\prime}(n_2w_n)-
s_l^{\prime}(n_1w_n)e_l(n_2w_n)\,,
  \label{W-T} \\
  P_l(n_1w_n, n_2w_n)&=&s_l(n_1w_n)e_l^{\prime}(n_2w_n)+
s_l^{\prime}(n_1w_n)e_l(n_2w_n)\,{,}
  \label{P-T}
\end{eqnarray}
 and we have introduced the dimensionless Matsubara frequencies
\begin{equation}
\label{M1} w_n=a \omega_n = 2 \pi n a T, \quad n=0,1,2, \ldots
\,{.}
\end{equation}
The prime on the summation sign in eq.~(\ref{U}) means that the
$n=0$ term is counted with half  weight. In what follows it turns
out to be important, for example, when removing the divergencies,
that we proceed from the expression for the internal energy
instead of from the analogous formula for free energy.

When considering the  low temperature  behavior of the
thermodynamic functions of a dielectric ball   the term
proportional to $T^3$ in our paper \cite{NLS} was lost. It was due
to the following. We have introduced the summation over the
Matsubara frequencies in eq.\ (3.20) under the sign of the
$R$-integral. Here we show how to do this summation in a correct
way.

In the $\Delta^2$-approximation the last term in eq.\ (3.20) from
the article \cite{NLS}
\begin{equation}
\label{eq-2} \overline {U}_W(T)=2 T\Delta n^2 \sum_{n=0}^\infty
\!{}^{'}w^2_n\int_{\Delta n}^2\frac{e^{-2w_nR}}{R}\,d  R{,} \quad
w_n=2 \pi na T
\end{equation}
can be represented in the following form
\begin{equation}
\label{eq-3} \overline {U}_W(T)=-2 T\Delta n^2
\sum_{n=0}^\infty\!{}^{'} w^2_n \,E_1(4w_n){,}
\end{equation}
where $E_1(x)$ is the exponential-integral function \cite{AS}. Now
we accomplish the summation over the Matsubara frequencies by
making use of the Abel-Plana formula
\begin{equation}
\label{eq-4} \sum_{n=0}^\infty\!{}^{'} f(n) =\int_0^\infty
f(x)\,d  x+i\int_0^\infty \frac{f(ix)-f(-ix)}{e^{2\pi x}-1}\;d  x{.}
\end{equation}
The first term in the right-hand side of this equation gives the
contribution independent of the temperature, and the net
temperature dependence is produced by the second term in this
formula. Being interested in the low temperature behavior of the
internal energy we substitute into the second term in eq.\
(\ref{eq-4}) the following  expansion of the function $E_1(z)$
\begin{equation}
\label{eq-5} E_1(z) =-\gamma -\ln z- \sum^\infty_{k=1}\frac{(-1)^k
z^k}{k\cdot k!},\quad |\arg z |<\pi {,}
\end{equation}
where $\gamma $ is the Euler constant \cite{AS}. The contribution
proportional to $T^3$ is produced by the logarithmic term in the
expansion (\ref{eq-5}). The higher powers of $T$ are generated by
the respective terms in the sum over $k$ in this formula $(t=2\pi
a T)$
\begin{equation}
\label{eq-6} \overline {U}_W(T)=\frac{\Delta n^2}{\pi a}\left (
-\frac{1}{96}+ \frac{\zeta_{\text{R}} (3)}{4\pi ^2} t^3 -\frac{1}{30}t^4
+\frac{8}{567} t^6 -\frac{8}{1125}t^8+{\cal O}(t^{10})\right ) {.}
\end{equation}
All these terms, safe for $2 \zeta (3) \Delta n^2 a^2T^3$, are
also reproduced by the last term in eq.\ (3.31) in our paper
\cite{NLS} (unfortunately additional factor 4 was missed there)
\[
\frac{\Delta n^2}{8}T\cdot 4\, t^2\int ^2_{\Delta n}\frac {d  R}{R}
\frac{\coth (tR)}{\sinh^2 (tR)}{.}
\]
Taking all this into account we arrive at the following low
temperature behavior of the internal Casimir energy of a dilute
dielectric  ball
\begin{equation}
\label{eq-7} U(T)= \frac{\Delta n^2}{\pi a}\left ( \frac{23}{384}
+\frac{\zeta_{\text{R}}(3)}{4\pi^2}t^3 -\frac{7}{360}t^4 +\frac{22}{2835}t^6
-\frac{46}{7875}t^8 +{\cal O}(t^{10})
 \right ){.}
\end{equation}
The relevant  thermodynamic relations give the following low
temperature expansions for free energy
\begin{equation}
\label{eq-8} F(T)=\frac{\Delta n^2}{\pi a}\left (
\frac{23}{384}-\frac{\zeta_{\text{R}} (3)}{8\pi ^2}t^3+\frac{7}{1080}t^4
 -\frac{22}{14175}t^6+\frac{46}{55125}t^8+{\cal O}(t^{10})
\right )
\end{equation}
and for entropy
\begin{equation}
\label{eq-9} S(T)=-\frac{\partial F}{\partial T}=\Delta n^2 \left
( \frac{3\zeta_{\text{R}} (3)}{4\pi ^2}t^2-\frac{7}{135}t^3
+\frac{88}{4725}t^5- \frac{736}{55125}t^7+ {\cal O}(t^9) \right
){.}
\end{equation}

The $T^3$ term in the free energy (\ref{eq-8}) does not give
contribution to the Casimir force exerted on the surface of a
dielectric ball, however, it proves to be important for insuring
the positive entropy (\ref{eq-9}) at low temperatures.

The range of applicability of the  expansions (\ref{eq-7}),
(\ref{eq-8}), and (\ref{eq-9}) can be roughly estimated in the
following way. The curve $S(T)$ defined by eq.\ (\ref{eq-9})
monotonically goes up when the dimensionless temperature $t =2\pi
a T$ changes from 0 to $t \sim 1.0$. After that  this curve
sharply goes down to the negative values of $S$. It implies  that
eqs.\ (\ref{eq-7}) -- (\ref{eq-9}) can be used in the region
$0\leq t < 1.0$. The $T^3$-term in eqs.\ (\ref{eq-7}) and
(\ref{eq-8}) proves to be principal because it gives the first
positive term in the low temperature expansion for the  entropy
(\ref{eq-9}). It is worth noting, that the exactly the same
$T^3$-term, but with opposite sign, arises in the high temperature
asymptotics of free energy in the problem at hand (see eq.\ (4.30)
in ref.~\cite{High-T}).

For large temperature $T$ we found~\cite{NLS}
\begin{equation}
\label{eq-10}
 U(T) \simeq  \frac{\Delta n^2}{8} T {,}\;\;
 F(T)  \simeq  -\frac{\Delta n^2}{8} T [\ln (aT)-c]{,}\;\;
 S(T)\simeq \frac{\Delta n^2}{8}[\ln (aT)+c+1] {,}
\end{equation}
where $c$ is a constant~\cite{BNP,Barton} $ c=\ln 4 +\gamma
-{7}/{8}\,{.}$ Analysis of eqs.\ (3.20) and (3.31) from the
paper~\cite{NLS} shows that there are only exponentially
suppressed corrections to the leading terms (\ref{eq-10}).

In the course of calculation of the thermodynamical functions of a
dilute dielectric ball  these functions were first obtained for a
compact ball with the same velocity of light inside and
outside~\cite{NLS}. There  we have derived the following {\it
exact}, in the $\xi^2$-approximation,  expression for the internal
energy
\begin{equation}
\label{UPf}
             U(T)=
\frac{\xi^2}{2} T\left [ t^2\frac{\coth
(2t)}{\sinh^2(2t)}+\frac{t}{\sinh^2(2t)} +\frac{1}{2}\coth (2t)
\right ]{.}
\end{equation}

In the small $T$ region eq.\ (\ref{UPf}) gives
\begin{equation}
\label{UPto0} U(T)=\frac{\xi^2}{\pi a}\left (
\frac{5}{32}+\frac{1}{90}t^4+\frac{8}{945}t^6-\frac{8}{525}t^8
+{\cal O}(t^{10})\right ){.}
\end{equation}

Integration of the thermodynamic relation
\begin{equation}
\label{td} U(T)=\frac{\partial}{\partial \beta}[\beta F(T)], \quad
\beta =T^{-1}
\end{equation}
 enables one to get the free energy
\begin{equation}
\label{tdint} F(T)= - T \int \frac{U(T)}{T^2}d  T + C\,T\,{,}
\end{equation}
where $C$ is a constant.
 Substituting the asymptotics (\ref{UPto0}) into eq.\ (\ref{tdint})
we obtain  the respective free energy in the  low temperature
region
\begin{equation}
\label{FPto0} F(T)=\frac{\xi^2}{\pi a}\left
(\frac{5}{32}-\frac{1}{135}t^4 -\frac{8}{4725} t^6 +
\frac{512}{3675} t^8+  {\cal O}(t^{10}) \right )\,{.}
\end{equation}
Here the  linear in $T$ term $CT$ has been dropped in view of
 the requirement that the entropy $S(T)$  should vanish at
$T=0$ gives~\cite{Klich-FMR}. Indeed, the thermodynamic relation
(\ref{4-14}) gives in this limit
\begin{equation}
\label{S} S(0)=\lim_{T\to 0} T^{-1} \left ( U(T)-F(T) \right
)=C=0\,{.}
\end{equation}
Hence, at low temperature the expansions both for the internal
energy (\ref{UPto0}) and for the free energy (\ref{FPto0}) involve
only even powers of the temperature beginning from $T^4$. At zero
temperature we have
\begin{equation}
\label{T0} U(0)=F(0)=E_{\text{ball}}=\frac{5\xi^2}{32 \pi a}\,{,}
\end{equation}
where $E_{\text{ball}} $ is the {\it exact}, in the
$\xi^2$-approximation, Casimir energy of a compact ball with the
same velocity of light inside and outside the
ball (see eq.\ (\ref{exact})).

 The entropy in the problem at hand is obtained by differentiation
of the free energy (\ref{FPto0}) (see eq.\ (\ref{4-14}))
\begin{equation}
\label{entropy} S(T)=\pi a \left ( \frac{4}{135}t^3
+\frac{96}{4725}t^5 -\frac{128}{3675}t^7 +{\cal O}(t^9) \right
){.}
\end{equation}

The exact formula (\ref{UPf}) leads to the following high
temperature asymptotics of the internal energy
\begin{equation}
\label{UPtoinfty} U(T)\simeq \frac{\xi ^2}{4}\,T, \quad T \to
\infty\,{.}
\end{equation}

Substituting this asymptotics into eq.\ (\ref{tdint}) we arrive at
the high temperature limit for the free energy $F(T)$
\begin{equation}
\label{FPtoinfty} F(T)\simeq - \frac{\xi^2}{4} T \left
[\ln (aT) + \alpha \right ], \quad T \to \infty .
\end{equation}
 The constant $\alpha $ will be calculated in subsect.\ \ref{HT-cb} (see eq.\
(\ref{eq4_17}))
\[
\alpha =\gamma +\ln 4 - \frac{5}{4}\,{.}
\]
This constant has  been found in ref.\ \cite{Klich-FMR} too. The asymptotics (\ref{eq-10})
(\ref{UPtoinfty}) and (\ref{FPtoinfty}) contain terms the Planck constant, thus it is pure
classical contributions (see also sect.~ \ref{HT}).

Summarizing we conclude that now there is a complete agreement
between the results of  calculation of  the Casimir thermodynamic
functions for a dilute dielectric ball carried out in the
framework of two different approaches:  by the mode summation
method~\cite{LSN,NLS} and by perturbation theory for quantized
electromagnetic field, when  dielectric ball is considered as a
perturbation in unbounded continuous surroundings~\cite{Barton}.

\section{Non-smoothness of the boundary and divergencies of vacuum energy}

\setcounter{equation}{0} \label{N-S}

\subsection{Semi-circular cylinder}
The spectrum of electromagnetic oscillations, as well as
oscillations of other fields, is determined by the  boundary form
and by the conditions imposed on the field functions on the
boundary. In view of this, one could anticipate that the
divergencies in vacuum energy are also connected with the boundary
geometry. In general it is true but  this relation turns out to be
very complicated and it is  still far from being clear. We are going to
show this by calculating the Casimir energy for a simple
configuration, namely, for a semi-circular cylindrical shall  by
making use of the zeta function technique. This shell is obtained
by crossing an infinite circular cylindrical shell by a plane
passing through the symmetry axes of the cylinder. All the
surfaces, including the infinite cutting plane, are assumed to be
perfectly conducting. Obviously it is sufficient to consider only
a half of this configuration (left or right) which we shall refer
to as a semi-circular cylindrical shell or, for sake of
shortening, as a semi-circular cylinder (see fig.~1). The internal boundary
value problem for this configuration is nothing else as a
semi-cylindrical waveguide. In the theory of
waveguides~\cite{HdP1} it is well known that a semi-circular
waveguide has the same eigenfrequencies as the cylindrical one but
without degeneracy (without doubling) and safe for one frequency
series (see below). Notwithstanding the very close spectra, the
zeta function technique does not give a finite  result for a
semi-circular cylinder unlike for a circular one (see
subsect.~\ref{SZF+VE-cyl}).

We start with considering the natural modes of electromagnetic
field for circular and semi-circular cylinders. The construction
of the solutions to the Maxwell equations with boundary conditions
given on closed surfaces proves to be nontrivial problem. Mainly
it is due to the vector character of the electromagnetic field
\cite{HdP1,Stratton,HdP2}. In the case of cylindrical symmetry the
electric ${\bf E}$ and magnetic ${\bf H}$ fields are expressed in
terms of the electric (${\bf \Pi}^{\prime}$) and magnetic (${\bf
\Pi}^{\prime\prime}$) Hertz vectors having only one non-zero
component
\begin{eqnarray}\label{P1}
 {\bf \Pi}^{\prime} & = & {\bf e}_z \Phi (r, \varphi)\,e^{\pm i k_z'
 z} \,,\\
 {\bf \Pi}^{\prime\prime} & = & {\bf e}_z \Psi (r, \varphi)\,e^{\pm i
 k_z'' z} \label{P2}\,.
\end{eqnarray}
Here the cylindrical coordinate system $r, \varphi, z$ is used
with $z$ axes directed along the cylinder axes. The common
time-dependent factor $e^{i\omega t}$ is dropped. The scalar
functions $\Phi (r, \varphi)$ and $\Psi (r, \varphi)$ are the
eigenfunctions of the two-dimensional transverse Laplace operator
and meet, respectively, the Dirichlet and Neumann conditions on
the boundary $\partial\Gamma$
\begin{equation}\label{D}
  ({\bbox{\nabla}}^2_{\bot}+\gamma^{'\,2})\Phi(r, \varphi)=0\,, \qquad
  \left. \Phi(r, \varphi)\right|_{\partial\Gamma}=0 \,,
\end{equation}
\begin{equation}\label{N}
  ({\bbox{\nabla}}^2_{\bot}+\gamma''^{2})\Psi(r, \varphi)=0\,, \qquad
 \left. \frac{\partial\Psi(r, \varphi)}{\partial n}\right|_{\partial\Gamma}
=0 \,,
\end{equation}
where $\bbox{\nabla}^2_{\bot}$ is the transverse part of the
Laplace operator
\begin{equation}\label{Ltr}
{\bbox{\nabla}}^2_{\bot}=\frac{\partial^2}{\partial r^2}
  +\frac{1}{r}\frac{\partial}{\partial r}
  +\frac{1}{r^2}\frac{\partial^2}{\partial\varphi^2}
\end{equation}
and
\begin{equation}\label{gamma-nonsmooth}
  \gamma^{'\, 2}=\omega^2-k_z^{'\, 2}\,, \qquad
  \gamma^{''\, 2}=\omega^2-k_z^{''\, 2}\,.
\end{equation}

First we consider a cylindrical shell. In this case the functions
$\Phi (r, \varphi)$ and $\Psi (r, \varphi)$ should be
$2\pi$-periodic in angular variable $\varphi$. As a result the
Dirichlet boundary value problem (\ref{D}) has the following
unnormalized eigenfunctions ($E$-modes)
\begin{equation}
  \Phi_{nm}(r, \varphi) =   {\sin \atop \cos} (n\varphi)
\cases{J_n(\gamma_{nm}' r), \quad r<a ,\cr H_n^{(1)}({\bar
\gamma}_{nm}'\, r), \quad
                        r>a, \cr} \label{efd}
\end{equation}
where $a$ is the cylinder radius, $J_n(x)$ are the Bessel
functions, $H^{(1)}(x)$ are the Hankel functions of the first
kind, and $\gamma'_{nm}$, $\bar{\gamma}_{nm}'$ stand for the roots
of the frequency equations
 \begin{eqnarray}
\label{fed}
   &J_n(\gamma_{nm}'\, a)=0, \quad
 H_n^{(1)}(\overline{\gamma}_{nm}'\, a)=0,&  \\ &n=0, 1,
 2, \ldots , \quad m=1, 2, \ldots \,.& \nonumber \end{eqnarray}

For the Neumann boundary value problem (\ref{N}) we have the
$H$-modes
\begin{equation}
  \Psi_{nm}(r, \varphi) ={\sin \atop \cos} (n\varphi)
  \cases{J_n(\gamma_{nm}'' r), \quad r<a, \cr
H_n^{(1)}(\bar{\gamma}{''}_{nm}\, r),
 \quad r>a, \cr} \label{efn}
\end{equation}
where $\gamma''_{nm}$ and  $\overline{\gamma}_{nm}''$ are the
roots of the equations
 \begin{eqnarray}
   &\left. \frac{\displaystyle d }{\displaystyle
d  r}J_n(\gamma''_{nm}\, r)\right|_{r=a}=0, \quad
   \left. \frac{\displaystyle d }{\displaystyle
d  r}H^{(1)}(\overline{\gamma}_{nm}^{\,''}\,
r)\right|_{r=a}=0\,,&
\label{fen} \\
 &n=0, 1, 2, \ldots , \quad m=1, 2, \ldots \,. & \nonumber
\end{eqnarray}
As usual, it is assumed that for $r>a$ the eigenfunctions should
satisfy the radiation condition.

It is important to note that each root
\begin{equation}\label{dr}
 \gamma'_{nm}, \quad \overline{\gamma}_{nm}^{\,'}, \quad
 {\gamma}_{nm}^{\,''}, \quad \overline{\gamma}_{nm}^{\,''}, \quad
 n\geq 1, \quad m\geq 1
\end{equation}
is doubly degenerate since, according to eqs.\ (\ref{efd}),
(\ref{efn}), because there are two eigenfunctions which are proportional
to either $\sin (n\varphi)$ or $\cos (n \varphi)$. The frequencies
with $n=0$
\begin{equation}\label{ndr}
 \gamma'_{0m}, \quad \overline{\gamma}_{0m}^{\,'}, \quad
 {\gamma}_{0m}^{\,''}, \quad \overline{\gamma}_{0m}^{\,''}, \quad
 m= 1, 2, \ldots
\end{equation}
are independent on $\varphi$, and the degeneracy disappears.

For given Hertz vectors ${\bf \Pi}'$ and ${\bf \Pi}''$ the
electric and magnetic fields are constructed by the formulas
 \begin{eqnarray}
 {\bf E}&= &\bbox{\nabla}\times \bbox{\nabla} \times {\bf \Pi}'\,,
 \quad {\bf H}=-i\omega \bbox{\nabla}\times {\bf \Pi}' \qquad
 (E\tx{-modes})\,,\ \nonumber \\
 \label{fields}
 {\bf E}& =& i\omega \bbox{\nabla} \times {\bf \Pi}''\,,
 \qquad {\bf H}=\bbox{\nabla}\times \bbox{\nabla}\times
 {\bf \Pi}'' \qquad
 (H\tx{-modes})\,.
\end{eqnarray}
It has been proved~\cite{Heyn} that the superposition of these
modes gives the general solution to the Maxwell equations in the
problem under consideration. An essential merit of using the Hertz
polarization vectors is that in this approach the necessity to
satisfy the gauge conditions does not arise.

\begin{figure}

\begin{center}

\epsfysize=7cm

\epsfbox{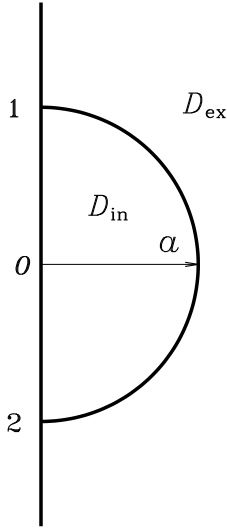}

\end{center}

\caption{\label{fig1} The cross section of an infinite
semi-circular cylindrical shell of radius $a$. All the surfaces
(bold-faced lines) are assumed to be perfectly conducting. At the
same time this picture presents the two-dimensional (plane)
version of the problem under consideration, i.e., \ the
semi-circular boundaries for massless fields defined on the
plane.}

\end{figure}























Now we turn to a waveguide which is obtained by cutting the
infinite cylindrical shell by a plane passing through the symmetry
axes of the cylinder (see fig.~1). All the surfaces are assumed to
be perfectly conducting. In this case the boundary value problems
(\ref{D}) and (\ref{N}) for the Hertz electric (${\bf \Pi}'$) and
magnetic (${\bf \Pi}''$) vectors have the following eigenfunctions
\begin{eqnarray}
 & \Phi_{nm}(r, \varphi)=\sin (n\varphi)\cases{
  J_n(\gamma_{nm}'\,r)\,, \quad r<a\,, \cr
         H_n^{(1)}(\overline{\gamma}_{nm}'\,r)\,, \quad r>a \,,
         \cr }& \label{fds}  \\
 &n=1, 2, \ldots \,, \qquad m=1, 2, \ldots \, & \nonumber
\end{eqnarray}
 and
\begin{eqnarray}\label{fns}
  &\Psi_{nm}(r, \varphi)=\cos (n\varphi) \cases{
  J_n(\gamma_{nm}''\,r)\,, \quad r<a\,, \cr
         H_n^{(1)}(\overline{\gamma}_{nm}''\,r)\,, \quad r>a\,,\cr} & \\
   &n=0, 1, 2, \ldots \,, \qquad m=1, 2, \ldots \,. & \nonumber
\end{eqnarray}
The frequencies $\gamma_{nm}'$, $\overline{\gamma}_{nm}'$,
$\gamma_{nm}''$, and $\overline{\gamma}_{nm}''$ are determined by
the same equations (\ref{fed}) and (\ref{fen}). However the new
spectral problem has two essential distinctions: i) the
frequencies (\ref{dr}) are now nondegenerate, and ii) two series
of eigenfrequencies
\begin{equation}\label{absent}
  \gamma_{0m}', \quad \overline{\gamma}_{0m}'\,, \quad m=1, 2,
  \ldots
\end{equation}
are absent. At first sight one could expect that such a change of
the spectrum cannot influence drastically on the ultraviolet
behavior of the relevant spectral density. However, as it will be
shown below, the zeta function for a semi-circular cylinder,
unlike for a circular one, does not provide a finite answer for
the Casimir energy in the problem in question.

In view of all above-mentioned the zeta function for
electromagnetic field obeying  the boundary conditions on the
surface of the semi-circular cylinder is the sum of two zeta
functions for scalar massless  fields satisfying the Dirichlet and
Neumann conditions on the lateral of this cylinder.

By making use of the contour integration and analytical
continuation discussed in subsect.~\ref{SZF+VE-cyl} one can
construct the spectral zeta functions for a semi-circular cylinder
also. In ref.~\cite{s-cir} it has been done first for scaler
massless fields obeying, on internal and external surfaces of a
semi-circle infinite cylinder, the Dirichlet and Neumann
conditions. As a result the following expressions were obtained
for relevant vacuum energies
\begin{equation}
\label{6-17}
E^{\text{D}}_{\text{s-cyl}}=\frac{1}{2}\zeta^{\text{D}}_{\text{s-cyl}}\left(
-\frac{1}{2} \right )=\frac{1}{2a^2}\left ( 0.000523 - 0.004974
\left .\frac{1}{s+1/2}\right|_{s\to -1/2} \right ){,}
\end{equation}
\begin{equation}
\label{6-18}
E^{\text{N}}_{\text{s-cyl}}=\frac{1}{2}\zeta^{\text{N}}_{\text{s-cyl}}\left(
-\frac{1}{2} \right )=\frac{1}{2a^2}\left ( -0.04345 - 0.0149
\left .\frac{1}{s+1/2}\right|_{s\to -1/2} \right ){.}
\end{equation}
The sum of these formulae gives the Casimir energy of
electromagnetic field for semi-circular cylinder
\begin{equation}\label{6-19}
E^{\text{EM}}_{\text{s-cyl}}=\frac{1}{2}\zeta^{\text{D+N}}_{\text{s-cyl}}\left(
-\frac{1}{2} \right )=\frac{1}{2a^2}\left (-0.0439 - 0.0199 \left
.\frac{1}{s+1/2}\right|_{s\to -1/2} \right ){.}
\end{equation}
Pole singularities in eqs. (\ref{6-17}), (\ref{6-18}) and
(\ref{6-19}) imply that the zeta function technique does not
provide a finite value of vacuum energy for this configuration and
further renormalization is needed.

The same situation takes place also in the two-dimensional version
of the problem under study, i.e., \ when a semi-circle is
considered on a plane. In fact, the relevant zeta functions can be
obtained in a straightforward way by making use the relation
(\ref{relation}). It gives~\cite{s-cir}
\begin{equation}
\label{6-20}
E^{\text{D}}_{\text{s-cir}}=\frac{1}{2}\zeta^{\text{D}}_{\text{s-cir}}\left(
-\frac{1}{2} \right )=\frac{1}{2a^2}\left ( 0.038127 - \left
.\frac{1}{256}\frac{1}{s}\right|_{s\to 0} \right ){,}
\end{equation}
\begin{equation}
\label{6-21}
E^{\text{N}}_{\text{s-cir}}=\frac{1}{2}\zeta^{\text{N}}_{\text{s-cir}}\left(
-\frac{1}{2} \right )=\frac{1}{2a^2}\left ( -0.237103 - 0.0062
\left .\frac{1}{s}\right|_{s\to0} \right ){.}
\end{equation}
The sum of these energies is also infinite.

Keeping in mind that the zeta function regularization does nor
supply a finite value for the Casimir energy also in the case of
spheres in even dimensional spaces (see subsect.~\ref{SZF+VE-cyl})
we have thus a quite long list of boundary conditions for which
the vacuum energy cannot be calculated by making use of the known
mathematical methods. It is worth elucidating the geometrical
peculiarities of the boundaries that are responsible for this
failure.  In order to do this we have to address the corresponding heat
kernel coefficients.

\subsection{Mutual cancellation of divergencies in vacuum energy}
The relevant analysis of the heat kernel coefficients (up to
$B_{5/2}$) has been accomplished in ref.~\cite{NPD}. We are
interested in the coefficients $B_{2}$ and $B_{3/2}$. As it has
been noted in subsect.~\ref{RVE-hkt}, the nonzero value of $B_{2}$
($B_{3/2}$) implies that it is impossible to derive  a finite value
for the vacuum energy in the problem at hand, when the dimension
of the configuration space $d=3$ ($d=2$). From the asymptotic
expansion (\ref{3-23}) it follows that the heat kernel
coefficients introduced in this way are the same for the
three-dimensional cylindrical-like boundaries and for the
corresponding plane problem obtained  by crossing the former
boundary by a transverse plane. Certainly, it is correct only for
the Laplace operator $-\Delta$. In fact, the respective
eigenvalues in these two problems are related by
\begin{equation}
\label{6-22} \lambda_n(d=3)=k^2+\lambda_n(d=2),\quad 0\leq
k<\infty{.}
\end{equation}
Hence
\begin{equation}
\label{6-23} K_{d=3}(t)=\int_{-\infty}^\infty\frac{d
k}{2\pi}e^{-k^2t}K_{d=2}(t)=\frac{1}{\sqrt{4\pi t}}K_{d=2}(t){.}
\end{equation}
Taking into account the definition (\ref{3-23}) one can easily
deduce form (\ref{6-23}) that the heat kernel coefficients
$B_{n/2}$ in the two eigenvalue boundary value problems mentioned
above, are equal.

\begin{table}
\label{table} \caption{The contribution of different parts of
boundary to heat kernel coefficients; D and N stands for the
Dirichlet and Neumann boundary conditions; the upper (lower) sign
refers to internal (external) region.}\begin{tabular}{cccc}
\hline
&&Curvature&Corners\\
\hline $B_{3/2}$&D&$\frac{{\displaystyle
\sqrt{\pi}\pi}}{{\displaystyle 64R}}$&$\pm\frac{{\displaystyle
\sqrt {\pi}}}
{{\displaystyle4R}}$\\
&&&\\
&N&$\frac{{\displaystyle 5\sqrt{\pi}\pi}}{{\displaystyle 64R}}$&
$\pm\frac{{\displaystyle 3\sqrt {\pi}}}{{\displaystyle 4R}}$\\
&&&\\
$B_{2}$&D&$\pm\frac{{\displaystyle 4\pi}}{{\displaystyle
315R^2}}$&$\frac{{\displaystyle \pi}}
{{\displaystyle 8R^2}}$\\
&&&\\
&N&$\pm\frac{{\displaystyle 4\pi}}{{\displaystyle 45R^2}}$&
$\frac{{\displaystyle 3\pi}}{{\displaystyle 8R^2}}$ \\
&&&\\
$B_{5/2}$&D&$\frac{{\displaystyle 37\sqrt {\pi} \pi
}}{{\displaystyle 8192R^3}}$&$\pm\frac{{\displaystyle
25\sqrt{\pi}}}
{{\displaystyle 96R^3}}$\\
&&&\\
&N&$\frac{{\displaystyle 269\sqrt{\pi}\pi}}{{\displaystyle
8192R^3}}$&
$\pm\frac{{\displaystyle 21\sqrt{\pi}}}{{\displaystyle 32R^3}}$\\
 \hline
\end{tabular}

\end{table}

The heat kernel coefficients we are interested in are contributed
by the following peculiarities of the boundary (see fig.~1): the
curvature of semi-circle 1 -- 2 and four right-angled corners at
points 1 and 2.  It is worthy to remind that we are considering
both internal ($D_{\text{in}}$) and external ($D_{\text{ex}}$)
regions. All these contributions are presented in table~I.  The
contributions of the curvature of the semi-circle 1--2  to the
coefficients $B_{2}$ for internal and external regions are of the
same absolute value but they have opposite signs. As a result they
are mutually cancelled in the net $B_2$ coefficient. Unlike this
the contributions to $B_2$ due to four right-angled corners at the
points 1 and 2 are added
\begin{equation}
\label{6-24} B^{\text{D}}_2=2\frac{\pi}{8R^2}=\frac{\pi}{4R^2},\quad
B^{\text{N}}_2=2\frac{3\pi}{8R^2}=\frac{3\pi}{4R^2},\quad
B^{\text{EM}}_2=B^{\text{D}}_2+B^{\text{N}}_2 =\frac{\pi}{R^2}{.}
\end{equation}
This fact has been noted at first time in ref.~\cite{Dowker}. In
the case of the coefficient $B_{3/2}$ the situation is opposite,
i.e., \ the contributions of the corners from internal and
external regions are mutually cancelled, while the contributions
of the curvature of the arc 1 -- 2 from internal and external
regions are added
\begin{equation}
\label{6-25} B^{\text{D}}_{3/2}=2\frac{\pi\sqrt \pi
}{64R}=\frac{\pi\sqrt\pi}{32R},\quad
B^{\text{N}}_{3/2}=2\frac{5\pi\sqrt \pi}{64R}=\frac{5\pi\sqrt \pi
}{32R},\quad  B^{\text{D+N}}_{3/2}=\frac{3\pi\sqrt \pi}{16R}{.}
\end{equation}

Different geometrical origins of the zeta function failure to
provide a finite value of the vacuum energy in the two- and
three-dimensional versions of the boundary value problem in
question probably imply the impossibility of obtaining a finite
and unique value of this quantity by taking advantage of the
atomic structure of the boundary~\cite{Candelas} or its quantum
fluctuations~\cite{FS1}. It is clear because any physical reason
of the Casimir energy divergences should be valid simultaneously
in the two- and three-dimensional versions of the boundary
configuration under consideration.

The alternating sign changes of contributions due to curvature and corners to heat kernel
coefficients are  presumably true  for all the coefficients $B_n$ with
$n=1,\;3/2,\;2,\ldots$, at least it is the case for $B_{1},\;B_{3/2},\;B_{2}$ and
$B_{5/2}$ (see table~I and also ref.~\cite{NPD}). The analogous situation takes place for
spheres in spaces of even and odd dimensions, namely, the curvature contributions to
respective heat kernel coefficients are mutually cancelled for odd dimension of ambient
space and they are added in spaces of even dimension (see subsect.~\ref{SZF+VE-cyl}). In
the former case the zeta regularization gives a finite value for vacuum energy, in the
latter situation it is not the case,  and further renormalization of vacuum energy is
needed in the problem at hand.

The corners at points 1 and 2 of the boundary considered here (see
fig.~1) are obtained by intersection of a straight line and an arc
of a circle possessing nonzero curvature ($1/a$). They contribute
to all the heat kernel coefficients starting with $n=1$. These
corners should be distinguished from those  formed by crossing two
{\it straight lines}, for example, from the corners of a
rectangle. Such corners contribute only to the coefficient $B_1$,
the contribution being the same for Dirichlet and Neumann boundary
conditions
\begin{equation} \label{6-26}c(\alpha)= \frac{\pi^2-\alpha
^2}{6\alpha}{,}
\end{equation}
where $\alpha$ is the angle  of the corner. Such corner
singularity of the boundary proves to be very close to the {\it
conical singularity}, which is important in many areas of
mathematical physics~\cite{Sommer,HowTr2}. Lately it has been
investigated in connection with studies of quantum fields on the
background of black holes~\cite{ZCV} and cosmic
strings~\cite{FS,Wedge}.

This general assertion concerning the corner contribution to the
heat kernel coefficients can be illustrated by a known  heat
kernel expansion for a rectangle with sides $a$ and $b$
\begin{eqnarray}
\label{eq4-14}
K(t)&=&\sum_{m=1}^{\infty}\sum_{n=1}^{\infty}\exp\left [- \left
(\frac{m^2}{a^2}+\frac{n^2}{b^2} \right )\pi^2t \right ] = \left
(\frac{a}{\sqrt{4\pi t}}-\frac{1}{2} \right ) \left
(\frac{b}{\sqrt{4\pi t}}-\frac{1}{2}
\right ) +\mbox{ES}  \nonumber \\
&=& \frac{ab}{4\pi t} -\frac{a+b}{4\sqrt{\pi t}}+\frac{1}{4}+
\mbox{ES}\,{,}
\end{eqnarray}
where ES denotes the exponentially small corrections as $t\to +0$
(see, for example,~\cite{W}). Here the scalar operator $-\Delta$
with Dirichlet boundary conditions is considered. For a rectangle
the coefficient $B_1$ is obviously equal to the contributions of
four right-angled corners. Indeed, the third term in the expansion
(\ref{eq4-14}) can be represented in the form
\[
\frac{1}{4}=\frac{1}{4\,\pi}\,B_1=\frac{1}{4\,\pi}\,4
\,c(\alpha=\pi/2)\,{,}
\]
where $c(\alpha)$ is given in eq.\ (\ref{6-26}). Besides these
corners the boundary of a rectangle has no other singularities,
therefore the heat kernel coefficients $B_n$ with $n=3/2,\;2,\;\ldots$ vanish in
this problem.

These rules for obtaining the heat kernel coefficients are
directly generalized to an arbitrary polygon with the angles
$\alpha_i$. The first two coefficients $B_0$ and $B_{1/2}$ are
defined by eq.\ (\ref{3-23a}), where $V$ is the area of the
polygon and $S$ is its perimeter. The third coefficient $B_1$ is
equal to the sum of the contributions due to the angles $\alpha_i$
\begin{equation}
\label{eq4-16a} B_1=\sum_{i}c(\alpha_i)\,{.}
\end{equation}
The rest of the coefficients $B_n,\;n\geq1$ vanish both for
internal and external regions. In particular, it implies that the
zeta function technique should provide  a finite value of the
Casimir energy for a polygon on a plane ($B_{3/2}=0$) and for a
cylindrical generalization of the polygon spectral problem
($B_2=0$). These subjects have been discussed earlier in
papers~\cite{Dow1}. The calculations of vacuum energy have been
accomplished only for parallelepiped
geometries~\cite{Lukosz,Rug,Mam-Tr,AmWolf,Caruso,Caruso-1,Actor,Actor-1,Li,Que}
and with allowance for only internal region. For such
configurations the spectrum of the Laplace operator is known
exactly and the zeta regularization works well. For example, the
Casimir energy for the cube with sides $a$ is negative
\begin{equation}
\label{cube}
E_{\text{cube}}=\frac{\pi}{2a}\sum_{n,m,k=1}^{\infty}\sqrt{n^2+m^2+k^2}\simeq
-\frac{0.015}{a}{.}
\end{equation}
However  Milton~\cite{Milton-JPhys} comments
these results in the following way: here "divergences  occur which
cannot be legitimately removed, which nevertheless are
artificially removed by zeta function methods. It is the view of
the author that such finite results are without meaning". With
regard to external region the wave equation is not separable
outside a cube or a rectangular solid.

Recently the image method has been extended to a set of geometries
with planar boundaries and without rectangular
angles~\cite{Ahmedov}. In this approach the Casimir energy of a
massless scalar field was calculated inside a triangle with angles
proportional to $\pi/N,\;N\geq3$. However the final expression for
the vacuum energy is not ready for numerical estimation. Further
this technique was applied for calculating the Casimir energy of a
conical wedge and a conical cavity~\cite{Ahmedov-1}.

\section{High  temperature behavior of the vacuum energy}

\setcounter{equation}{0} \label{HT}  The influence of temperature on the Casimir effect was an
important topic since its first experimental demonstration
\cite{Sparnaay} which had been done at room temperature. It was first
shown in ref.~\cite{Mehra} that the temperature influence was just
below what had been measured (see, also, paper~\cite{Brown}). Now it
is expected that the temperature contributions will be seen in the
upcoming series of experiments~\cite{thermal}. The temperature
dependence of the vacuum energy proves to be important practically in
all problems where the Casimir effect is taken into account, for
example, in hadron physics (quark deconfinement in the framework of
bag models~\cite{bag}) and in cosmology~\cite{Od-2}.

 The Casimir calculations at finite temperature is a
nontrivial problem specifically for boundary conditions with nonzero
curvature. Investigation of the high temperature limit, i.e., , the
classical limit, in this problem is of independent
interest~\cite{FMR}. For this goal a powerful method of the zeta
function technique and the heat kernel expansion can be used. It is
important that for obtaining the high temperature asymptotics of the
thermodynamic characteristics it is sufficient to know the heat
kernel coefficients and the determinant   for the {\it spatial part}
of the  operator governing the field dynamics.
 This is an essential merit
of this approach.

\subsection{Heat kernel coefficients and high temperature
expansions} In quantum field theory, finite temperature effects
can be described at equilibrium in the Matsubara formalism by
imposing periodic (resp. antiperiodic for fermions) boundary
conditions in the imaginary time coordinate~\cite{Kapusta}. Let
the dynamics of quantum field $\varphi(t,x)$ be defined, as
before, by eq. (\ref{3-3}). The Helmholtz free energy  $F$ of this
field is determined by the functional integral~\cite{Kapusta,Feynman}
\begin{equation}
\label{4-1} e^{-\beta F}=\int D\varphi \exp\left (-\int d
t\,d  x\, \varphi^*(t,x)L_T\varphi(t,x) \right ){.}
\end{equation}
Here $\beta$ is inverse temperature $\beta =T^{-1}$. For
simplicity the Boltzmann constant $k_{\text {B}}$ is assumed to be
equal to 1. Therefore  the temperature $T$ is measured in energy
units. For the Euclidean time and for the Euclidean version of the
field $\varphi (t,x)$ we use the former notations. The operator
$L$ in eq. (\ref{4-1}) is the Euclidean version of the full
differential operator in the field equation~(\ref{3-3})
\begin{equation}
\label{4-2} L_T=L-\frac{\partial^2}{\partial t^2}
\end{equation}
with the eigenfunctions
\begin{equation}
\label{4-3} \phi_{mn}(t,x)=e^{i\Omega_mt}\varphi_n(x)
\end{equation}
and the eigenvalues
\begin{equation}
\label{4-4} L_T\phi_k(t,x)=\lambda^T_k\phi_k(t,x),\quad
k=\{m,n\},\quad \lambda^T_k=\Omega_m^2+\omega_n^2{,}
\end{equation}
where $\Omega_m=2\pi m T/\hbar, \quad m=0,\pm 1,\pm 2,\ldots $ are
the Matsubara frequencies, and $\omega_n$ are defined in eq.
(\ref{3-4}).

Functional integration in eq. (\ref{4-1}) yields
\begin{equation}
\label{4-5} e^{-\beta F}=(\det L_T)^{-1/2}{.}
\end{equation}
The determinant of the operator $L_T$ is expressed in terms of its
spectral zeta function
\begin{equation}
\label{4-6}
\zeta_{T}(s)=\sum_k(\lambda^T_k)^{-s}=\sum\limits_{m=-\infty}^{\infty}
\sum\limits_{n} \left(\Omega_m^2+\omega_n^2\right)^{-s}{.}
\end{equation}
Indeed
\begin{equation}
\label{4-7HT} \det L_T=\prod_k\lambda_k^T, \quad \ln(\det
L_T)=\sum_k \ln \lambda _k^T{.}
\end{equation}
Differentiation of the zeta function $\zeta _T(s)$ at the point
$s=0$ gives
\begin{equation}
\label{4-8HT} \zeta'_T(0)=-\sum_k\ln \lambda^T_k{.}
\end{equation}
>From eqs. (\ref{4-6}), (\ref{4-7HT}) and (\ref{4-8HT}) we deduce
finally
\begin{equation}
\label{4-9} F=-\frac{T}{2}\,\zeta'_{T}(0).
\end{equation}

The characteristics of the quantum field (\ref{3-3}) at zero
temperature are determined   by the zeta function $\zeta (s)$
associated with the operator $L$ (see eq. (\ref{3-9})). From the
mathematical point of view the zeta function $\zeta(s)$
corresponding to the space part of the full wave equation
(\ref{3-3}) is, undoubtedly, a simpler object  than the complete
zeta function $\zeta_{T}(s)$ because the definition (\ref{4-6})
involves an additional sum over the Matsubara frequencies. Here a
natural question arises whether one can gain knowledge of the
quantum field at nonzero temperature possessing only the zeta
function $\zeta (s)$. In ref.\ \cite{DK} it was shown that
proceeding from the zeta function $\zeta (s)$ one can deduce the
high temperature asymptotics of the thermodynamic functions such
as Helmholtz free energy, internal energy, and entropy. Let us
remind briefly the derivation of these asymptotics. By making use
of the formula (\ref{3-22}) the zeta  function~(\ref{3-21}) can be
represented in the form
\begin{equation}
\label{4-10} \zeta_{T}(s)=\frac{1}{\Gamma(s)}\int_0^\infty dt \,
t^{s-1}\sum_{m=-\infty}^{\infty}e^{-\Omega^2_m t}\sum_{k}
e^{-\omega^2_k t}{.}
\end{equation}
The term with $m=0$ in this formula gives the zeta function
(\ref{3-9}). In the remaining terms we  substitute the heat kernel
$K(t)$ of  the operator $L$ by its asymptotic
expansion~(\ref{3-23}). As a result we arrive at the following
asymptotic representation for the complete zeta function
$\zeta_{T}(s)$
\begin{eqnarray}
\zeta_{T}(s)&\simeq &\zeta(s) \nonumber \\
&& +\frac{2}{(2\pi)^{3/2}}\sum_{n=0,1/2,\dots}B_n
\left(\frac{\hbar}{2 \pi T}\right)^{2 s-3+2
n}\frac{\Gamma(s-3/2+n)}{\Gamma(s)}\,\zeta_{\text R}(2 s+2 n-3),
\label{4-11}
\end{eqnarray}
where $\zeta_{\text{R}}(s)$ is the Riemann zeta
function~(\ref{3-10a}). Taking the derivative   of the right hand
side of eq.\ (\ref{4-11}) at the point $s=0$ and substituting the
result into eq.\ (\ref{4-9}) one obtains the high temperature
expansion for the free energy
\begin{eqnarray}
F(T) &\simeq&-\frac{T}{2}\zeta'(0)+B_0\frac{T^4}{\hbar^3}\,
\frac{\pi^2}{90}-B_{1/2}\,\frac{T^3}{4\pi^{3/2} \hbar^2 }
\zeta_{\text {R}}(3) -\frac{B_1}{24}\frac{T^2}{\hbar}
+\frac{B_{3/2}}{(4\pi)^{3/2}}\,T\,\ln\frac{\hbar}{T}\nonumber\\
&&-\frac{B_2}{16\pi^2}\,\hbar\, \left[\ln\left(\frac{\hbar}{4 \pi
T}\right)+\gamma\right]-\frac{B_{5/2}}{(4\pi)^{3/2}}\frac{\hbar^2}{24
T}\nonumber\\ &&-T\sum_{n\geq
3}\frac{B_n}{(4\pi)^{3/2}}\left(\frac{\hbar}{2 \pi
T}\right)^{2n-3}\,\Gamma(n-3/2)\,\zeta_{\text{R}}(2 n-3)\,{,} \quad
T\to \infty {.} \label{4-12}
\end{eqnarray}
Here $\gamma$ is the Euler constant. The argument of the logarithms in expansion
(\ref{4-12}) are dimensional, but upon collecting similar terms with account for the
logarithmic ones in $\zeta'(0)$ it is easy to see that finally the logarithm function has
a dimensionless argument, at least for $B_2=0$. It is worth noting that the auxiliary
evolution variable $\tau$ in eq.\ (\ref{3-23}) has the dimension [length]$^2$.

The asymptotic expansions for the internal energy $U(T)$ and the
entropy $S(T)$ are deduced from eq.\   (\ref{4-12}) employing the
thermodynamic relations
\begin{eqnarray}
U(T)&=&-T^2\frac{\partial}{\partial T}\left(T^{-1}F(T)\right),
\label{4-13}\\ S(T)&=&T^{-1}\left(U(T)-F(T)\right)=
 - \frac{\partial F}{\partial T}{.}
\label{4-14}
\end{eqnarray}
Substituting the expansion (\ref{4-12}) into eqs.\ (\ref{4-13})
and (\ref{4-14}) one arrives at the asymptotics
\begin{eqnarray}
U(T)&\simeq&B_0\frac{T^4}{\hbar^3}\,\frac{\pi^2}{30}+
B_{1/2}\,\frac{T^3}{\hbar^2} \,\frac{\zeta_{\text
{R}}(3)}{2\,\pi^{3/2}}+B_1\, \frac{T^2}{24\,\hbar}\,
+\frac{B_{3/2}}{(4\pi)^{3/2}}T\nonumber\\
&&-B_2\,\frac{\hbar}{16\pi^2} \left[\ln\left(\frac{\hbar}{4\pi
T}\right)+\gamma+1\right]
-\frac{B_{5/2}}{(4\pi)^{3/2}}\,\frac{\hbar^2}{12\,T} \nonumber\\
&& -\frac{T}{4\,\pi^{3/2}}
\sum_{n\geq3}\,B_n\,\left(\frac{\hbar}{2\pi T}
\right)^{2n-3}\,\Gamma(n-1/2)\,\zeta_{\text {R}}(2n-3),
\label{4-15}\\
S(T)&\simeq& \frac{1}{2}\zeta'(0)+B_0\,\frac{T^3}{\hbar^3}\,
\frac{2\,\pi^2}{45}+B_{1/2}\,\frac{T^2}{\hbar^2}\,\frac{3}{4}\,
\frac{\zeta_{\text {R}}(3)}{\pi^{3/2}}
+B_1\,\frac{T}{12\,\hbar}\nonumber\\ &&
+\frac{B_{3/2}}{(4\pi)^{3/2}}\left(1-\ln\frac{\hbar}{T}\right)
-B_2\frac{\hbar}{16 \pi^2 T
}-\frac{B_{5/2}}{(4\pi)^{3/2}}\,\frac{\hbar^2}{24\, T^2 }
\nonumber\\ &&-\frac{1}{4 \,\pi^{3/2}
}\,\sum_{n\geq3}B_n\,\left(\frac{\hbar}{2\pi T}\right)^{2n-3}
 (n-2)\,\Gamma(n-3/2)\,\zeta_{\text {R}}(2n-3)\,{.}
\label{4-16}
\end{eqnarray}
In eq.\ (\ref{4-15}) the term proportional to $B_2$ contains the
logarithm of dimensional quantity: $[\hbar/T]=[\text{time}]^{-1}$.
This is the result of the arbitrariness arising in from the
ultraviolet divergences in the case of $B_2\not=0$ (see ref.\
\cite{Bordag-db} for a more detailed discussion).  Unlike this
situation, collecting the logarithm functions in the
$B_{3/2}$-term and in $\zeta'(0)$ in eq.~(\ref{4-16}) leads to a
dimensionless argument of the logarithm in the final expression.

It is worth noting that the zeta determinant of the operator $L$,
i.e., \ $\zeta'(0)$, does not enter the asymptotic expansion for
the internal energy~(\ref{4-15}). Therefore this high temperature
expansion is completely  defined only by the heat kernel
coefficients. In view of this, the first term in the asymptotics
of the free energy in eq.\ (\ref{4-12}) is referred to as a pure
{\it entropic} contribution. Its physical origin is  till now not
elucidated.

\subsection{Perfectly conducting parallel plates in vacuum}

\label{HT-plates}  First we demonstrate the application of the high
temperature expansions (\ref{4-12}), (\ref{4-15}) and
(\ref{4-16}) to a simple problem of electromagnetic field confined
between two perfectly conducting parallel plates in vacuum.

Substituting eq.\ (\ref{eq3_4}) into eq.\ (\ref{3-24}) we obtain
for
 perfectly conducting   parallel plates only one nonzero coefficient~$B_0$
\begin{equation}
\label{B0} B_0=2\,\frac{V}{c^2},  \label{eq3_10}
\end{equation}
where $V=L_x \,L_y\,a$ is the volume of the space  bounded by the
plates.\footnote{For obtaining the vanishing $B_{1/2}$ coefficient
it is important to take into account the second term in eq.\
(\ref{eq3_4}) which depends on the photon mass~$\mu$.} In subsect.
\ref{SZF+VE-plates} it was noted that for parallel conducting
plates electromagnetic field is reduced to two massless scalar
fields obeying Dirichlet and Neumann conditions on internal
surface of the plates. As a result we have the multiplier 2 in eq.
(\ref{B0}) and vanishing coefficient $B_{1/2}$ (see eq.
(\ref{3-23a})). It should be noted here that we are considering
only electromagnetic field confined between the plates and do not
take into account that field outside the plates.

>From eqs.\ (\ref{4-15}) and (\ref{eq3_10}) it follows that the
density of internal energy  has the following high temperature
asymptotics
\begin{equation}
\frac{U(T)}{V}\simeq 4\,\frac{\sigma}{c}\,T^4, \quad T\to\infty
{,} \label{eq3_11}
\end{equation}
where $\sigma$ is the Stefan-Boltzmann constant
\begin{equation}
\sigma=\frac{\pi^2\,k_{\text{B}}^4}{60\,c^2\,\hbar^3}{.}
\label{eq3_12}
\end{equation}
Recall that in our formulae we put $k_{\text{B}}=1$, that is, the
temperature is measured in energy units. The transition to degrees
is performed by the substitution $T \rightarrow k_{\text{B}}\,T$.

When calculating the high temperature asymptotics of the free
energy (\ref{4-12}) and the entropy (\ref{4-16}) one needs to
derive $\zeta'(0)$ for the zeta function (\ref{eq3_4}). Keeping in
mind that $\zeta_{\text{R}}(-2)=0$ it is convenient to use here the
Riemann reflection formula~\cite{GR}
\begin{equation}
2^{1-s}\,\Gamma(s)\,\zeta_{\text{R}}(s)\,\cos(\pi\,s/2)
=\pi^2\zeta_{\text{R}}(1-s) \label{eq3_13}
\end{equation}
which yields
\begin{equation}
\zeta_{\text{R}}(2
s-2)\mathop{\simeq}_{s\to0}-s\,\frac{\zeta_{\text{R}}(3)}{2\pi^2}+{\cal
O}(s^2). \label{eq3_14}
\end{equation}
>From here we deduce
\begin{equation}
\zeta'(0)=\frac{L_x\,L_y}{4\,\pi\,a^2}\,\zeta_{\text{R}}(3)=
\frac{V}{4\,\pi\,a^3}\,\zeta_{\text{R}}(3). \label{eq3_15}
\end{equation}
Insertion of eqs.\   (\ref{eq3_10}) and (\ref{eq3_15}) into eq.\
(\ref{4-12}) gives the following high temperature behavior for the
density of free energy
\begin{equation}
\frac{F}{V}\simeq-\frac{T}{8 \,\pi\,
a^3}\,\zeta_{\text{R}}(3)-\frac{T^4}{c^3\,\hbar^3}\,
\frac{\pi^2}{90}.\label{eq3_16}
\end{equation}
As was noted above, we are considering only  electromagnetic field
between the plates. Therefore when calculating the Casimir forces
one should drop the last term in eq.~(\ref{eq3_16}) since its
contribution is cancelled by the pressure of the black body
radiation on the outward surfaces of the plates. As a result the
high temperature asymptotics of the Casimir force, per unit
surface area,  attracting two perfectly conducting plates in
vacuum~is
\begin{equation}
{\cal F}\simeq  -\frac{T}{4 \pi a^3} \zeta_{\text{R}}(3).
\label{eq3_17}
\end{equation}
Usually in the Casimir calculations the contribution of the free
black body radiation is subtracted from the very
beginning~\cite{PMG}.

It is interesting  to  note that  the Casimir force (\ref{eq3_17})
and the first term on the right hand side of eq.\ (\ref{eq3_16})
are pure classical quantities because they do not  involve the
Planck constant $\hbar$.  These classical asymptotics seem to be
derivable without appealing to the notion of quantized
electromagnetic field. The classical limit  of the theory of the
Casimir effect  is discussed in a recent paper~\cite{FMR}.

Employing eqs.\ (\ref{4-14}) and (\ref{eq3_16}) one arrives at the
high temperature  behavior of the entropy density
\begin{equation}
\frac{S(T)}{V}\simeq\frac{\zeta_{\text{R}}(3)}{8 \pi
a^3}+\frac{2\,T^3\,\pi^2}{45\,c^3\,\hbar^3}. \label{eq3_18}
\end{equation}
Corrections to eqs. (\ref{eq3_11}),
(\ref{eq3_16}) and (\ref{eq3_18}) are exponentially small.

\subsection{Sphere}
We consider electromagnetic field
subjected to three types of boundary conditions  on the surface of
a sphere: i) an infinitely thin and perfectly conducting spherical
shell; ii) the surface of a sphere delimits two material media
with the same velocity of light; iii) a dielectric ball placed in
unbounded dielectric medium. In order to obtain the heat kernel
coefficients determining the high temperature asymptotics
(\ref{4-12}), (\ref{4-15}) and (\ref{4-16}) it is convenient to
use the explicit representation of the relevant spectral zeta
functions in terms of the Riemann zeta function. These formulae
were derived in subsect.~\ref{SZF+VE-sph} by taking into account
the first two terms of the uniform asymptotic expansion for the
product of the modified Bessel functions $I_{\nu}(\nu
z)\,K_{\nu}(\nu z)$.

\subsubsection{Perfectly conducting spherical shell}
The corresponding spectral zeta function is given in eq.\
(\ref{shell3}). The terms omitted in this equation are of the form
\begin{equation}
q_k(s) \left[\left(2^{2
(k+s)+1}-1\right)\,\zeta_{\text{R}}(2\,k+2\,s+1)-2^{2\,(k+s)+1}\right],\quad
k=2,3,4, \dots , \label{eq4_3}
\end{equation}
where $q_k(s)$  stand for some polynomials in~$s$.

Analysis of eqs.\ (\ref{shell3}) and (\ref{qpol}) shows that the
zeta function  for a perfectly conducting spherical shell enables
one to find the exact values of the first six heat kernel
coefficients, namely:
\begin{equation}
B_0=0, \quad B_{1/2}=0,\quad B_1=0, \quad B_{3/2}=2\,\pi^{3/2},
\quad B
_2=0,\quad B_{5/2}=\frac{\pi^{3/2}}{20}\,\frac{c^2}{a^2}.
\label{eq4_4}
\end{equation}
Taking into account the structure of the  omitted terms
(\ref{eq4_3}) it is easy to see that
\begin{equation}
B_j=0,\qquad j=3,4,5,  \dots\, {.} \label{eq4_5}
\end{equation}
Having obtained the heat kernel coefficients   (\ref{eq4_4}) and
(\ref{eq4_5}) we are in position to construct the high temperature
asymptotics of the internal energy of electromagnetic field by
making use of eq.~(\ref{4-15})
\begin{equation}
U (T)\simeq\frac{T}{4}-\left(\frac{c\,\hbar}{R}\right)^2
\frac{1}{1920\,T}+{\cal O}(T^{-3}). \label{eq4_6}
\end{equation}
Applying the technique developed in ref.\ \cite{BKE} more terms to
this expansion can be easily added.

In order to write the asymptotic expansions (\ref{4-12}) and
(\ref{4-16}) the derivative of the zeta function at the point
$s=0$ should be calculated. Equation (\ref{shell3}) gives an
approximate value for $\zeta'(0)$
\begin{equation}
\zeta'_{\text{shell}} (0)=\frac{\gamma}{2}+\ln 2
+\frac{7}{16}\,\zeta_{\text{R}}(3)-\frac{9}{8}+\frac{1}{2}\,\ln\frac{a}{c}
=0.38265 +\frac{1}{2}\,\ln\frac{a}{c}\,{.} \label{eq4_7}
\end{equation}
The terms omitted in (\ref{shell3}) will render precise only the
first term in the final form of this expression, while the second
term $(1/2)\,\ln(a/c)$ will not change. The exact value of
$\zeta'_{\text{shell}} (0)$ is calculated  in  Appendix
\ref{appb-sph}
\begin{eqnarray} \zeta'_{\text{shell}}
(0)&=&\frac{1}{2}-\frac{\gamma}{2}+\frac{7}{6}\ln
2+6\,\zeta'_{\text{R}}(-1)
+\left(-\frac{5}{8}+\frac{1}{2}\,\ln\frac{a}{c}+\ln
2+\frac{\gamma}{2} \right) \nonumber\\
&=&0.38429+\frac{1}{2}\ln\frac{a}{c}. \label{eq4_8}
\end{eqnarray}
It is worth noting that the  expression in the round parentheses,
being multiplied by $\xi^2$, is exactly the  value of $\zeta
'_{\text{ball}} (0)$ for a compact ball with continuous velocity of
light on its surface (see eq.\ (\ref{eq4_16}) in the next
subsect.). As a result we have the following high temperature
asymptotics of the free energy and the entropy in the problem in
question
\begin{eqnarray}
F (T)&\simeq&-\frac{T}{4}\,\left(\ln\frac{a\,T}{\hbar c
}+0.76858\right)\,-\,\left(\frac{\hbar\,c}{a}\right)^2\,
\frac{1}{3840\,T}+{\cal O }(T^{-3}), \label{eq4_9}\\
S (T)&\simeq&0.44215 +
\frac{1}{4}\,\ln\frac{a\,T}{\hbar\,c}-\frac{1}{3840}\,
\left(\frac{\hbar\,c}{a\,T}\right)^2+{\cal O}(T^{-4})\,{.}
\label{eq4_10}
\end{eqnarray}
The expression (\ref{eq4_9}) exactly reproduces the  asymptotics
obtained in ref.\ \cite{BD} (1978) by making use of the multiple
scattering technique (see eq.~(8.39) in that paper). We have not
calculated the coefficient $B_{7/2}$, therefore we do not know the
sign of the $T^{-3}$-correction in (\ref{eq4_9}). In ref.\
\cite{BD} it is noted that this term is  negative.

In eqs.\ (\ref{eq4_6}), (\ref{eq4_9}) and (\ref{eq4_10}) the
large expansion parameter is actually a dimensionless
`temperature' $\tau =aT/(\hbar c)$. Therefore the same formulae
describe the behavior of the thermodynamic functions when $a\to
\infty$ and temperature $T$ is fixed.

The high temperature asymptotics of the thermodynamic functions
derived by making use of the general expansions (\ref{4-12}),
(\ref{4-15}) and (\ref{4-16}) contain the terms independent of the
Planck constant $\hbar$ or, in other words, {\it classical}
contributions (see eqs. (\ref{eq4_6}), (\ref{eq4_9}) and
(\ref{eq4_10})). This is also true for the high temperature limit
of the Casimir force calculated per unit area of a sphere
\begin{equation}
{\cal F} (T)\simeq-\frac{1}{4\pi R^2\,}\frac{\partial F
(T)}{\partial a}=\frac{T}{16\pi a^3}-\left(\frac{\hbar
c}{a}\right)^2\frac{1}{4\pi a^3}\,\frac{1}{1920\,T}+{\cal
O}(T^{-3}). \label{eq4_11}
\end{equation}

The leading classical term in the asymptotics (\ref{eq4_11})
describes the Casimir force that seeks to expand the sphere. The
quantum correction in this formula stands for the Casimir pressure
exerted on the sphere surface.

    In eqs.\ (\ref{eq4_6}), (\ref{eq4_9}) and (\ref{eq4_10}) the
Stefan-Boltzmann terms proportional to $T^4$ are absent because
the contribution of the Minkowski space was subtracted from the
very beginning in our calculations (see eq.\ (\ref{zeta})). As a
result we obtain the vanishing heat kernel coefficient $B_0$
which, in general case, is determined by the volume of the system under
study (see eq. (\ref{B0})). Therefore our results describe only
the {\it deviation} from the Stefan-Boltzmann law caused by the
perfectly conducting sphere.

The vanishing of the coefficients $B_{1/2}$ and $B_1$ in the
problem at hand can be  explained by taking into account the
general properties of the heat kernel coefficients~\cite{Bordag}
and by making use of the results obtained in ref.\ \cite{BKE}. As
known \cite{Stratton} the solutions to the Maxwell equations with
allowance for a perfectly conducting sphere are expressed in terms
of the two scalar functions that satisfy the Laplace equation with
the Dirichlet and Robin boundary conditions on internal and
external surfaces of the sphere. In view of this one can write
\begin{equation}
\label{eq4_11a} B_n=B_{n+}^{\text{D}}+  B_{n-}^{\text{D}}+
B_{n+}^{\text{R}} + B_{n-}^{\text{R}} {,}\quad n=1/2,\,1,\, \ldots
\,{,}
\end{equation}
where the subscript  plus (minus) corresponds to internal (external)
region and the rest notations are obvious. In ref.\ \cite{BKE} it
was found that
\begin{eqnarray}
\label{eq4_11b} &B_{1/2+}^{\text{D}}=-2\pi^{3/2}a^2=
B_{1/2-}^{\text{D}}, \quad
 B_{1/2+}^{\text{R}}= 2\pi^{3/2}a^2= B_{1/2-}^{\text{R}},&\nonumber \\
&B_{1\pm}^{\text{D}}=\pm\frac{{\displaystyle 8\pi }}{{\displaystyle
3}}a, \quad  B_{1\pm}^{\text{R}}= \mp\frac{{\displaystyle 16 \pi
}}{{\displaystyle 3}}a{.}&
\end{eqnarray}
As a result we have for electromagnetic field inside and outside a
perfectly conducting sphere
\begin{equation}
\label{eq4_11c} B_{1/2}=B_1=0\,{.}
\end{equation}

   Having
calculated the corrections to the Stefan-Boltzmann law one should
naturally discuss the possibility of their detection.  The ratio
of the leading term in eq.\ (\ref{eq4_6}) to the internal energy
of black body radiation in unbounded space given by the
Stefan-Boltzmann law (\ref{eq3_11}) is proportional to $\tau
^{-3}$.  Already for $\tau \sim 10$  the corrections prove to be
of order $10^{-3}$. The same value of $\tau $ can be reached by
varying the scale of length $a$ in the problem under consideration
or by respective choice of the temperature~$T$. Keeping in mind
the value of the conversion coefficient $c\hbar
=197.326\rm{MeV}\cdot\rm{fm} = 0.229\rm{K}\cdot \rm{cm}$
\cite{PDG} we obtain the following estimations. For $a\sim
10^{-13}\rm{cm}$ (a typical hadron size) the temperature $T$
should satisfy the inequality $T\gg 200$\rm{MeV} in order to apply
the asymptotics found.  For $a\sim 1$\rm{cm} we have $T\gg
0.229$\rm{K} and for $a\sim 7\cdot 10^{10}$\rm{cm} (radius of the
Sun) the range of applicability of the asymptotics at hand extends
practically to any temperature value $T\gg 10^{-10}$\rm{K}. Here
we shall not go into the details of a concrete experimental
equipment that enables one to observe the calculated corrections
to the Stefan-Boltzmann law confining ourselves to the estimations
presented above.

\subsubsection{Compact ball with equal velocities of light inside
and outside} \label{HT-cb} The spectral zeta function for this
configuration is given in eqs. (\ref{con}) and (\ref{pol}). It
affords the exact heat kernel coefficients up to $B_{5/2}$
\begin{eqnarray}
&B_0=0, \quad B_{1/2}=0,\quad B_1=0,\quad
B_{3/2}=2\,\pi^{3/2}\,\xi^2, \quad B_2=0, & \nonumber \\
& B_{5/2}= \xi ^2\frac{{\displaystyle c^2}}{{\displaystyle a
^2}}\pi\sqrt \pi p(-1) =0.&\label{eq4_14}
\end{eqnarray}
With allowance of the structure of the omitted terms in eq.\
(\ref{con}) we can again deduce that
\begin{equation}
\label{eq4_14a} B_j=0,\quad j=3,4,5,\ldots \, {.}
\end{equation}
Substitution of  these coefficients into eq.\ (\ref{4-15}) gives
the  following high temperature behavior of the internal energy in
the problem under consideration
\begin{equation}
U(T)\simeq \xi^2 \,\frac{T}{4}+{\cal O}(T^{-3})\,{.}
\label{eq4_15}
\end{equation}

The value of $\zeta'_{\text{ball}}(0)$ is calculated in  the Appendix
\ref{appb-ball}
\begin{equation}
\zeta'(0)=\xi^2 \left(-\frac{5}{8}+\frac{1}{2}\,\ln\frac{a}{c}+\ln
2+\frac{\gamma}{2} \right)
=\xi^2\left(0.35676+\frac{1}{2}\ln\frac{a}{c}\right).
\label{eq4_16}
\end{equation}
It is this  value that is supplied by eq.\ (\ref{in}) with
allowance for that $p(-1)=0$.

By making use of  eqs.\ (\ref{4-12}), (\ref{eq4_14}) and
(\ref{eq4_16}) we deduce the high temperature asymptotics for free
energy
\begin{eqnarray}
F(T)&=&-\xi^2\,\frac{T}{4}\,\left(\gamma+\ln4-\frac{5}{4}\right)+
\frac{\xi^2}{4}\,T\,\ln\frac{\hbar \,c}{a\,T}+{\cal O}(T^{-3}), \nonumber \\
&=&-\xi^2\,\frac{T}{4}0.71352+ \frac{\xi^2}{4}\,T\,\ln\frac{\hbar
\,c}{a\,T}+{\cal O}(T^{-3}){.} \label{eq4_17}
\end{eqnarray}
The entropy in the present case has the following high temperature
behavior
\begin{eqnarray}
S(T)&=&\frac{\xi^2}{4}\,\left(1+\gamma+\ln
4-\frac{5}{4}-\ln\frac{\hbar\,c}{a\,T}\right)+{\cal O}(T^{-4}),
\nonumber\\
&=&\frac{\xi^2}{4}\,\left(1.71352-\ln\frac{\hbar\,c}{a\,T}\right)+
{\cal O}(T^{-4}).
 \label{eq4_18}
\end{eqnarray}
The asymptotics (\ref{eq4_15}) and (\ref{eq4_17}) completely
coincide with the analogous formulae obtained in subsect.\
\ref{ddb-T} by the mode summation method combined with the
addition theorem for the Bessel functions (see also ref.\
\cite{Klich-FMR}).

The exact expression  for the internal energy in the problem at
hand (see eq.\ (\ref{UPf})) gives only exponentially suppressed
corrections to the leading term (\ref{eq4_15})
\begin{equation}
\label{eq4_18a} U(T)\simeq \xi^2\frac{T}{4}\left [
1+2(4t^2+4t+1)e^{-4t} \right ]{.}
\end{equation}
 The asymptotics (\ref{eq4_18a}) implies in particular
that in reality in eq.\ (\ref{eq4_15}) there are no corrections
proportional to the inverse powers of the temperature $T$. From
here it follows immediately that all the heat kernel coefficients
with integer and half integer numbers equal or greater than 3
should vanish
\begin{equation}
\label{eq4_18b}
              B_j=0, \quad j=3,\, 7/2,\, 4,\, 5/2,\, 5,\, \ldots
\end{equation}
(compare with eq.\ (\ref{eq4_14a})). In view of this the sign
denoting the omitted terms in eqs.\ (\ref{eq4_15}),
(\ref{eq4_17}) and (\ref{eq4_18}) should be substituted  by
${\cal O} (e^{-8 \pi aT})$.

\subsubsection{Dielectric ball in unbounded dielectric medium} The
zeta function for electromagnetic field in the background of a
pure dielectric ball ($\mu_1=\mu_2 =1,\;\; \varepsilon_1\not=
\varepsilon _2$) has not been obtained in an explicit form. In
ref.\ \cite{Bordag-db} the heat kernel coefficients up to $B_2$ in
this problem were found.  Here we use the results of this paper
confining ourselves to the $\Delta n^2$-approximation, where
$\Delta n=n_1-n_2=n_1\,n_2\,(c_2-c_1)/c\simeq(c_2-c_1)/c$, $n_i$
and $c_i$ are the refractive index and the velocity of light
inside ($i=1$) and outside ($i=2$) the ball, and $c$ is the
velocity of light in the vacuum: $n_i=\sqrt{\varepsilon _i}, \quad
c_i=c/n_i, \quad i=1,2$. It is assumed that $c_1$ and $c_2$ differ
from $c$ slightly, therefore $c_2-c_1$ and $\Delta n$ are small
quantities. In view of this we have
\begin{eqnarray}
\label{eq4_19} &B_0=\frac{{\displaystyle8}}{{\displaystyle 3}} \pi
a^3\frac{{\displaystyle c_2^3-c_1^3}}{{\displaystyle
c_1^3c_2^3}}\simeq 8\pi\frac{{\displaystyle a^3}}{{\displaystyle
c^3}}\,(\Delta n+2\,\Delta n^2),
&  \nonumber \\
&B_{1/2}=- 2\pi^{3/2}a^2\frac{{\displaystyle
(c_1^2-c_2^2)^2}}{{\displaystyle c_1^2c_2^2(c_1^2+c_2^2)}}
\simeq-4\,\pi^{3/2}\,\frac{{\displaystyle a^2}}{{\displaystyle c^2}}\Delta n^2,& \nonumber \\
&B_1\simeq 0, \quad B_{3/2}=\pi^{3/2}\frac{{\displaystyle
(c_1^2-c_2^2)^2}}{{\displaystyle (c_1^2+c_2^2)^2}}
\simeq\pi^{3/2}\Delta n^2, \quad B_2\simeq 0. &
\end{eqnarray}
The coefficients $B_1$ and $B_2$ equal zero only  in the $\Delta
n^2$-approximation considered here. In the general case they
contain terms proportional to $\Delta n^k$, where $k\geq3$.

Allowance for one more term in  the uniform asymptotic expansion
of the modified Bessel functions, as compared with the
calculations in ref.\ \cite{Bordag-db}, gives the next heat kernel
coefficient
\begin{equation}
\frac{B_{5/2}}{(4\pi)^{3/2}}=\frac{25}{2688}\frac{c^4}{a^2}\Delta
n^4. \label{eq4_20}
\end{equation}
Correcting the mistake made in \cite{Neapol} we state that this
coefficient has no contributions proportional to $\Delta n^2$, and
in the $\Delta n^2$-approximation one has to put
\begin{equation}
B_{5/2}\simeq 0. \label{eq4_20a}
\end{equation}

Making use of the technique developed  in ref.\ \cite{BGKE,BGKE-1}
one obtains the following expression for the derivative of the zeta
function for a pure dielectric ball at the point $s=0$ (see ref.\
\cite{High-T})
\begin{equation}
\zeta'(0)=\frac{\Delta
n^2}{4}\left(-\frac{7}{8}+\ln\,\frac{a}{c}+\ln 4 + \gamma \right).
\label{eq4_21}
\end{equation}

Before turning to the construction of the high temperature
asymptotics in the problem at hand by making use of the general
formulae (\ref{4-12}), (\ref{4-15}) and (\ref{4-16}) a physical
remark should be done. When considering the electromagnetic field
in the background of a dielectric body in the formalism of quantum
electrodynamics of continuous media, as a matter of fact one is
dealing with a system consisting of two objects: electromagnetic
field plus a continuous dielectric body. It is important that this
body is described (phenomenologically) only by respective
permittivity without introducing into the Hamiltonian special
additional dynamical variables. As a result the zeta function and
the relevant heat kernel coefficients calculated in this formalism
also describe both electromagnetic field and dielectric body. When
we are interested in the Casimir thermodynamic functions in such
problems we have obviously to separate in the general expressions
the contributions due to the dielectric body
itself~\cite{Barton-5th}.



Let us turn to such separation procedure in the high temperature
asymptotics for a dielectric ball. Following the reasoning of
refs.\ \cite{Barton-T,Barton-dis} we divide  the Helmholtz free
energy of a material body with volume $V$ and the  surface area
$S$ into the parts
\begin{equation}
F=V\, f+S\,\sigma+F_{\text{Cas}}, \label{eq4.22}
\end{equation}
where $f$ is the free energy of a unit volume of a ball, $\sigma$
denotes the surface tension, and $F_{\text{Cas}}$ is refereed to as
the Casimir free energy of electromagnetic field  connected with
this body and having the temperature $T$.
 In this way we obtain the following high
temperature behavior of the  free energy $F(T)$ in the problem at
hand
\begin{equation}
\label{eq4_23a} F(T)\simeq B_0\frac{T^4}{\hbar
^3}\frac{\pi^2}{90}- B_{1/2}
\frac{T^3}{4\pi^{3/2}\hbar^2}\zeta_{\text{R}}(3)+F_{\text{Cas}}(T),
\end{equation}
where $B_0$ and $B_{1/2}$ are defined in eq.\ (\ref{eq4_19}) and
\begin{equation}
F_{\text{Cas}}(T)\simeq-\frac{\Delta n^2}{8}\,T\,\left(\ln
\frac{4\,T\,a}{\hbar\,c}+\gamma-\frac{7}{8}\right)+
{\cal O}(T^{-2}). \label{eq4_23}
\end{equation}
The high temperature asymptotics for the Casimir internal energy
and for  the Casimir entropy can be derived by making use of the
respective thermodynamical relations (\ref{4-15}), (\ref{4-16})
\begin{eqnarray}
U_{\text{Cas}}(T)&\simeq&\frac{\Delta n^2}{8}\,T+
{\cal O}(T^{-2}){,}
\label{eq4_24}\\
S_{\text{Cas}}(T)&\simeq&\frac{\Delta n^2}{8}\left(
\frac{1}{8}+\gamma+\ln\,\frac{4\,a\,T}{\hbar\,c}\right)+
{\cal O}(T^{-3})\,{.} \label{eq4_25}
\end{eqnarray}

It is worth comparing  these results with analogous asymptotics
obtained by different methods. In our paper \cite{NLS} at the
beginning of calculations the first term of expansion of internal
energy (\ref{eq4_24}) was derived. The subsequent integration of
the thermodynamic relation (\ref{4-13}) gave the correct
coefficient of the logarithmic term in the asymptotics of free
energy (\ref{eq4_23}) (see eq.\ (\ref{eq-10})). In paper
\cite{Barton-T} Barton managed to deduce the asymptotics
(\ref{eq4_23}) -- (\ref{eq4_25}). One should keep in mind that our
parameter $\Delta n$ corresponds to $2\pi\alpha\, n$ in the
notations of ref.~\cite{Barton}.

The asymptotics (\ref{eq4_23})--(\ref{eq4_25}) contain  the
$a$-independent terms. As far as we know the physical meaning of
such  terms remains unclear.

Preliminary analysis of a complete expression for the internal
energy of a dielectric ball (see eqs.\ (3.20) and (3.31) in ref.\
\cite{NLS}) shows that probably there are only exponentially
suppressed corrections to the leading term (\ref{eq4_24}). In that
case in addition to eq.\ (\ref{eq4_20a}) all the heat kernel
coefficients with number greater than 3 should vanish in the
$\Delta n^2$-approximation.

\subsection{Cylinder}
\label{HT-cyl} The calculation of the vacuum energy of
electromagnetic field with boundary conditions defined on a cylinder,
to say nothing  of the temperature corrections, turned out to be
technically a more involved problem than the analogous one for a
sphere. Therefore the Casimir problem for a cylinder has been
considered only in a few
papers~\cite{BD,MNN,deraad,romeo,NP-cyl,BP,LNB}. We again examine
three cases: i) perfectly conducting  cylindrical shell; ii) solid
cylinder with $c_1=c_2$; iii) dielectric cylinder when $c_1\neq c_2$.
Here we shall use the results of our previous papers \cite{BP,LNB}.
\subsubsection{Perfectly conducting cylindrical shell}
\label{HT-cyl-shell}
 The zeta function for this problem is
given by eqs.\ (\ref{zetasc}), (\ref{3.20}), (\ref{Z3con}) and
(\ref{Z3N}).

The function $Z_1(s)$ is defined in the strip $-3/2<\tx{Re}\,
s<1/2$, while the functions $Z_2(s)$ and $Z_3(s)$ are analytic
functions in the whole complex plane $s$ except for the points,
where $\Gamma(s)$ and $\zeta_{\text{R}}(s)$ have simple poles. In
order to find the heat kernel coefficients $B_0$, $B_{1/2}$, and
$B_{1}$ trough the relation (\ref{3-24}) one needs the zeta
function defined in the region  $1/2+\varepsilon\leq \rm{ Re }
s\leq 3/2 +\varepsilon$  with $\varepsilon$ being a positive
infinitesimal. However in this region eq.\ (\ref{3.20}) is not
applicable directly due to the bad behavior of the integral at the
upper limit. In the most simple way we can overcome this
difficulty as in the case of perfectly conducting plates by
introducing the photon mass $\mu$ at the very beginning of the
calculation and making then the analytic continuation  of the zeta
function to the points $s=1/2,\,1,\,3/2$. Upon taking the residua
at these points one should  put $\mu=0$.

With regard to all this and using the relation (\ref{3-24}) we
find the heat kernel coefficients
\begin{equation}
B_0=0,\quad B_{1/2}=0,\quad B_1=0, \quad B_2=0. \label{eq5_5}
\end{equation}
The vanishing  heat kernel coefficient $B_2$  implies that the
zeta regularization gives a finite value  for the vacuum energy in
the problem at hand (see eq.\ (\ref{zetasc1}) and
ref.~\cite{MNN}). The coefficient $B_{3/2} $ is determined by the
function $Z_2(s)$ only (see eq.\ (\ref{Z3con}))
\begin{equation}
\frac{B_{3/2}}{(4\pi)^{3/2}}=\frac{3}{64\,a}.
\end{equation}
The coefficient $B_{5/2}$ is defined by the function $Z_3(s)$
given in eq.\ (\ref{Z3N})
\begin{equation}
\frac{B_{5/2}}{(4\pi)^{3/2}}=\frac{153}{8192}\frac{c^2}{a^3}.
\end{equation}
The calculation of the next heat kernel coefficients
$B_3,\;B_{7/2},\dots$ would demand a knowledge of the additional
terms in the  expansion of the spectral zeta function in the
problem under consideration in terms of the Riemann zeta function.
These terms are proportional to $\zeta_{\text{R}}(2 k+2 s+1)$ with
$k=2,3,\dots\,$, and may be obtained employing the technique
presented in subsect. \ref{SZF+VE-cyl}. Analyzing the position of
poles for these Riemann zeta functions it is easy to show that, as
well as in the spherical case, we have
 \[ B_j=0,\quad j=3,\,4,\,5\dots \,.\]

The zeta determinant entering the high temperature asymptotics of
Hlemholtz free energy (\ref{4-12}) and entropy (\ref{4-16}) is
calculated in the Appendix~\ref{appc-shell}
\begin{equation}
\zeta'(0)=\frac{0.45847}{a}+\frac{3}{32\,a}\,\ln\frac{a}{2\,c}.
\label{eq5_6}
\end{equation}

Now we are able to construct the high temperature expansions of
the thermodynamic functions in the problem under consideration.
For the free energy we have
\begin{equation}
F(T)\simeq-0.22924\,\frac{T}{a}-\frac{3\,T}{64\,a}\,
\ln\frac{a\,T}{2\,\hbar\,c}-\frac{51}{65536}\,\frac{\hbar^2\,c^2}{a^3\,T}
+{\cal O}(T^{-3}). \label{eq5_7}
\end{equation}
When comparing eq.\  (\ref{eq5_7}) with results of other authors
one should remember that all the thermodynamic quantities that we
obtained in this section  are related to a cylinder of unit
length. The high temperature asymptotics  of the electromagnetic
free energy in presence of perfectly conducting cylindrical shell
was investigated  in ref.\ \cite{BD}. To make the comparison handy
we rewrite  their result as follows
\begin{equation}
F(T)\simeq-0.10362\,\frac{T}{a}-\frac{3\,T}{64\,a}\,
\ln\frac{a\,T}{2\,\hbar \,c}. \label{eq5_8}
\end{equation}
The discrepancy between the  terms linear  in $T$ in eqs.\
(\ref{eq5_7}) and (\ref{eq5_8}) is due to the double scattering
approximation used in ref.\ \cite{BD} (see also the next
subsection). Our approach provides  an opportunity to calculate
the exact value of this term (see eq.\  (\ref{eq5_7})).

And finally, making use of the general formulae (\ref{4-15}) and
(\ref{4-16}) we derive
\begin{eqnarray}
U(T)&\simeq&\frac{3\,T}{64\,a}-\frac{153}{98304}\,\frac{c^2\,\hbar^2}{a^3\,T}+
{\cal O}(T^{-3}){,} \label{eq5_9}\\
S(T)&\simeq&\frac{0.27612}{a}+\frac{3}{64\,a}\,\ln\frac{R\,T}{2\,\hbar\,c}-
\frac{153}{196608}\,\frac{c^2\,\hbar^2}{a^3\,T^2}+{\cal
O}(T^{-4}). \label{eq5_10}
\end{eqnarray}

\subsubsection{Compact cylinder with $c_1=c_2$ and with $c_1\neq
c_2$} \label{HT-cylc1=c2} Here we consider the boundary conditions
for electromagnetic field of two types: i) a compact infinite
cylinder with uniform velocity of light on its lateral surface,
ii) a pure dielectric cylinder with $c_1\neq c_2$.

The zeta function for the former configuration is given by eqs.
(\ref{3.25}), (\ref{Z1lin}), (\ref{Z2}) and (\ref{Z3}). It enables
one to find the following heat kernel coefficients
\begin{eqnarray}
&B_0=0, \quad B_{1/2}=0,\quad B_{1}=0,\quad B_{3/2}
=\frac{{\displaystyle 3\pi\sqrt \pi \xi^2}}{{\displaystyle 8\,a}}, \quad B_2=0, & \nonumber \\
&B_{5/2}=\xi^2\frac{{\displaystyle c^2}}{{\displaystyle
a^3}}\frac{{\displaystyle 45\pi\sqrt \pi}}{{\displaystyle 1024}},
\quad B_j=0,\;\;j=3,4,5, \ldots \,{.}& \label{eq5_11}
\end{eqnarray}
As before we are considering the $\xi^2$-approximation. The heat
kernel coefficients (\ref{eq5_11}) lead to the following high
temperature behavior of the internal energy in the problem at hand
\begin{equation}
U(T) = \frac{3\xi^2T}{64\,R}\left( 1-
\frac{5}{512}\frac{c^2\hbar^2}{R^2T^2} \right ) + {\cal
O}(T^{-3})\,{.} \label{eq5_13}
\end{equation}
The corresponding zeta determinant is calculated in Appendix
\ref{appc-compact}
\begin{equation}
\zeta'(0)= \frac{\xi^2}{a}\left( 0.20699+
\frac{3}{32}\,\ln\frac{a}{2\,c}\right). \label{eq5_14}
\end{equation}
Now we can write the high temperature asymptotics for free energy
\begin{equation}
F(T)= -\xi^2\frac{T}{a}\left [0.10350 +\frac{3}{64}\ln
\frac{Ta}{2\hbar c} +\frac{15}{65536}\frac{c^2\hbar^2}{a^2 T^2}
\right ] +{\cal O}(T^{-3}) \label{eq5_15}
\end{equation}
and for entropy
\begin{equation}
S(T)=\frac{\xi^2}{a} \left [ 0.10350 + \frac{3}{64}\left (1+ \ln
\frac{a T}{2\hbar c} \right ) - \frac{15}{65536}\frac{c^2\hbar
^2}{T^2 a^2} \right ]+{\cal O}(T^{-4})\,{.} \label{eq5_16}
\end{equation}

  Putting in these equations $\xi^2=1$ we arrive at the double scattering
approximation for a perfectly conducting cylindrical shell (see
eq.\ (\ref{eq5_8})). A slight distinction between the linear in
$T$ terms in eq. (\ref{eq5_8}) and eq.\ (\ref{eq5_15}) is due to a
finite error inherent in the numerical methods employed in both
the approaches.

In the case of a pure dielectric cylinder
($\mu_1=\mu_2=1,\;\varepsilon_1\neq\varepsilon_2$) the first four
heat kernel coefficients are different from zero even in the
dilute approximation~\cite{BP} (small difference between the
velocities of light inside and outside the cylinder)
\begin{eqnarray}
B_0=-\frac{6\,\pi\,a^2}{c_2^4}\,(c_1-c_2)+
\frac{12\,\pi\,a^2}{c_2^5}\,(c_1-c_2)^2, \quad
B_{1/2}=-\frac{2\,\pi^{3/2}\,a}{c_2^4}\,(c_1-c_2)^2, \nonumber\\
B_1=\frac{8\,\pi}{c_2^2}\,(c_1-c_2)-\frac{14\,\pi}{3\,c_2^3}\,(c_1-c_2)^2,
\quad B_{3/2}=\frac{3\,\pi^{3/2}}{16 a\, c_2^2}\,(c_1-c_2)^2, \nonumber \\
B_2=0,  \quad
 \frac{B_{5/2}}{(4\pi)^{3/2}}=
\frac{857}{61440}\frac{(c_1-c_2)^2}{a^3} \label{eq5_17}.
\end{eqnarray}
It should be noted that the coefficient $B_2$ vanishes only in the
$(c_1- c_2)^2$-approximation. As a matter of fact $B_2$ contains
nonvanishing $(c_1-c_2)^3$-terms and those of higher
order~\cite{BP}. Therefore the zeta regularization provides a
finite answer for the vacuum energy of a pure dielectric cylinder
only in the $(c_1-c_2)^2$-approximation even at zero temperature.

The contribution to the asymptotic expansions of the first three
heat kernel coefficients  should be involved into the  relevant
phenomenological parameters in the general expression of the
classical energy of a dielectric cylinder (in the same way as it
has been done for a pure dielectric ball). By making use of the
coefficients $B_{3/2}$ and $B_{5/2}$ we get the high temperature
asymptotics of the internal energy in the problem at hand
\begin{equation}
U(T)=  \Delta n^2  \frac{3}{128}\frac{T}{R}\left ( 1-
\frac{857}{17280} \frac{c^2\hbar^2}{T^2R^2} \right ) +{\cal
O}(T^{-2})\,{.}
\end{equation}
where $\Delta n=n_1-n_2\simeq(c_2-c_1)/c$.

In view of  considerable technical difficulties we shall not
calculate the zeta function determinant for a pure dielectric
cylinder. We recover the respective asymptotics of free energy by
integrating the thermodynamic relation (\ref{4-13}) and of entropy
by using the relation (\ref{4-14}). Pursuing this way we introduce
a new constant of integration $\alpha $ that remains undetermined
in our consideration
\begin{equation}
F(T)=-\Delta n^2\frac{3 }{128}\frac{T}{a}\left (\alpha + \ln
\frac{aT}{\hbar c} +\frac{857}{34560} \frac{c^2\hbar^2}{T^2a^2}
\right )+{\cal O}(T^{-2})\,{,}
\end{equation}
\begin{equation}
S(T)=\Delta n^2\frac{3}{128} \left (1+ \alpha + \ln\frac{aT}{\hbar
c}- \frac{857}{34560}
 \frac{c^2\hbar^2}{T^2 a^2}
\right )+{\cal O}(T^{-3})\,{.}
\end{equation}
\subsection{Summary of sect. \ref{HT}}
We have demonstrated efficiency and universality
of the high temperature expansions in terms of the heat kernel
coefficients for the Casimir problems with spherical and
cylindrical symmetries. All the known results in this field are
reproduced  in a uniform approach and in addition  a few new
asymptotics are derived (for a compact ball with $c_1=c_2$ and for
a pure dielectric infinite cylinder).

As the next step in the development of this approach one can try
to retain  the terms exponentially  decreasing when $T\to \infty
$. These corrections are well known, for example, for
thermodynamic functions of electromagnetic field in the presence
of perfectly conducting parallel plates \cite{PMG} (see also eq.\
(\ref{eq4_18a})). In order to reveal  such terms, first of all the
exponentially decreasing corrections should be retained in the
asymptotic expansion (\ref{3-23}) for the heat kernel.

It is worth noting that in the framework of the method employed
the high temperature asymptotics can also be constructed in the
problems when the zeta regularization does not provide a finite
value of the vacuum energy at zero temperature, i.e., \ when the
heat kernel coefficient $B_2$ does not vanish. Such problems were
considered, for example, in ref.~\cite{Balian} by making use of
the asymptotic energy densities (see also paper~\cite{Francia}).

     In ref.\ \cite{FMR} it was argued that in the high temperature
limit  the behavior of the Casimir  thermodynamic quantities
should be the following. In the case of disjoint boundary pieces
the free energy tends to minus infinity, the entropy approaches a
constant, and the internal energy vanishes. Contributions to the
Casimir thermodynamic quantities due to each  individual connected
component  of the boundary exhibits logarithmic deviations in
temperature from the behavior just described. In our consideration
we were obviously dealing with  an individual connected component
of the boundary (a sphere or cylinder). Our results corroborate
the relevant conclusions  of ref.\ \cite{FMR} concerning the free
energy and entropy. However the internal energy in our
calculations tends to infinity like $T$ instead to vanish, this
increase being caused by the respective logarithmic terms in the
high temperature asymptotics of  free energy.

  It is worth developing the methods that enable one to reveal the
{\it low temperature asymptotics} of Casimir energy by making use of
the spectral zeta functions and heat kernel technique.

\section{Conclusions}

\label{Con} The reality of  Casimir forces and consequently the
reality of relevant vacuum energy of quantized fields is now well
founded  experimentally. Therefore  the task of the theory in this
area  is to develop mathematically consistent methods for
calculating these quantities. Unfortunately we have  to assert
that till now one didn't succeed  in casting the definition of
the Casimir energy  (\ref{3-2}) on a rigorous mathematical footing
for boundaries of an arbitrary geometry. It is this reason that
causes the controversies persisting here for a long time (see, for
example,
papers~\cite{Milton-book,Milton-contr,Milton-0401117,Graham,Brevik-contr,Hagen-contr,Fulling}
and references therein).

In accordance with  the general concept of the quantum field theory, the
subtraction  procedure (\ref{3-2}) could be rigorously specified
in the framework of proving the renormalizability  of the quantum
field model under study. However it is very unlikely to implement
this by a straightforward generalization of the renormalizability
of quantum electrodynamics treated in unbounded Minkowski space.
Point is that the boundaries or nonhomogeneous characteristics of
the configuration space drastically complicate the
propagators~\cite{Bordag-rad} which are the kernels of the
Schwinger-Dayson integral equations for complete Green's
functions. The proof  of the renormalizability is substantially
based on the analysis of these equations~\cite{IZuber,B+Shir}.

  The following consideration is also important here. The proof
of renormalization theorem is carried out, as a matter of fact, in
the framework of  perturbation theory because the complete kernels of
the Schwinger-Dayson equations are known only in the form of
perturbation series. In order to anticipate the convergence of
resulting  series one has to assume the interaction in the theory
under consideration to be small. The boundaries can also be treated
as an interaction with appropriate classical fields. However the
latter  cannot be for sure considered as a weak one~\cite{Jaffe}. In
this situation application to the calculation of the vacuum energy of
the renormalization group technique is interesting~\cite{Od-4,Cher}

To our opinion, the present  status of the Casimir
calculations can be appropriately described by the following words
due to Milton~\cite{Milton-0401117}: "Obviously we are still at
the early stages of understanding quantum field theory. The nature
of divergences in vacuum energy calculations is still not
understood. However, there are a few established peaks that rise
above the murky clouds of ignorance, and we should not abandon
them lightly because the rest is obscure."

In this situation the spectral zeta function method and heat kernel
technique are distinguished here because they relay on sound
mathematics, reproduce the results obtained in other approaches and
are applicable, at least in principle, to a broad range of physical
problems.

An urgent task in the Casimir studies is the development of  methods
that enable one to do calculations for boundaries without high
symmetry. It is worth noting the proposal to employ in this field
the world line formalism in QFT~\cite{Gies-wl} and optical approach to
the Casimir effect~\cite{optic}. Certainly, one can anticipate here
only {\it approximate} solutions. In order to preserve the rigorous
treatment of the divergencies we suppose that it is  worthy to use here
the chain of relations between the Casimir energy, the spectral zeta
function (\ref{3-11}) and the relevant heat kernel (\ref{3-22}). For
the latter one can construct the {\it integral} equations~\cite{comp}
which provide us at least with perturbation series. The procedure of
analytic continuation should be done here for each term of these
series separately.

\acknowledgments This review was prepared during the visits of one
of the authors (V.V.N.) to Salerno University. He would like to
thank G.~Scarpetta and G.~Lambiase for the kind hospitality
extended to him. Some results on the Casimir calculations
discussed here were obtained in joint works with I.~G.~Pirozhenko,
M.~Bordag and J.~Dittrich. This collaboration is acknowledged with
 gratitude.  V.V.N. was supported in part by the Russian Foundation
for Basic Research (grant No.~03-01-00025). The financial support
of INFN is acknowledged. G.L. and G.S. were supported by PRIN03 \& PRIN05.

\appendix

\section{Analysis of the divergencies generated by $W_{\lowercase{l}}^2$}

\label{appA} Here we reveal an important relation between linear

and quadratic in $\Delta n$ terms in $W_l^2$ defined in eq.\

(\ref{W}).

Let us put
\begin{equation}\label{AW1}
  x_1=y\left(1+\frac{\Delta n}{2}\right)\,,\quad
  x_2=y\left(1-\frac{\Delta n}{2}\right)\,,\quad
  \Delta x=\Delta n\,y\,.
\end{equation}
The Taylor expansion yields
\begin{eqnarray}
  W_l(x_1,
  x_2)&=&s_l(x_1)e_l^{\prime}(x_2)-s_l^{\prime}(x_1)e_l(x_2) \nonumber
  \\
  &=&-1+\left(2s_l^{\prime}e_l^{\prime}-s_le_l^{\prime\prime}-
             s_l^{\prime\prime}e_l\right)\,\frac{\Delta
             x}{2}\nonumber \\
 & &+\left[\frac{1}{2}\left(s_le_l^{\prime\prime\prime}-
        s_l^{\prime\prime\prime}e_l\right)
        +\frac{3}{2}\left(s_l^{\prime\prime}e_l^{\prime}-
         s_l^{\prime}e_l^{\prime\prime}
        \right)\right]\,\frac{\Delta x^2}{4}+O(\Delta x^3)\,. \label{AW2}
\end{eqnarray}
For brevity we have dropped the argument $y$ of the function $s_l$
and $e_l$, and have used the value of the Wronskian
\begin{equation}\label{AW3}
  W\{s_l(y), e_l(y)\}=s_le_l^{\prime}-s_l^{\prime}e_l=-1\,.
\end{equation}
By making use of the equation for the Riccati--Bessel functions
\begin{equation}\label{AW4}
  w_l^{\prime\prime}(y)-L(l, y)\,w_l(y)=0\,,\quad L(l, y)\equiv
  1+\frac{l(l+1)}{y^2}\,,
\end{equation}
we obtain
\begin{eqnarray}
 s_l^{\prime\prime\prime}e_l-s_le_l^{\prime\prime\prime}&=&L(l,
 y)\,, \nonumber \\
 s_l^{\prime\prime}e_l^{\prime}-s_l^{\prime}e_l^{\prime\prime}&=&-
            L(l, y)\,.  \label{AW5}
\end{eqnarray}
Substitution of (\ref{AW5}) into (\ref{AW2}) gives
\begin{equation}\label{AW6}
 W_l(x_1. x_2)=-1+[s_l^{\prime}e_l^{\prime}-L(l, y)s_le_l]\Delta
 x-\frac{1}{2}\,L(l, y)\,\Delta x^2 +O(\Delta x^3)\,.
\end{equation}
Squaring eq.\  (\ref{AW6}) one gets
\begin{equation}\label{AW7}
  W_l^2(x_1, x_2)=1+A_l\,\Delta n+B_l\,\Delta n^2+O(\Delta n^3)\,,
\end{equation}
where
\begin{eqnarray}
  A_l&=&y(s_l''e_l+s_le_l''-2s_l'e_l')=2y\left[2L(l,
  y)s_le_l-\frac{1}{2}\,(s_le_l)^{\prime\prime}\right]\,, \label{AW8}\\
  B_l&=&y^2L(l, y)+\frac{1}{4}\,A_l^2\,. \label{AW9}
\end{eqnarray}

In terms of these notations we can write
\begin{equation}\label{AW10}
  \ln \left(W_l^2-\frac{\Delta n^2}{4}\,P_l^2\right)=
  A_l\Delta n+\left(B_l-\frac{AW_l^2}{2}\right)\Delta n^2-
  \frac{\Delta n^2}{4}\,P_l^2+O(\Delta n^3)\,.
\end{equation}
The terms quadratic in $\Delta n$ in eq.\  (\ref{AW10}) exactly
reproduce the function $F_l(y)$ in eq.\  (9) of the paper \cite{BMM}.
It is this function that affords the whole finite value of the
Casimir energy in the problem under consideration. Unlike the papers
\cite{BMM,BM,Mar} we didn't introduce the contact terms in the
definition of the Casimir energy and nevertheless we have reproduced
the key function $F_l(y)$. It implies that the contact terms do not
give a contribution into the finite part of the Casimir energy in
this problem.  They  merely cancel the terms $A_l\Delta n$ in eq.\
(\ref{AW10}).

Now we show, without invoking the contact terms, that the $A_l$
terms in eq.\ (\ref{AW10}) do not contribute into the vacuum
energy.

Using  eq.\ (\ref{addition}) with $\theta =0$ we introduce the notation
\begin{equation}
\sum_{l=1}^\infty (2l+1)s_l(yr)e_l(y\rho)+1
=\frac{yr\rho}{|r-\rho|} e^{-y|r-\rho|}\equiv \overline {\cal
D}(r,\rho,y)\,{.} \label{AW11}
\end{equation}
Taking into account the explicit form of the coefficients $A_l$
defined in eq.\ (\ref{AW8}) one can write
\begin{equation}
\Delta n\sum_{l=1}^\infty (2l+1)A_l=\left . y\Delta n \left(
\frac{\partial ^2}{\partial r^2}-2\frac{\partial ^2}{\partial r
\partial \rho}
+\frac{\partial^2}{\partial \rho^2} \right ) \overline{\cal
D}(r,\rho,y)\right |_{r=\rho=1}+1\,{.} \label{AW12}
\end{equation}
When $r=\rho=1$ the derivatives of the function ${\cal \overline
D}$ in eq.\ (\ref{AW12}) tend  to infinity. Therefore a
preliminary regularization should be introduced here in order to
put our consideration on a rigorous mathematical footing. To this
end we define the right-hand side of eq.\ (\ref{AW12}) in the
following way
\begin{equation}
\Delta n\sum_{l=1}^\infty (2l+1)A_l=\left .\Delta n
 \lim_{\varepsilon \to 0}
\left( \overline{\cal D}_{rr} -2\overline{\cal D}_{r\rho}
+\overline{\cal D}_{\rho \rho} \right)\right
|_{r=1+{\varepsilon}/{2} \atop \rho=1-{\varepsilon}/{2}} +1\,{,}
\label{AW13}
\end{equation}
where the positive constant $\varepsilon $ is a regularization
parameter. From the explicit form of the function $\overline {\cal
D}(r,\rho,y)$ (see eq.\ (\ref{AW11})) it follows immediately
\begin{equation}
\left .\lim_{\varepsilon \to 0} \left( \overline{\cal D}_{rr}
-\overline{\cal D}_{\rho \rho} \right)\right
|_{r=1+{\varepsilon}/{2} \atop \rho=1-{\varepsilon}/{2}} =0\,{.}
\label{AW14}
\end{equation}
The analogous limit for the differences
\begin{equation}
 \overline{\cal D}_{rr} -\overline{\cal D}_{r \rho} \quad \tx{ and }
 \quad
 \overline{\cal D}_{\rho \rho} -\overline{\cal D}_{r\rho}
\label{AW15}
\end{equation}
also vanishes. Hence  in the regularization introduced above the
sum under consideration has the following value
\begin{equation}
\Delta n\sum_{l=1}^\infty (2l+1)A_l=1\,{.} \label{AW16}
\end{equation}
It implies immediately that the term linear in $\Delta n$, which
encounters eq.\ (\ref{AW10}) does not contribute to the vacuum
energy $E_2$ defined in eq.\ (\ref{E2}).

  Now we show that the contributions to the Casimir energy given by
$\sum_lB_l$ and by $(1/4)\sum_lA_l^2$ are the same. In other
words, $y^2L(l, y)$ in eq.\  (\ref{AW9}) does not give any finite
contribution to the vacuum energy. In order to prove this, we
consider the expression
\begin{equation}\label{AW17}
  I=\sum_{l=1}^{\infty}\nu\int_0^{\infty}y^2\,dy\,,\quad
  \nu=l+\frac{1}{2}\,{.}
\end{equation}
Instead of the cutoff regularization we shall use here the
analytical regularization presenting (\ref{AW17}) in the following
form
\begin{eqnarray}
  I &=& \lim_{s\to 0}\sum_{l=1}^{\infty}\nu\int_0^{\infty}y^{2-s}d  y=
       \lim_{s\to 0}\sum_{l=1}^{\infty}\nu^{4-s}
       \int_0^{\infty}z^{2-s}d  z\nonumber \\
    &=& \lim_{s\to 0}\lim_{\mu^2\to 0}\sum_{l=1}^{\infty}\nu^{4-s}
       \int_0^{\infty}(z^2+\mu^2)^{1-s/2}d  z\,. \label{AW18}
\end{eqnarray}
Here the change of integration variable $y=\nu z$ is done and the
photon mass $\mu$ is introduced. Further we have
\begin{eqnarray}
 I&=&\lim_{s\to 0}\lim_{\mu^2\to 0}[(2^{-4+s}-1)\zeta
 (s-4)-2^{-4+s}]\,\frac{\mu^{3-s}}{2}{
 \frac{\Gamma {\left(\displaystyle\frac{1}{2}\right)}
 \Gamma\left(\displaystyle
  {-\frac{3}{2}+\frac{s}{2}}\right)}{\Gamma\left(
 \displaystyle\frac{s}{2}-1\right)}} \nonumber
 \\
 &=& -\frac{\pi}{24}\lim_{s\to 0}\lim_{\mu^2\to 0}
 \frac{\mu^2}{\Gamma\left (\displaystyle\frac{s}{2}-1
\right )}\to 0\,. \label{AW19}
\end{eqnarray}

In view of all this, we are left with the following scheme for
calculating the Casimir energy of a dielectric ball in the $\Delta n^2$--approximation.  First, the $\Delta
n^2$--contribution should be find, which is given by the sum
$\sum_lW_l^2$. Upon changing its sign to the opposite one, we
obtain the contribution generated by $W_l^2$, when this function
is in the argument of the logarithm. Obviously, this result would
be deduced directly if one could find in a closed form the sum
$\sum_lW_l^2W_l^2$ \cite{Klich}.  This assertion can be represented
by a symbolic formula
\begin{equation} \ln \left(W_l^2-\frac{\Delta
n^2}{4}\,P_l^2\right) \sim -\Delta n^2 B_l- \frac{\Delta
n^2}{4}\,P_l^2+O(\Delta n^3)\,{.} \label{AW20}
\end{equation}
The sign $\sim $ means here the equality subject to the
regularizations described above are employed.

\section{Zeta function determinants for spherically symmetric  boundaries}
 \label{appb}
\subsection{Perfectly conducting sphere}
\label{appb-sph}   We shall use here the integral representation
for the zeta function given in eq. (\ref{zetareal}) with
$\xi^2=1$.
 The analytic continuation of this expression to the region
$\tx{ Re }s<0$ is performed by adding and subtracting from the
integrand its uniform asymptotics at large $\nu$
\begin{equation}
\label{A3} \sigma^2_l(\nu \,z)\simeq\frac{t^6(z)}{4\nu^2}, \qquad
t(z)=\frac{1}{\sqrt{1+z^2}}\,{.}
\end{equation}
As a result we obtain
\begin{equation}
\label{A4} \zeta (s)= Z(s)+ \zeta_{\text{as}}(s),
\end{equation}
where
\begin{eqnarray}
Z(s)&=&\left (\frac{a}{c}\right )^{2s}\frac{\sin (\pi s)}{2\pi}
\sum_{l=1}^{\infty} \nu^{1-2s}\int _0^{\infty} \frac{d
z}{z^{2s}}\frac{d }{d  z }\left\{\ln[1-\sigma_{l}^2(\nu
z)]+\frac{1}{4\nu^2}\frac{1}{(1+z^2)^3}\right\},
\label{A5}\\
\zeta_{\text{as}}(s)&=&\left (\frac{a}{c} \right )^{2s}\frac{3\sin
(\pi s)}{4\pi}\sum_{l=1}^{\infty}\nu^{-1-2s}\int_{0}^{\infty}
d  z\,z^{1-2s}\,t^8(z) \nonumber \\
&=&\frac{1}{4}\left (\frac{a}{c} \right )^{2s}s(1+s)(2+s) \left [
(2^{1+2s}-1)\zeta_{\text{R}}(1+2s)-2^{1+2s} \right ].
 \label{A6}
\end{eqnarray}

When calculating $\zeta '(0)$ one can put in eq. (\ref{A5}) $s=0$
everywhere except for $\sin (\pi s)$, the latter function  being
substituted simply by $\pi s$. In view of this the integral in
eq.\ (\ref{A5}) is evaluated easy if one takes into account the
limits
\begin{equation}
\label{A7} \lim _{z\to 0}\sigma ^2_l(\nu z)=\left[ \frac{\Gamma
(\nu)}{2\Gamma(\nu+1)} \right ] ^2=\frac{1}{4\nu^2}, \quad
\lim_{z\to 0} \frac{t^6(z)}{4\nu^2}= \frac{1}{4\nu^2}
\end{equation}
and the asymptotics (\ref{A3}) at large $z$. As a result we obtain
\begin{equation}
\label{A8} Z'(0)=-2\sum_{l=1}^{\infty}\nu \left [\ln\left
(1-\frac{1}{4\nu^2} \right )+\frac{1}{4\nu^2} \right ]{.}
\end{equation}
Differentiation of eq.\ (\ref{A6}) with respect to $s$ at the
point $s=0$ gives
\begin{equation}
\label{A9} \zeta '_{\text{as}}(0)=-\frac{5}{8}+\frac{1}{2}\ln a +\ln
2+\frac{\gamma}{2}\,{.}
\end{equation}
In order to calculate  the sum over $l$ in eq.\ (\ref{A8}) we
consider  an auxiliary sum
\begin{equation}
S(q)=-\sum_{l=1}^{\infty}2\,\nu \left
[\ln\left(1-\frac{q^2}{4\nu^2}\right)+\frac{q^2}{4\,\nu^2}\right],
\quad S(0)=0,\quad S(1)=Z'(0), \label{A10}
\end{equation}
where $q$ is a parameter. Derivative of this sum with respect to
$q$ can be rewritten in the form
\begin{equation}
\label{A11} S'(q)=-\frac{q}{2}\sum_{l=1}^{\infty}\left [
\frac{1}{l+1/2}-\frac{1}{l+(1+q)/2}+\frac{1}{l+1/2}-
\frac{1}{l+(1-q)/2} \right ]{.}
\end{equation}
The summation in eq.\ (\ref{A11})  is accomplished  by making use of
the following relations~\cite{GR}
\begin{eqnarray}
\sum_{k=1}^{\infty}\left (\frac{1}{y+k}-\frac{1}{x+k}
\right ) =\frac{1}{x}-\frac{1}{y}+\psi (x)-\psi(y), \nonumber \\
\psi (x+1) =\psi (x) +\frac{1}{x}, \quad \psi \left (\frac{1}{2}
\right )=-\gamma -2\ln 2, \label{A12}
\end{eqnarray}
where $\psi (x)$ is the digamma function (the Euler $\psi $
function): $\psi (x) =(d  /d  x)\ln \Gamma (x)$. This gives
\begin{equation}
\label{A13} S'(q)=q\,(2-\gamma -2 \ln 2) -\frac{q}{2} \left [ \psi
\left (\frac{3}{2}+\frac{q}{2} \right ) + \psi \left
(\frac{3}{2}-\frac{q}{2} \right ) \right ]{.}
\end{equation}
Now we integrate the both sides of eq.\ (\ref{A13}) over $a$ from
$0$ to $1$ by making use of `Maple'
\begin{equation}
\label{A14} S(1)=Z'(0)=
\frac{1}{2}-\frac{\gamma}{2}+\frac{7}{6}\ln 2 +6\,\zeta
'_{\text{R}}(-1)\,{.}
\end{equation}
>From eqs.\ (\ref{A4}), (\ref{A9}) and (\ref{A14}) it follows that
\begin{eqnarray}
\zeta'(0)&=&\frac{1}{2}-\frac{\gamma}{2}+\frac{7}{6}\ln
2+6\,\zeta'_{\text{R}}(-1)+\left(-\frac{5}{8}+\frac{1}{2}\ln
\frac{a}{c}+\ln
2+\frac{\gamma}{2} \right )\nonumber\\
&=& -\frac{1}{8}+\frac{13}{6}\ln 2+6\zeta'_{\text{R}}(-1)+
\frac{1}{2}\ln \frac{a}{c}= 0.38429+\frac{1}{2}\ln \frac{a}{c}.
\label{A999}
\end{eqnarray}

\subsection{Material ball with $c_1=c_2$}
\label{appb-ball}  The complete zeta function in this problem is
given in eq. (\ref{zetareal}) (without expanding in powers of the
parameter $\xi^2$ ). Adding and subtracting under the integral
sign in this equation the uniform asymptotics of the integrand at
large $\nu$ we get
\begin{eqnarray}
\zeta(s)&=&\left (\frac{a}{c}\right )^{2s}\frac{\sin(\pi s)}{2\pi}
\sum_{l=1}^{\infty} \nu^{1-2s}\int_0^{\infty} \frac{d
z}{z^{2s}} \frac{d }{d  z }\left\{\ln[1-\xi^2\sigma_{l}^2(\nu
z)]+\frac{\xi^2}{4\nu^2}\frac{1}{(1+z^2)^3}\right\}\nonumber\\
&&+ \xi^2\zeta_{\text{as}}(s),
 \label{A17}
\end{eqnarray}
where the function $\zeta_{\text{as}}(s)$ was introduced in eq.\
(\ref{A6}). Proceeding in the same  way as in the previous
subsection we obtain for the derivative of the function $\zeta
(s)$ at the point $s=0$
\begin{equation}
\zeta'(0)=S(\xi^2) +\xi ^2 \zeta'_{\text{as}}(0),
 \label{A18}
\end{equation}
where the function $S(\xi ^2)$ is defined in eq.\ (\ref{A10}). For
small values of the argument $\xi^2$  we deduce from eq.\
(\ref{A10})
\begin{equation}
\label{19}
S(\xi^2)=\frac{\xi^4}{16}\sum_{l=1}^{\infty}\frac{1}{\nu ^3}=
\frac{\xi^4}{16}\left [7\,\zeta_{\text{R}}(3)-8\right ]+{\cal
O}(\xi^6).
\end{equation}
Therefore restricting ourselves to the first order of $\xi^2$ we
arrive at the final result
\begin{equation}
\zeta'(0)=\xi^2\,\zeta'_{\text{as}}(0)
=\xi^2\,\left(-\frac{5}{8}+\frac{1}{2}\ln \frac{a}{c}+\ln
2+\frac{\gamma}{2} \right ). \label{A20}
\end{equation}
\section{Zeta function determinants for
 cylindrically symmetric  boundaries}
 \label{appc}
\subsection{Perfectly conducting cylindrical shell}
\label{appc-shell} The spectral zeta function for this
configuration has been considered in subsect.\ \ref{SZF+VE-cyl}.
This function was used there for obtaining a finite value for the
relevant Casimir energy. For this the value of the zeta function
at the point $s=-1/2$ was calculated. Now we are interested in the
value of $\zeta'(0)$. Therefore  it is convenient to represent the
initial formula (\ref{3.11}) with $\xi^2=1$ in a slightly
different way as compared with the subsect.\ \ref{SZF+VE-cyl},
namely
\begin{eqnarray}
\zeta(s)&=&\frac{a^{2s-1}}{2\sqrt{\pi}c^{2s}
\Gamma(s)\,\Gamma(3/2-s)}\int _0^{\infty}d  y\,y^{1-2s}
\frac{d }{d  y}\ln[1-\mu _0^2(y)]\nonumber\\
&&+\frac{a^{2s-1}}{\sqrt{\pi}c^{2s}
\Gamma(s)\,\Gamma(3/2-s)}\sum_{n=1}^{\infty}n^{1-2s} \int
_0^{\infty}d  y\,y^{1-2s} \frac{d }{d  y}\ln[1-\mu
_n^2(ny)], \label{B1}
\end{eqnarray}
where $\mu _n(y)$ was defined in eq.\ (\ref{3.12}). The first term
on the right hand side of eq.\ (\ref{B1}) is an analytic function
of the complex variable $s$ in the strip $-1/2< \tx{ Re } s <
1/2$. Therefore there is no need in analytic continuation of this
expression when calculating $\zeta '(0)$. As regard to the second
term  in eq.\ (\ref{B1}) its analytic continuation to the region
$\tx{ Re } s<0$ can be accomplished in a standard way. We add and
subtract here the uniform asymptotics of the integrand when $n$
tends to infinity
\begin{equation}
\ln[1-\mu ^2_n(ny)]\simeq-\frac{y^4\,t^6(y)}{4\,n^2}+ {\cal
O}(n^{-4}),\quad t(y)=\frac{1}{\sqrt{1+y^2}}. \label{B2}
\end{equation}
As a result we obtain
\begin{eqnarray}
\zeta(s)&=&\frac{a^{2s-1}}{2\sqrt{\pi}\,c^{2s}
\Gamma(s)\,\Gamma(3/2-s)} \int_0^{\infty} \frac{d
y}{y^{2s-1}}\frac{d  }{d   y
}\ln[1-\mu ^2_{0}(y)]\nonumber\\
&&+\frac{a^{2s-1}}{\sqrt{\pi}c^{2s}
\Gamma(s)\,\Gamma(3/2-s)}\sum_{n=1}^{\infty}n^{1-2s} \int
_0^{\infty} \frac{d  y}{y^{2s-1}}\frac{d  }{d   y
}\left\{\ln[1-\mu ^2_{n}(ny)]+\frac{y^4\,t^6}{4\,n^2}\right\}
\nonumber\\&&-\frac{a^{2s-1}}{32\sqrt{\pi}c^{2s}} (1-2s)(3-2
s)\zeta_{\text{R}}(2s+1) \frac{\Gamma(1/2+s)}{\Gamma(s)}. \label{B3}
\end{eqnarray}
Keeping in mind the behavior of the gamma  function at the  origin
$\Gamma (s)\simeq s^{-1}$  one can easily find the derivative of
$\zeta (s)$ at the point $s=0$
\begin{eqnarray}
\lefteqn{\zeta'(0)= \frac{1}{\pi a}\int _0^{\infty}d
y\,y\frac{d }{d  y}
\ln[1-\mu _0^2(y)] }\nonumber  \\
&&+\frac{2}{\pi\,a}\sum_{n=1}^{\infty}n \int _0^{\infty}d
y\,y\frac{d }{d  y} \left\{\ln[1-\mu _n^2(n\,y)]+\frac{y^4
t^6}{4n^2}\right\}
+\frac{1}{32a}\,\left(3\gamma-4-3\ln\frac{2c}{a}\right).
\label{B4}
\end{eqnarray}
Unlike the spherically symmetric boundaries,  the integration is
not removed in the formula obtained for $\zeta '(0)$. Therefore
the first two terms in eq.\ (\ref{B4}) can be  calculated only
numerically
\begin{equation}
\label{B5} -\frac{1}{\pi a}\int _0^{\infty}dy \ln[1-\mu _0^2(y)]
=\frac{0.53490}{a}{.}
\end{equation}
Applying the FORTRAN subroutine that approximates the Bessel
functions by Chebyshev's polynomials we evaluate the first 30
terms in the sum in eq.\ (\ref{B4})
\begin{equation}
\label{B6} -\frac{2}{\pi\,a}\sum_{n=1}^{\infty}n \int
_0^{\infty}dy \left\{\ln[1-\mu _n^2(n\,y)]+\frac{y^4
t^6}{4n^2}\right\}=-\frac{0.00554}{a}{.}
\end{equation}
Finally  gathering together all these results we have
\begin{equation}
\zeta'(0)=\frac{0.45847}{a}+\frac{3}{32\,a}\ln\frac{a}{2c}.
\label{B7}
\end{equation}

\subsection{Compact infinite cylinder with $c_1=c_2$}
\label{appc-compact} Now we turn to a compact cylinder placed into
unbounded medium such that the velocity of light is uniform on the
lateral surface of the cylinder. Proceeding as in the case of a
cylindrical shell we rewrite the initial equation (\ref{3.11}) in
the linear approximation with respect to $\xi^2$ in the form
\begin{eqnarray}
\zeta(s)&=& -\frac{a^{2s-1}\,\xi^2}{2\sqrt{\pi}c^{2s}
\Gamma(s)\Gamma(3/2-s)}\int_0^{\infty}d  y\,y^{1-2s}
\frac{d }{d  y}\mu _0^2(y)\nonumber\\
&&-\frac{a^{2s-1}\xi^2}{\sqrt{\pi}c^{2s}
\Gamma(s)\Gamma(3/2-s)}\sum_{n=1}^{\infty} \int _0^{\infty}d
y\,y^{1-2s} \frac{d }{d  y}\mu _n^2(y), \label{B9}
\end{eqnarray}
where $\mu_n(y)$ is defined in eq.\ (\ref{3.12}). The analytic
continuation to the region $\tx{ Re }s<0$ is needed only for the
second term in eq.\ (\ref{B9}). Adding and subtracting here the
uniform asymptotics of the integrand for large~$n$
\begin{equation}
-\mu ^2_n(ny)\simeq-\frac{y^4t^6(y)}{4n^2} +{\cal O}(n^{-4}),
\label{B10}
\end{equation}
we obtain
\begin{eqnarray}
\zeta'(0)&=& -\frac{\xi^2}{\pi a}\int _0^{\infty}d
y\,y\frac{d }{d  y} \mu _0^2(y)+\frac{2\xi^2}{\pi
a}\sum_{n=1}^{\infty}n \int _0^{\infty}d  y\,y\frac{\drm}{\drm
y}
\left[-\mu _n^2(n\,y)+\frac{y^4\,t^6}{4\,n^2}\right]\nonumber\\
&&+\frac{\xi^2}{32a}\left(3\gamma-4-3\ln\frac{2\,c}{a} \right) {.}
\label{B11}
\end{eqnarray}
The first two terms in eq.\ (\ref{B11}) can again be calculated
only numerically
\begin{equation}
\label{B12} \frac{\xi^2}{\pi a}\int_{0}^{\infty}\drm y\,
\mu_0^2(y)=\frac{\xi^2}{a}\,0.28428{,}
\end{equation}
\begin{equation}
\label{B13} -\frac{2\xi^2}{\pi
a}\sum_{n=1}^{\infty}n\int_{0}^{\infty}\drm y \left [
-\mu^2_n(ny)+\frac{y^2t^6}{4n^2} \right ]
=-\frac{\xi^2}{a}\,0.00640{.}
\end{equation}
The final result reads
\begin{eqnarray}
\zeta'(0)&=&\frac{\xi^2}{a}\left[0.28428-0.00640+
\frac{1}{32}\left(3\gamma-4-3\ln\frac{2c}{a}\right)\right]\nonumber\\
&=& \frac{\xi^2}{a}\left(0.20699+
\frac{3}{32}\ln\frac{a}{2c}\right). \label{B14}
\end{eqnarray}


\begin{thebibliography}{99}
\bibitem{Casimir} Casimir~H.~B.~G.,\ {Proc. K. Ned. Akad. Wet.}\ {\bf 51}\ (1948)\ {793}.
\bibitem{Sparnaay}{Sparnaay~M.~J.,} {Physica (Utrecht)} {\bf 24}
 (1958) {751}; {Physics in Making}, edited by
{A.~Sarlemijn and M.~J.~Sparnaay} (North-Holland, Amsterdam)
1989.
\bibitem{dielectric} {Tabor~D., Winterton~R.~H.~S.,} {Proc.
 R.~Soc.\ London A} {\bf 312} (1969) {435};
 {Israelachvili~J.~N.,  Tabor~D.,}
{Proc.
 R.~Soc. London A} {\bf 331} {(1972)} {19}; {Van Blokland~P.~H.~G.~M.,
Overbeek~J.~T.~G.,} {J. Chem. Soc. Faraday Trans.} {\bf 74} (1978)
 2637.
\bibitem{Lamoreaux} {Lamoreaux~S.~K.,} {Phys. Rev. Lett.} {\bf 78} {(1997)} {5};
{Comm. Mod. Phys. D: At. Mol. Phys.} {\bf 2} {(2000)} {247}.
\bibitem{Mohideen} {Mohideen~U.,  Roy~A.,} {Phys. Rev.
Lett.} {\bf 81} {(1998)} {4549}; {Roy~A., Mohideen~U.,} {Phys. Rev.
Lett.} {\bf 82} {(1999)} {4380}; {Harris~B.~M., Chen~F.,
Mohideen~U.,} {Phys. Rev. A} {\bf 62} {(2000)} {052109}; {Roy~A.,
Lin~C-Y, Mohideen~U.,} {Phys. Rev. D} {\bf 60} {(1999)} {111101};
{Comm. Mod. Phys. D: At. Mol. Phys.} {\bf 2} {(2000)} {263}.
\bibitem{chan1}{Chan~H.~B. et al.,} {Science} {\bf 291} {(2001)} {1941}.
\bibitem{roberto}{Bressi~G., Carugno~G., Onofrio~R.,  Ruoso~G.,}
{Phys. Rev. Lett.} {\bf 88} {(2002)} {041804}.
\bibitem{Serry}{Serry~F.~M. et al.,} {J.~Microelectromechanical
 Syst.} {\bf 4} {(1995)} {193}.
\bibitem{chan2}{Chan~H.~B. et al.,} {Phys. Rev. Lett.} {\bf 87} {(2001)} {1801}.
\bibitem{thermal} {Chen~F., Klimchitskaya~G.~L.,
Mohideen~U.,  Mostepanenko~V.~M.,} {Phys. Rev. Lett.} {\bf 90}
{(2003)} {160404}.
\bibitem{confront}
{Chen~F., Klimchitskaya~G.~L., Mohideen~U., Mostepanenko~V.~M.,}
{Phys. Rev. A} {\bf 69} {(2004)} {022117}.
\bibitem{GEsp} {Bimonte~G., Calloni~E., Esposito~G., Milano~L. and Rosa~L.,}
arXive: quant-ph/0406188 v3.
\bibitem{Boyer} {Boyer~T.~H.,} {Phys. Rev.} {\bf 174} {(1968)} {1764}.
\bibitem{Davies}{Davies~B.,} {J. Math. Phys.} {\bf 13} ({1972}) {1324}.
\bibitem{BD} {Balian~R., Duplantier~B.,} {Ann. Phys. (N.Y.)} {\bf 104} {(1977)} {300};
  {\it ibid.} {\bf 112} {(1978)} {165}.
\bibitem{MRS}{Milton~K.~A., DeRaad (Jr.)~L.~L., Schwinger~J.,}
{Ann. Phys. (N.Y.)} {\bf 115} {(1978)} {388}.
\bibitem{piro} {Nesterenko~V.~V. , Pirozhenko~I.~G.,} {Phys. Rev.
D} {\bf 57} {(1997)} {1284}.
\bibitem{bowers} {Bowers~M.~E.,  Hagen~C.~R.,} {Phys. Rev. D} {\bf 59} {(1999)} {025007}.
\bibitem{MNN} {Milton~K.~A.,  Nesterenko~A.~V.,  Nesterenko~V.~V.,} {Phys.
Rev. D} {\bf 59} {(1999)} {105009}.
\bibitem{deraad} {DeRaad (Jr.)~L.~L., Milton~K.,} {Ann. Phys.
(N.Y.)} {\bf 136} {(1981)} {229}; {DeRaad (Jr.)~L.~L.,} {Fortschr.
Phys.} {\bf 33} {(1985)} {117}.
\bibitem{romeo}{Gosdzinsky~P., Romeo~A.,} {Phys. Lett.
B} {\bf 441} {(1998)} {265}.
\bibitem{Lif} {Lifshitz~E.~M.,} {Zh. Eksp. Teor.
Fiz.} {\bf 29} {(1955)} {894};
              [{Sov. Phys. JETP} {\bf 2} {(1956)} {73}];
{Lifshitz~E.~M., Pitaevskii~L.~P.}, {Statistical Physics}
(Pergamon, Oxford) 1980, Part~2.
\bibitem{Kampen} {van Kampen~N.~G., Nijboer~B.~R.~A., Schram~K.,}
{Phys. Lett. A} {\bf 26} {(1968)} {307}.
\bibitem{SDeRM} {Schwinger~J., DeRaad (Jr.)~L.~L. and Milton~K.~A.,}
{Ann. Phys. (N.Y.)} {\bf  115} {(1978)} {1}.
\bibitem{BNP} {Brevik~I., Nesterenko~V.~V. and  Pirozhenko~I.~G.,}
 {J.~Phys. A} {\bf 31} {(1998)} {8661}.
\bibitem{NP-cyl} {Nesterenko~V.~V. and Pirozhenko~I.~G.,}
{Phys. Rev. D} {\bf 60} {(1999)} {125007}.
\bibitem{BC} {Brevik~I., Kolbenstvedt~H.,} {Ann. Phys.
(N.Y.)} {\bf 143} {(1982)} {179}; {\it ibid.} {\bf 149} {(1983)}
{237}; {Can. J. Phys.} {\bf 62} {1984} {805}; {\it ibid.} {\bf 63}
{(1985)} {1409}.
\bibitem{Brevik+Nyland}
{Brevik~I. and  Nyland~G.~H.,} {Ann. Phys. (N.Y.)} {\bf 230} {(1994)}
{321}.
\bibitem{BC1} {Brevik~I. and  Einevoll~G.,} {Phys. Rev. D} {\bf 37} {(1988)} {2977}.
\bibitem{BC2} {Brevik~I. and Clausen~I.,} {Phys. Rev. D} {\bf 39} {(1989)} {603}.
\bibitem{BC3} {Brevik~I.} {J.~Phys.\ A} {\bf 20} {(1987)} {5189}.
\bibitem{BY} {Brevik~I. and Yousef~T.~A.,}
{J.~Phys. A} {\bf 33} {(2000)} {5819}.
\bibitem{Klich} {Klich~I.,} {Phys. Rev. D} {\bf 61} {(2000)} {025004}.
\bibitem{Klich-2} {Klich~I.,} {Phys. Rev. D} {\bf 64} {(2001)} {045001}.
\bibitem{Klich-Romeo} {Klich~I. and Romeo~A.,} {Phys. Lett. B} {\bf 476} {(2000)} {369}.
\bibitem{Klich-FMR}
{Klich~I., Feinberg~J., Mann~A. and Revzen~M.,}
{Phys. Rev. D} {\bf 62} {(2000)} {045017}.
\bibitem{Milton} {Milton~K.~A.,} {Ann. Phys. (N.Y.)} {\bf 127} {(1980)} {49}.
\bibitem{MNg2} {Milton~K.~A. and  Ng~Y.~J.,} {Phys. Rev.
E} {\bf 57} {(1998)} {5504}.
\bibitem{Barton}  {Barton~G.,} { J.~Phys. A} {\bf 32} {(1999)} {525}.
\bibitem{BMM}{Brevik~I.,  Marachevsky~V.~N. and
Milton~K.~A.,}
 {Phys. Rev. Lett.} {\bf 82} {(1999)} {3948}.
\bibitem{BM}{Brevik~I. and  Marachevsky~V.~N.,}
 {Phys. Rev. D} {\bf 60} {(1999)} {085006}.
 \bibitem{LSN} {Lambiase~G., Scarpetta~G. and
Nesterenko~V.~V.,} {Mod. Phys. Lett.} {\bf 16} {(2001)} {1983}.
\bibitem{Mar} {Marachevsky~V.~N.,} arXive: hep-th/9909210, Preprint
SPb-IP-99-11; {Phys. Scripta} {\bf 64} {(2001)} {2005}.
\bibitem{HB-stat}{H{\o}ye~J.~S. and Brevik~I.,} {J.~Stat. Phys.} {\bf 100} {(2000)} {223}.
\bibitem{HBA-stat-QFT}{H{\o}ye~J.~S.,  Brevik~I. and
Aarseth~J.~B.}  {Phys. Rev. E} {\bf 63} {(2001)} {051101}.
\bibitem{NLS} {Nesterenko~V.~V., Lambiase~G. and
Scarpetta~G.,} {Phys. Rev. D} {\bf 64} {(2001)} {025013}.
\bibitem{Barton-T} {Barton~G.,}
{Phys.\ Rev.\ A} {\bf 64} {(2001)} {032103}.
\bibitem{Barton-cyl} {Barton~G.,} {J. Phys., A} {\bf 34} {(2001)} {4083}.
\bibitem{BP} {Bordag~M. and Pirozhenko~I.~G.,}
{Phys. Rev., D} {\bf 64} {(2001)} {025019}.
\bibitem{C-P+Milton} {Cavero-Pelaez~I. and  Milton~K.~A.,} arXiv: hep-th/0412135.
\bibitem{RomeoMilton} {Romeo~A. and  Milton~K.~A.,} arXive: hep-th/0504207.
\bibitem{derjaguin}{Derjaguin~B.~V. and Abriksova~I.~I.,} {Sov. Phys.
JETP} {\bf 3} {(1957)} {819}; {Derjaguin~B.~V.,} {Sci. Am.} {\bf 203}
{(1960)} {47}.
\bibitem{Gies} {Gies~H., Langfeld~K. and Moyaerts~L.,}
{JHEP} {\bf 0306} {(2003)} {018}.
\bibitem{schaden1}{Schaden~M. and Spruch~L.,}
\bibitem{saha} {Saharian~A.~A.,}
{Phys. Rev. D} {\bf 63} {(2001)} {125007}; {Int. J. Mod. Phys. A}
{\bf 19} {(2004)} {4301}.
\bibitem{sahats} {Saharian~A.~A.,} ICTP preprint, IC/2000/14; hep-th/0002239.
\bibitem{Ahmedov-close} {Ahmedov~H. and Duru~I.~H.,}
{J.~Math. Phys.} {\bf 44} {(2003)} {5487}.
\bibitem{Jaffe-1} {Jaffe~R.~L. and Scardicchio~A.,e} arXiv: hep-th/0501171.
\bibitem{Milton-JPhys} {Milton~K.~A.,}
 {J.~Phys. A} {\bf 37} {(2004)} {R209}.
{Phys. Rev. Lett.} {\bf 84} {(2000)} {459}.
\bibitem{Bordag} {Bordag~M., Mohideen~U. and
Mostepanenko~V.~M.} {Phys. Rep.} {\bf 353} {(2001)} {1}.
\bibitem{Milton-book} {Milton~K.~A.,}
{\it The Casimir Effect: Physical Manifestations of Zero-Point
Energy}, (World Scientific, Singapoore) 2001.
\bibitem{MT} {Mostepanenko~V.~M. and Trunov~N.~N.,} {\it The
Casimir effect and its applications} (Clarendon, Oxford) 1997.
\bibitem{GMM} {Grib~A.~A., Mamaev~S.~G. and
Mostepanenko~V.~M.}, {\it Vacuum Quantum Effects in Strong Fields}
(Friedman Laboratory Publishing,  St.  Petersburg) 1994.
\bibitem{ElRomeo} {Belifante~F.~J.,}
{Am. J. Phys.} {\bf 55} {(1987)}{134}; {Elizalde~E. and Romeo~A.,}
{Am.\ J.~Phys.} {59} {(1991)} {711}.
\bibitem{Milonni-book}{Milonni~P.~W.},
{\it The Quantum Vacuum: An Introduction to Quantum
Electrodynamics} (Academic Press, San Diego) 1994.
\bibitem{Kardar} {Kardar~M. and Golestanian~R.,}
{Rev. Mod. Phys.} {\bf 71} {(1999)} {1233}.
\bibitem{PMG} {Plunien~G., Muller~B. and Greiner~W.,}
 {Phys. Rep.} {\bf 134} {(1986)} {87}.
\bibitem{Roberts} {Roberts~M.~D.}, {\it Vacuum Energy},
arXive: hep-th/0012062.
\bibitem{Bordag-W3}  {\it Leipzig 1995, Quantum Field Theory Under the
Influence of External Conditions}, Proceedings of the 3d Workshop
on Quantum Field Theory under the Influence of External Conditions
(Leipzig, Germany, 18 -- 22 Sep 1995), edited by {M.~Bordag}
(Teubner, Stutgart) 1996.
\bibitem{Bordag-W4}  {\it The Casimir effect 50 years later},
Proceedings of the 4th Workshop on Quantum Field Theory under the
Influence of External Conditions (Leipzig, Germany, 14 -- 18 Sep
1998), edited by {M.~Bordag} (World Scientific, Singapore) 1999.
\bibitem{Bordag-W5} Proceedings of the 5th Workshop on Quantum
Field Theory under the Influence of External Conditions (Leipzig,
Germany, 10-14 Sep 2001), edited by {M.~Bordag} {\it Int. J. Mod.
Phys. A} {\bf 17} (2002) No.~6 and 7.
\bibitem{Bordag-W6} {\it Norman 2003, Quantum Field Theory Under the Influence of
External Conditions}, Proceedings of the 6th Workshop on Quantum
Field Theory under the Influence of External Conditions (Norman,
Oklahoma, USA, 15-19 Sep 2003), edited by {K.~A.~Milton} (Rinton
Press, Paramus, NJ, USA) 2004.
\bibitem{Planck} {Planck~M.,}
 {Ann. d. Phys.} {\bf 37} {(1912)} {642};
Milonni~P.~W. and Shih M.-L., Amer. J. Phys. {\bf 59} (1991) 684.
\bibitem{ES} {Einstein~A.,  Stern~O.,}
 {Ann. d. Phys.} {\bf 40} {(1913)} {551}.
\bibitem{DW} {Debye~P.,}
 {Ann. d. Phys.} {\bf 43} {(1914)} {49}.
\bibitem{Mulliken} {Mulliken~R.~S.,}
 {Nature} {\bf 114} {(1924)} {349}.
\bibitem{Einstein} {Einstein~A.,}
 {Ann. d. Phys.} {\bf 22} {(1907)} {180}.
\bibitem{Debye} {Debye~P.,}
{Ann. d. Phys.} {\bf 39} {(1912)} {789}.
\bibitem{MatthewsSalam}
{Matthews~P.~T. and Salam~A.,}
 {Phys. Rev.} {\bf 94} {(1954)} {185}.
\bibitem{Gunn} {Gunn~J.~C.,}
{Rep. Prog. Phys.} {\bf 18} {(1955)} {127}.
\bibitem{Dirac} {Dirac~P.~A.~M.},
 {\it Lectures on Quantum Field Theory} (Published by Belfer Graduate
School of Science, Yeshiva University, New York) 1967.
\bibitem{Schwinger-LMP} {Schwinger~J.,}
{Lett. Math. Phys.   {\bf 1} {(1975)} {43}; {\it  ibid.} {\bf 24}
{(1992)} {59}; {\it ibid.} {\bf 24} {(1992)} {227}.
\bibitem{Milonni-PRA} {Milonni ~P.~W.,}
{Phys. Rev. A  } {\bf 25} {(1982)} {1315}.
\bibitem{Milonni-PS} {Milonni ~P.~W.,}
{Phys. Scripta T  } {\bf  21} {(1988)} {102}.
\bibitem{Jaffe} {Jaffe~R.~L.}
preprint MIT-CTP-3614; arXiv:
  hep-th/0503158.
\bibitem{Barash} {Barash~Yu.~S.,}
{\it Van der Waals Forces} (Nauka, Moscow) 1988.
\bibitem{Hertz} {Hertz Heinrich}, {\it Die Principien der Mechanik
in neuem Zusammenhange dargestellt}, Gesam. Werke, Bd. III
(Leipzig) 1910.
\bibitem{Sommerfeld}{Sommerfeld A.,}  {\it Vorlesungen \"uber Theoretische Physik}
Bd.~I, {Mechanik}, vierte Auflage (Dieterich'sche
Verlagsbuchhandlung, Wiesbaden) 1949.
\bibitem{Marshall} {Marshall~T.~W.,}
{Proc. R. Soc. Lond. A} {\bf    276} {(1963)} {475}.
\bibitem{Boyer-1} {Boyer~T.~H.,}
{Phys. Rev. } {\bf   182} {(1969)} {1374}.
\bibitem{MB-exp} {Marshall~T.~W.,}
{Pros. Camb. Philos. Soc.} {\bf    61} {(1965)} {537}; {Nuovo
Cimento} {\bf  38} {(1965)} {206}; {Henry~L.~L. and Marshall~T.~W.,}
{Nuovo Cimento} {\bf  41} {(1966)} {188}; {Boyer~T.~H.,} {Phys. Rev.}
{\bf  186} {(1969)} {1304}; {Phys. Rev. D } {\bf   1} {(1970)}
{1526}; {\it ibid.} {\bf 1} (1970) 2257}.
\bibitem{IZuber} {Itzykson~C. and Zuber~J.-B.,}
{\it Quantum Field Theory} (McGraw-Hill, New York) 1980.
\bibitem{Birrell-Davies} {Birrell~N.~D. and Davies~P.~C.~W.,}
{\it Quantum Fields in Curved Space} (Cambridge University Press,
Cambridge, UK) 1982.
\bibitem{Kirsten} {Kirsten~K.,}
{\it Spectral Functions in Mathematics and Physics} (Chapman \&
Hall/CRC, Boca Raton, FL, USA) 2001.
\bibitem{Od} {Elizalde~E., Odintsov~S.~D., Romeo~A.,
Bytsenko~A.~A. and Zerbini~S.,}, {\it Zeta Regularization Technique
with Applications} (World Scientific, Singapore) 1994.
\bibitem{Ten} {Elizalde~E.},
{\it Ten Physical Applications of Spectral Zeta Functions}
(Springer, Berlin) 1995.
\bibitem{Vassilevich} {Vassilevich~D.~V.,}
{Phys. Rept.} {\bf 388} {(2003)} {279}.
\bibitem{Santangelo} {Santangelo~E.~M.,}
{Theor. Math. Phys.} {\bf  131} {(2002)} {527}; hep-th/0104025.
\bibitem{BVW} {Blau~S.~K., Visser~M. and Wipf~A.,}
{Nucl. Phys. B} {\bf 310} {(1988)} {163}.
\bibitem{Bytsenko} {Bytsenko~A.~A., Cognola~G., Vanzo~L. and Zerbini~S.}
{Phys. Rep.} {\bf 266} {(1996)} {1}.
\bibitem{Bytsenko-1} {Bytsenko~A.~A., Cognola~G., Elizalde~E., Moretti~V. and
Zerbini~S.} {\it Analytic Aspects of Quantum Fields} (World
Scientific, Singapore) 2003.
\bibitem{Od-3} {Elizalde~E., Odintsov~S.~D. and Romeo~A.} {J. Math.
Phys.} {\bf 37} {(1996)} {1128}.
\bibitem{DC} {Dowker~J.~S. and  Critchley~R.}
{Phys. Rev. D  } {\bf  13} {(1976)} {3224}.
\bibitem{Hawking} {Hawking~S.~W.,}
{Commun. Math. Phys.  } {\bf  55} {(1977)} {133}.
\bibitem{WW}{Whittaker~E.~T. and Watson~G.~N.,}
{\it A Course of Modern Analysis} (Cambridge University Press)
1969.
\bibitem{blackholes}  {Cognola~G, Vanzo~L. and
Zerbini~S.,} {Phys. Rev. D  } {\bf  52} {(1995)} {4548}; {Zerbini~S.,
Congola~G. and Vanzo~L.,} {Phys. Rev. D} {\bf    54} {(1996)} {2699};
{Cheeger~J.,} {J. Diff. Geom.  } {\bf  18} {(1983)} {576}.
\bibitem{Od-1} {Milton~k., Odintsov~S.~D. and Zerbini~S.,}
{Phys. Rev.,  D} {\bf 65} {(2002)} {065012}.
\bibitem{cosmicstr} {Frolov~V.~P. and Serebriany~E.~M.,}
{Phys. Rev., D} {\bf  35} {(1987)} {3779}.
\bibitem{wedge} {Nesterenko~V.~V., Lambiase~G. and Scarpetta~G.,}
{Ann. Phys. (N.Y.) } {\bf   298} {(2002)} {403}.
\bibitem{CVZ} {Cognola~G., Vanzo~L. and
 Zerbini~S.}, {J. Math. Phys.}{\bf 33} {(1992)} {222}.
\bibitem{BS} {Beneventano~C.~G. and Santangelo~E.~M.}, {Int. J. Mod. Phys. A}
{\bf 11} {(1996)} {2871}.
\bibitem{GR} {Gradshteyn~I.~S. and Ryzhik~I.~M.,}
{\it Table of Integrals, Series, and Products}, 6th ed. (Academic
Press, New York) 2000.
\bibitem{Moss} {Moss~I.~G.,}
{Class. Quantum Grav. } {\bf    6} {(1989)} {759}; {Moss~I.~G. and
Poletti~S.~J.,} {Phys. Lett. B } {\bf   333} {(1994)} {326}.
\bibitem{Esposito}
{D'Eath~P.~D. and Esposito~G.~V.~M.,} {Phys. Rev. D} {\bf    43}
{(1991)} {3234}; {D'Eath~P.~D. and Esposito~G.~V.~M.,} {Phys. Rev. D}
{\bf 44} {(1991)   1713}; {Esposito~G.}, {\it Quantum Gravity,
Quantum Cosmology and Lorentzian Geometries} (Springer, Berlin) 1994.
\bibitem{GEKK} {Esposito~G. and  Kirsten~K.,}
{Phys. Rev., D} {\bf 66} {(2002)} {085014}.
\bibitem{NPD} {Nesterenko~V.~V., Pirozhenko~I.~G. and Dittrich~J.,}
{Class. Quantum Grav. } {\bf   20} {(2003)} {431}.
\bibitem{Milton-3} See  the article due to
{\sc Milton~K.~A.} in ref.~\cite{Bordag-W4} pp. 20 -- 36.
\bibitem{NP}{Nesterenko~V.~V. and Pirozhenko~I.~G.,}
{Phys. Rev. D } {\bf   57} {(1998)} {1284}.
\bibitem{HdP1}
{Borgnis~E., Papas~C.~H.}, {\it Electromagnetic Waveguides and
Resonators}, In {\it Encyclopedia of Physics}, edited by
{S.~Fl\"uge} (Springer Verlag, Berlin) 1958; v.\ XVI, pp.\
285--422.
\bibitem{Jackson} {Jackson~J.~D.},
{\it Classical Electrodynamics}, 2nd edn (Wiley, New York) 1975.
\bibitem{RR} {Rubinstein~I.,  Rubinstein~L.}
{\it Partial Differential Equations in Classical
 Mathematical Physics}
(Cambridge University Press, Cambridge, UK) 1998.
\bibitem{Stratton} {Stratton~J.~A.,}
{\it Electromagnetic Theory}, (McGraw--Hill, New York) 1941.
\bibitem{Debye-ball} {Debye~P.,}
 {Ann. d. Phys.  } {\bf  30} {(1909)} {57}.
\bibitem{LNB} {Lambiase~G., Nesterenko~V.~V. and Bordag~M.,}
{J.~Math. Phys. } {\bf   40} {(1999)} {6254}.
\bibitem{Adler} {Adler~S. and  Piran~T.,}
{Rev. Mod. Phys.  } {\bf  56} {(1984)} {1}.
\bibitem{FGK}{Fishbane~P., Gasiorowics~S. and Kaus~P.,}
{Phys. Rev. D }{\bf   36} {(1987)} {251}; {\it ibid.} {\bf 37}
{(1988)} {2623}.
\bibitem{Bag} {Chodos~A., Jaffe~R.~L., Thorn~C.~B. and Weisskopf~V.,}
{Phys. Rev. D }{\bf   9} {(1974)} {3471};
 {Hasenfratz~P. and Kuti~J.,} {Phys.
Rep. C } {\bf   40} {(1978)} {75}; {Hiller~J.~R.,} {Ann. Phys. (N.Y.)
} {\bf 144} {(1982)  { 58}; {Phys. Rev. D } {\bf   30} {(1984)}
{1520}.
\bibitem{Milton-bag} {Milton~K.~A.,}
{Phys. Rev. D } {\bf   22} {(1980)} {1441}; {\it ibid.} {\bf 22}
{(1980)} {1444}.
\bibitem{Bender} {Bender~C.~M. and Hays~P.,}
{Phys. Rev. D } {\bf   14} {(1976)} {2622}.
\bibitem{BEKL}   Bordag~M., Elizalde~E.,  Kirsten~K. and Leseduarte~ S.,
{Phys. Rev. D} {\bf    56} (1997) {4896}.
\bibitem{EBK}{Elizalde~E., Bordag~M. and  Kirsten~K.,}
{J. Phys. A }{\bf    31} {(1998)} {1743}.
\bibitem{LP}{Landau~L.~D. and Peierls~R.,}
{Zs. Physik  } {\bf  62} {(1930)} {188}.
\bibitem{Bordag-95} {Bordag~M.,}
{J.~Phys. A  } {\bf  28} {(1995)} {755}.
\bibitem{ELR}{Elizalde~E., Leseduarte~S. and Romeo~A.,}
{J. Phys. A } {\bf   26} {(1993)} {2409}.
\bibitem{LR}{Leseduarte~S. and Romeo~A.,}
{J. Phys. A} {\bf    27} {(1994)} {2483}.
\bibitem{Magnus} {Magnus~W., Oberhettinger~F. and Soni~R.~P.,}
{\it Formulas and Theorems for the Special Functions of
Mathematical Physics}, 3rd ed. (Springer Verlag, New York) 1966.
\bibitem{Olver} {Olver~F.~W.~J.,}
{Phil. Trans. Royal Soc. London A} { \bf    247} {(1954)} {328}.
\bibitem{AS} {Abramowitz~M. and Stegun~I.~A.} (eds)
{\it Handbook of Mathematical Functions} (New York, Dover) 1972.
\bibitem{Cas-electron} {Casimir~H.~B.~G.,}
{Physica } {\bf   19} {(1956)} {846}.
\bibitem{MNg1} {Milton~K.~A. and Ng~Y.~J.,}
{Phys. Rev. E  } {\bf  55} {(1997)} {4207}.
\bibitem{cylinder+circle}
{Nesterenko~V.~V. and Pirozhenko~I.~G.,} {J.~Math. Phys. } {\bf   41}
{(2000)} {4521}.
\bibitem{LesRom} {Leseduarte~ S. and  Romeo~A.,}
{Ann. Phys. (N.Y.) } {\bf   250} {(1996)} {448}.
\bibitem{Sen} {Sen~S.,}
{J.~Math. Phys. } {\bf   22} {(1981)} {2868}; {\it ibid.} {\bf 25}
{(1984)} {2000}.
\bibitem{BenderMilton}  {Bender~C.~M. and  Milton~K.~A.,}
{Phys. Rev. D } {\bf   50} {(1994)} {6547}; {Milton~K.~A.,} {Phys.
Rev. D} {\bf  55} {(1997)} {4940}.
\bibitem{CEK} {Cognola~G., Elizalde~E. and Kirsten~K.,}
{J. Phys. A } {\bf   34} {(2001)} {7311}.
\bibitem{Bordag-db}{Bordag~M., Kirsten~K. and Vassilevich~D.,}
{Phys.\ Rev.\ D } {\bf   59} {(1999)} {085011}.
\bibitem{MSS} {Milonni~P.~W., Schaden~ M. and Spruch~ L.,}
{Phys.\ Rev.\ A} {\bf    59} {(1999)} {4259}.
\bibitem{Barton-dis} {Barton~G.,}
{J.~Phys.~A } {\bf   34} {(2001)} {4083}; {\it ibid.} {\bf 34}
{(2001)} {5781}.
\bibitem{sono} {Schwinger~J.,}
{\it Proc. Natl. Acad. Sci. USA}, {\bf 89} {(1992)} 4091, 11118;
               {\bf 90} (1993)  958,  2105,  4505, 7285;
               {\bf 91} (1994) 6473.
\bibitem{High-T} {Bordag~M., Nesterenko~V.~V. and Pirozhenko~I.~G.,}
{Phys. Rev. D } {\bf   65} {(2002)} {045011}.
\bibitem{HdP2} {Phillips~M.,}
{\it Classical Electrodynamics}, in {Encyclopedia of Physics},
edited by {S.~Fl\"uge} (Springer Verlag, Berlin) 1962, v.~IV, p.\
1-108.
\bibitem{Heyn} {Heyn~E.,}
{Math. Nachrichten  } {\bf 13} {(1955)} {25}.
\bibitem{s-cir} {Nesterenko~V.~V., Lambiase~G. and Scarpetta~G.,}
{J. Math. Phys. } {\bf   42} {(2001)} {1974}.
\bibitem{Dowker} {Dowker~J.~S.,} preprint MUTP/2000/2;
arXive:  hep-th/0006138.
\bibitem{Candelas} {Candelas~P.,}
{Ann. Phys. (N.Y.) } {\bf   143} {(1982)} {241}; {\it ibid.} {\bf
167} {(1986)} {257};
\bibitem{FS1} {Ford~L.~H. and Svaiter~N.~F.,}
{Phys. Rev. D } {\bf   58} {(1998)} {065007}.
\bibitem{Sommer} {Sommerfeld~A.,}
{Proc. London Math. Soc.} {\bf    28} {(1897)} {417};
 {Carslaw~H.~S.}
 {Proc. London Math. Soc. } {\bf   20} {(1898)} {121};
 {Fedosov~B.~V.,}
{Sov. Mat. Dokl. } {\bf   4} {(1963)} {1092}; {\it ibid.} {\bf  5}
{(1964)} {988}; {Dowker~J.~S.,} {J. Phys. A } {\bf   10} {(1977)}
{115}; {Phys. Rev. D } {\bf    18} {(1978)} {1856}; {Cheeger~J.,} {J.
Diff. Geom.} {\bf  18} {(1983)} {575}.
\bibitem{HowTr2} {Howls~C.~J.and  Trasler~S.~A.,}
{J. Phys. A } {\bf   31} {(1998)} {1911}.
\bibitem{ZCV}
{Cognola~G., Kirsten~K. and Vanzo~L.,} {Phys. Rev. D } {\bf   49}
{(1994)} {1029}; {Zerbini~S., Cognola~G. and Vanzo~L.,} {Phys. Rev.
D} {\bf  54} {(1996)} {2699}; {Fursaev~D.~V.,} {Phys. Lett. B }{\bf
334} {(1994)} {53}; {Class. Quantum Grav } {\bf   11} {(1994)}
{1431}; {\it ibid.} {\bf 14} {(1997)} {1059}.
 \bibitem{FS} {Frolov~V.~P. and Serebriany~E.~M.,}
 {Phys. Rev. D } {\bf   35} {(1987)} {3779}.
 \bibitem{Wedge}{Neterenko~V.~V., Lambiase~G. and Scarpetta~G.,}
 {Ann. Phys. (N.Y.)  } {\bf  298} {(2002)} {403}.
\bibitem{W}  {Waechter~R.~T.,}  {Proc.\ Camb.\
Phil. Soc. } \bf   72} {(1972)} {439}.
\bibitem{Dow1} {Dowker~J.~S.,}
{Class. Quantum Grav. } {\bf    11} {(1994)} {557}; {Commun. Math.
Phys.} {\bf  162} {(1994)} {633}.
\bibitem{Lukosz} {Lukosz~W.,} {Physica  } {\bf  56} {(1971)} {109};
{Z. Phys.  } {\bf  258} {(1973)} {99}; {\it ibid.} {\bf 262} {(1973)}
{327}.
\bibitem{Rug} {Ruggiero~J.~R., Zimerman~A.~H. and Villani~A.,}
{Rev. Bras. Fis. } {\bf   7} {(1977)} {663}; {J.~Phys.~A } {\bf   13}
{(1980)} {761}.
\bibitem{Mam-Tr} {Mamaev~S.~G.  and Trunov~V.~V.,}
{Theor. Math. Phys. } {\bf   38} {(1979)} {228}.
\bibitem{AmWolf} {Ambj{\o}rn~J. and Wolfram~S.,} {Ann. Phys. (N.Y.) } {\bf   47} {(1983)} {1}.
\bibitem{Caruso} {Caruso~F., Neto~N.~P., Svaiter~B.~F. and Svaiter~N.~F.,}
{Phys. Rev. D } {\bf   43} {(1991)} {1300}.
\bibitem{Caruso-1} {Caruso~F., De Paola~R. and Svaiter~N.~F.,}
{Int. J. Mod. Phys. A } {\bf   14} {(1999)} {2077}.
\bibitem{Actor} {Actor~A.~A.,} {Ann. Phys. (N.Y.) } {\bf   230} {(1994)} {303}.
\bibitem{Actor-1} {Actor~A.~A. and Bender~I.,}  {Phys. Rev. D} {\bf    52} {(1995)} {3581}.
\bibitem{Li} {Li~X., Cheng~H. and Zhai~X.,}
{Phys. Rev. D } {\bf   56} {(1997)} {2155}.
\bibitem{Que} Queiroz~H., da Silva~J.~C., Khanna~ F.~C.,
Revzen~M. and Santana~A.~E., arXive: hep-th/0311246.
\bibitem{Ahmedov} {Ahmedov~H. and Duru~I.~H.,}
{J.~Math. Phys.} {\bf    45} {(2004)} {965}.
\bibitem{Ahmedov-1}   {Ahmedov~H. and Duru~I.~H.,}
{ J.~Math. Phys.} {\bf  46} (2005) 022303; {\it ibid.} 022304.
\bibitem{Mehra} {Mehra~J.,}
{Physica } {\bf   37} {(1967)} {145}.
\bibitem{Brown}
{Brown~L.S. and Maclay~G.~J.,} {Phys. Rev. } {\bf   184} {(1969)}
{1272}.
\bibitem{bag}{Francia~M.~De.,} {Phys. Rev. D}, {\bf 50} {(1994)} {2908}.
\bibitem{Od-2} {Brevik~I., Milton~K., Nojiri~S. and Odintsov~S.~D.,}
{Nucl. Phys., B}{\bf 599} {(2001)} {305}; {Brevik~I., Milton~K., and
Odintsov~S.~D.,} {Ann. Phys. (N.Y.),} {\bf 302} {(2002)} {120}.
\bibitem{FMR}{Feinberg~J., Mann~M. and
Revzen~M.,}
{Ann.\ Phys. (N.Y.),} {\bf 288} {(2001)} {103}.
\bibitem{Kapusta} {Kapusta~J.~L.,}
{\it Finite temperature field theory} (Cambridge University Press,
Cambridge, England) 1992.
\bibitem{Feynman} {Feynman~R.~P.,} {\it Statistical Mechanics}
(W.~A.~Benjamin, Inc. Advanced Book Program, Reading, Massachusetts) 1972.
\bibitem{DK} {Dowker~J.~S. and Kennedy~G.,}
 {J.~Phys.  A }  {\bf 11} {(1978)} {895}.
\bibitem{BKE} {Bordag~M., Kirsten~K. and  Elizalde~E.,}
{J.~Math. Phys. } {\bf   37} {(1996)} {895}.
\bibitem{PDG} Particle Data Group, {Groom~D. et al.,}
{\it Eur. Phys. J.~C}, {\bf 15}, No. 1\,\&\,4 (2000) 73.
\bibitem{Neapol} {Bordag~M., Nesterenko~V.~V. and Pirozhenko~I.~G.,}
 {Nucl. Phys. Proc. Suppl.   } {\bf  104} {(2002)} {228}.
 \bibitem{BGKE} {Bordag~M., Geyer~B., Kirsten~K. and Elizalde~E.,}
{Commun. Math. Phys. } {\bf   179} {(1996)} {215}.
 \bibitem{BGKE-1} {Bordag~M., Kirsten~K. and Dowker~S.,}
{Commun. Math. Phys. } {\bf   182} {(1996)} {371}.
\bibitem{Barton-5th} See the contribution by { Barton~G.}  in
ref.\ \cite{Bordag-W5}. {Marachevsky~V.~N.,} {Int. J.\ Mod. Phys. A}
{\bf  17} {(2002)} {786}; {Rezaeian~A.~H., Saharian~A.~A.,} {Class.
Quantum Grav. } {\bf   19} {(2002)} {3625}.
\bibitem{Balian}
{Balian~R. and Bloch~C.,} {Ann. Phys. (N.Y.)} {\bf    60} {(1970)}
{401}; {\it ibid.} {\bf 63} {(1971)} {592}; {\it ibid.} {\bf 64}
{(1971)} {271}; {\it ibid.} {\bf 69} {(1972)} {76}.
\bibitem{Francia}
{De~Francia~M.,} {Phys. Rev. D  } {\bf 50} {(1994)} {2908}.
\bibitem{Milton-contr}
{Milton~K.~A.} {Phys. Rev. D } {\bf   68} {(2003)} {065020}.
\bibitem{Milton-0401117} {Milton~K.~A.,}
{\it Finite Casimir Energies in Renormalizable Quantum Field
Theory}, Invited talk given at `Marcel Grossmann X' (Rio de
Janeiro, July 2003); hep-th/0401117.
\bibitem{Graham} {Graham~N., Jaffe~R.~L., Khemani~V.,
Quandt~M., Scandurra~M. and Weigel~H.,} {Phys. Lett. B} {\bf    572}
{(2003)} {196}.
\bibitem{Brevik-contr}
{Brevik~I., Jensen~B. and Milton~K.~A.,} {Phys. Rev. D} {\bf    64}
{(2001)} {088701}.
\bibitem{Hagen-contr} {Hagen~C.~R.,} {Phys. Rev. D } {\bf   61} {(2000)} {065005};
{Eur. Phys. J. C} {\bf    19} {(2001)} {677}; arXive: hep-th/0004079.
\bibitem{Fulling} {Fulling~S.~A.,}
{J. Phys. A  } {\bf  36} {(2003)} {6857}.
\bibitem{Bordag-rad} {Bordag~M., Robaschik~D. and Wieczorek~E.,}
{Ann. Phys. (N.Y.) } {\bf   165} {(1985)} {192}.
\bibitem{B+Shir} {Bogoliubov~N.~N., Shirkov~D.~V.,}
{\it Introduction to the Quantum Fields} (Interscience, New York)
1959.
\bibitem{Od-4}
{Brevik~I., Granda~L. and Odintsov~S.~D.,} {Phys. Lett., B} {\bf 367}
{(1996)} {206}.
\bibitem{Cher}
Cherednikov~I.~O., {Acta. Phys. Slov.,} {\bf 52} {(2002)} {221};
arXive: hep-th/0206245.
\bibitem{Gies-wl} {Gies~H., Langfeld~K., Moyaerts~L.,}
{JHEP} {\bf  0306} {(2003)} {018}.
\bibitem{optic} Jaffe~R.~L., Scardicchio~A.,
{Phys. Rev. Lett.}\ {\bf 92}\ (2004)\ {070402}.
\bibitem{comp} Pirozhenko~I.~G., Nesterenko~V.~V., Bordag~M., J.~Math.\ Phys., {\bf 46}
 (2005) 042305.
\end{thebibliography}
\end{document}